\documentclass[review]{elsarticle}

\usepackage{amssymb}
\usepackage{lineno,hyperref}
\usepackage{caption}
\usepackage{subcaption}
\modulolinenumbers[1]
\usepackage{setspace} 
\usepackage{graphicx}
\usepackage[margin=1in]{geometry}
\usepackage{nicematrix}
\usepackage{multirow}

\journal{Nuclear Fusion}

\begin{document}
	
\begin{frontmatter}

\title{The effect of neon seeding on plasma edge transport in EAST}


\author[affiliation1,affiliation2,affiliation3]{Dieter Boeyaert \corref{mycorrespondingauthor}}
\cortext[mycorrespondingauthor]{Corresponding author}
\ead{boeyaert@wisc.edu}

\author[affiliation2]{Stefano Carli}

\author[affiliation2]{Wouter Dekeyser}


\author[affiliation3,affiliation4]{Sven Wiesen}

\author[affiliation5]{Liang Wang}

\author[affiliation5]{Fang Ding}

\author[affiliation5]{Kedong Li}

\author[affiliation3,affiliation5]{Yunfeng Liang}

\author[affiliation2]{Martine Baelmans}

\author{EAST-Team$^\textrm{1}$\footnote{$^\textrm{1}$See appendix of B.N. Wan et al., Nucl. Fusion 59 (2019) 112003}}

\address[affiliation1]{University of Wisconsin-Madison, Department of Nuclear Engineering and Engineering Physics, 1500 Engineering Drive, Madison, WI 53706, United States of America}
\address[affiliation2]{KU Leuven, Department of Mechanical Engineering, Celestijnenlaan 300, 3001 Leuven, Belgium}
\address[affiliation3]{Forschungszentrum Jülich GmbH, Institut für Energie- und Klimaforschung – Plasmaphysik, 52425 Jülich, Germany}
\address[affiliation4]{DIFFER - Dutch Institute for Fundamental Energy Research, De Zaale 20, 5612 AJ Eindhoven, the Netherlands}
\address[affiliation5]{Institute of Plasma Physics, Chinese Academy of Sciences, Hefei 230031, People’s Republic of China}

\begin{abstract}
The effect of neon seeding on different transport mechanisms in EAST is investigated by analyzing SOLPS-ITER simulations. By evaluating the agreement between experimental observations and the performed simulations, four simulations are selected for a detailed analysis. In this analysis, it is shown that the presence of neon reduces the influence of drifts on the simulated profiles. In the simulation results, double peaked profiles/profiles with two valleys are observed at the divertor targets which can be explained by the parallel drift velocities These drifts move particles from the outboard towards the inboard side and, in that way, also increase the ionization sources at the inboard side. It is shown that Ne$^+$ leaks towards the core making it difficult to perform experiments which contain as much neon as in the SOLPS-ITER simulations. In fact, the level of neon in the experiments is limited by the HL backtransition which takes place if higher order states of neon ionize in the core and cause in that way too much core radiation. Furthermore, the analysis of the radiated power profiles suggests that the presence of other radiators besides neon is important to bring the experiments into detachment. The ionization of deuterium is the most important neutral reaction present in the simulations. The amount of ionized deuterium is decreased when large amounts of neon are present and the anomalous transport is modified. Therefore, it is concluded that for the analyzed simulations, neon increases the radiated power fraction, decreases the deuterium ionization, increases the neutral friction, but does not manage to cause significant influence of deuterium recombination. As a consequence, volumetric recombination only plays a minor role in the studied simulations.
\end{abstract}


\end{frontmatter}


\section{Introduction}

For future fusion devices, it is a challenge to control the heat and particle exhaust 
\cite{pacher2007modelling}. In order to keep the wall loads limited, a detached plasma state is required \cite{wiesen2017plasma}. Plasma is detached by modifying edge transport as such that the plasma particles lose a large part of the energy and momentum they are carrying before interacting with the divertor targets. The reduced temperature of the plasma makes that the plasma facing components (PFC) are shielded (detached) from the hot plasma by a small neutral layer
\cite{fenstermacher1999physics, leonard2018plasma, krasheninnikov2017physics}. 

\begin{figure}
	\centering
	\medskip
	\includegraphics[width=4cm]{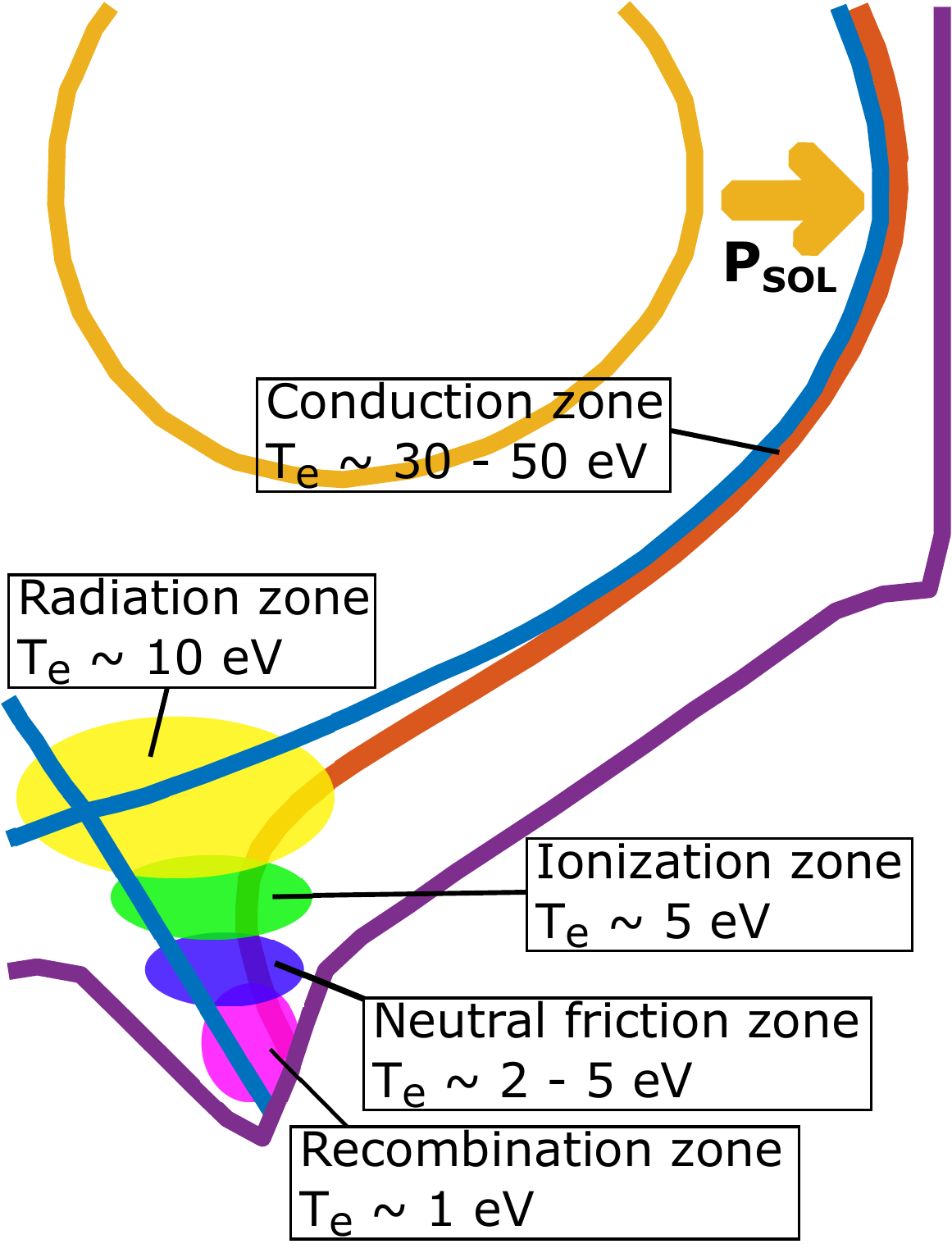}
	\caption{The path that the plasma particles and power follow while traveling from the core towards the divertor targets. The physical processes which take place on its way determine the plasma state.}
	\label{fig:detachment_physics}
\end{figure}

Figure \ref{fig:detachment_physics} shows the physics which happens when a particle travels from the upstream location at the outer midplane (OMP) -- where most plasma particles escape from the core -- to the downstream location at the divertor targets
. In case no interaction with other particles is taking place, plasma transport between up- and downstream is determined by convection and conduction. When the plasma quantities at the OMP and at the divertor targets are similar, the plasma transport is mainly dominated by convection \cite{stangeby2000plasma}. When the density is increased, the particle flux towards the divertor targets increases as well making the conduction dominant over the convection. During detachment, the upstream plasma density is getting saturated causing that the flux towards the divertor targets starts to decrease again, it experiences the so-called rollover. Several physic effects play a role in this rollover as indicated in figure \ref{fig:detachment_physics}: radiation, ionization, neutral friction and volumetric recombination. The energy loss due to radiation is caused by impurities. 
By employing impurities radiating around the X-point, the plasma temperatures are reduced when the particles enter the divertor region. The temperature of the plasma determines where the ionization of neutrals (resulting from plasma wall interaction - PWI) is taking place. Due to the impurity radiation, the ionization zone forms in between the X-point and the divertor target. This results in a poloidal plasma flow towards the targets. As the neutral density keeps increasing when coming closer to the divertor target, neutral friction becomes important. Due to friction between the ions and neutrals, the plasma particles lose parallel momentum, the flow velocity towards the targets is reduced, and the rollover can happen. Close to the divertor target, where the low flow velocity causes that the transit time is similar to the recombination time, volumetric recombination takes place. Depending on the plasma temperatures and densities along this path, all these transport mechanisms happen, or only some of them play a role.

Extrinsic impurity seeding is often used to induce detachment as it helps to increase the radiation losses \cite{kallenbach2013impurity}. Impurities with a lack of surface chemistry like noble gases or low-Z elements are key candidates as they radiate at temperatures which are achieved in the scrape-off layer (SOL) and pedestal of the plasma.

An interesting low-Z element, is neon (Ne). Recently, large tokamaks like JET have shown to handle Ne seeding in H-mode in a metallic environment \cite{gloggler2019characterisation} and Ne is considered for ITER and DEMO \cite{kallenbach2013impurity}. Apart from JET, EAST (Experimental Advanced Superconducting Tokamak) is the only tokamak with a tungsten divertor on which Ne-seeding detachment experiments have been performed successfully without adding other radiators. EAST is a superconducting tokamak located at ASIPP in Hefei, China \cite{wu2007overview}. It has a major radius of $\mathrm{1.7 \, m}$ and a minor radius of $\mathrm{0.4 \, m}$. During the experimental campaign of 2019, EAST was equipped with a tungsten upper divertor and a carbon lower divertor. 

The current paper analyzes more in detail the effect of Ne seeding on the different plasma edge transport mechanisms of figure \ref{fig:detachment_physics}. As experimental data are limited, especially to investigate the influence of neon on the different processes, the focus is rather on SOLPS-ITER simulations. SOLPS-ITER is a 2D plasma edge code coupling the B2.5 code solving the Braginskii equations for the ions and electrons, to the EIRENE code solving the Boltzmann equations for the neutrals \cite{wiesen2015new,bonnin2016presentation}. Two discharges, a reference discharge in H-mode (shot number 87626) and a similar discharge with the maximum added amount of Ne to keep the discharge in H-mode (shot number 87628), are taken as a starting point. The experimental data are used to constrain the simulation setup, where the remaining analysis is mainly based on the simulations themselves. 

Section \ref{sec:case_desbription} describes the investigated experiments and the setup for the SOLPS-ITER modeling. The up- and downstream profiles are discussed in section \ref{sec:up_downstream_profiles}. A description of the effect of drifts on the plasma quantities in section \ref{sec:drifts} is followed by a study of the effect of neon on the radiated power fraction in section \ref{sec:radiated_power}. Section \ref{sec:neutral_transport} discusses the impact of neutrals on the plasma transport. The final section discusses the overall effects of neon seeding on the plasma before coming to the summary and conclusion.


\section{Case description}
\label{sec:case_desbription}

\begin{figure}
	\centering
	\begin{subfigure}{6cm}
		\centering
		\medskip
		\includegraphics[width=6cm]{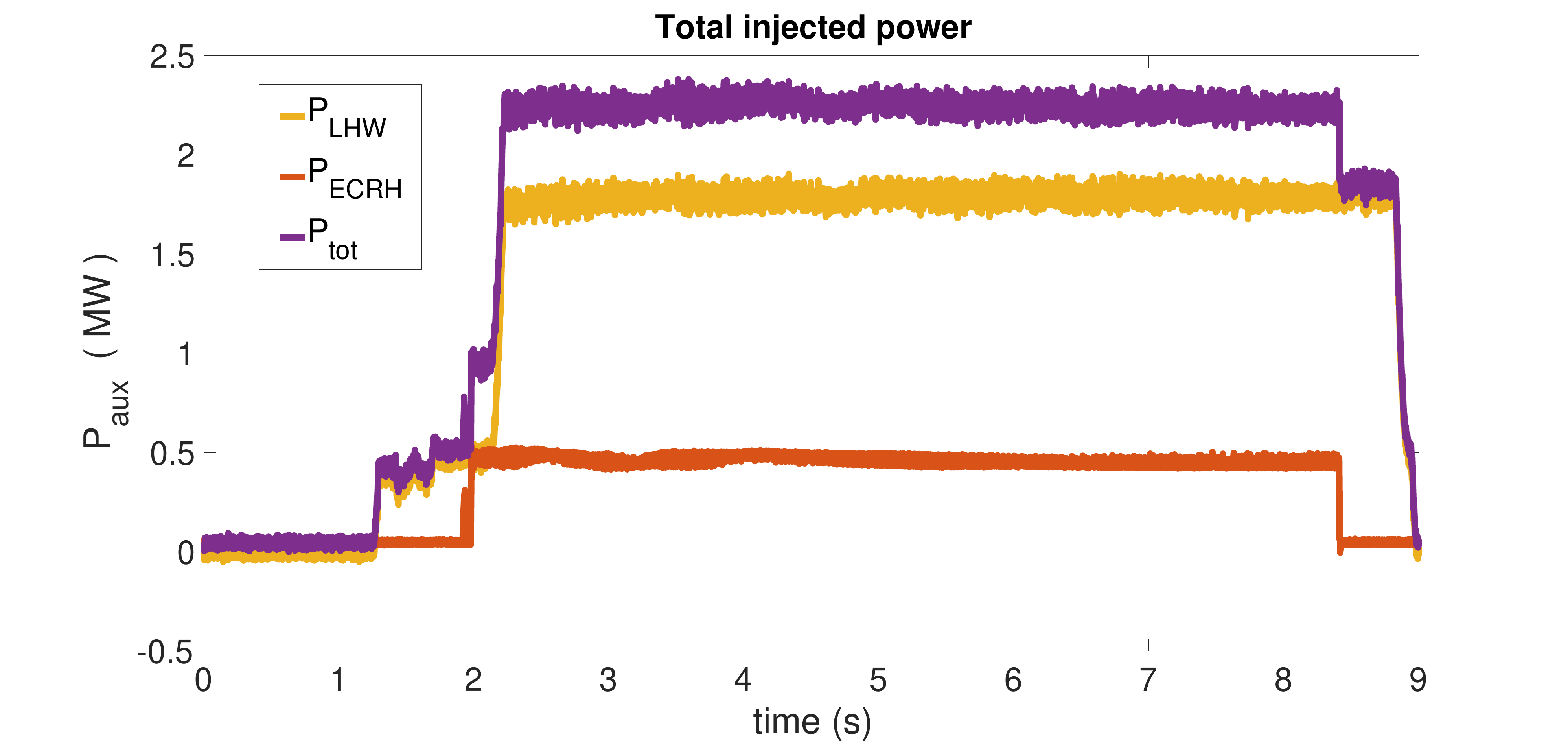}
		\caption{The injected power during the discharges. A total power of $\sim \mathrm{2.25 \, MW}$ was injected existing of $\sim \mathrm{1.80 \, MW}$ LHW and $\sim \mathrm{0.45 \, MW}$ ECRH.}
		\label{subfig:power}
	\end{subfigure}
	\begin{subfigure}{6cm}
		\centering
		\medskip
		\includegraphics[width=6cm]{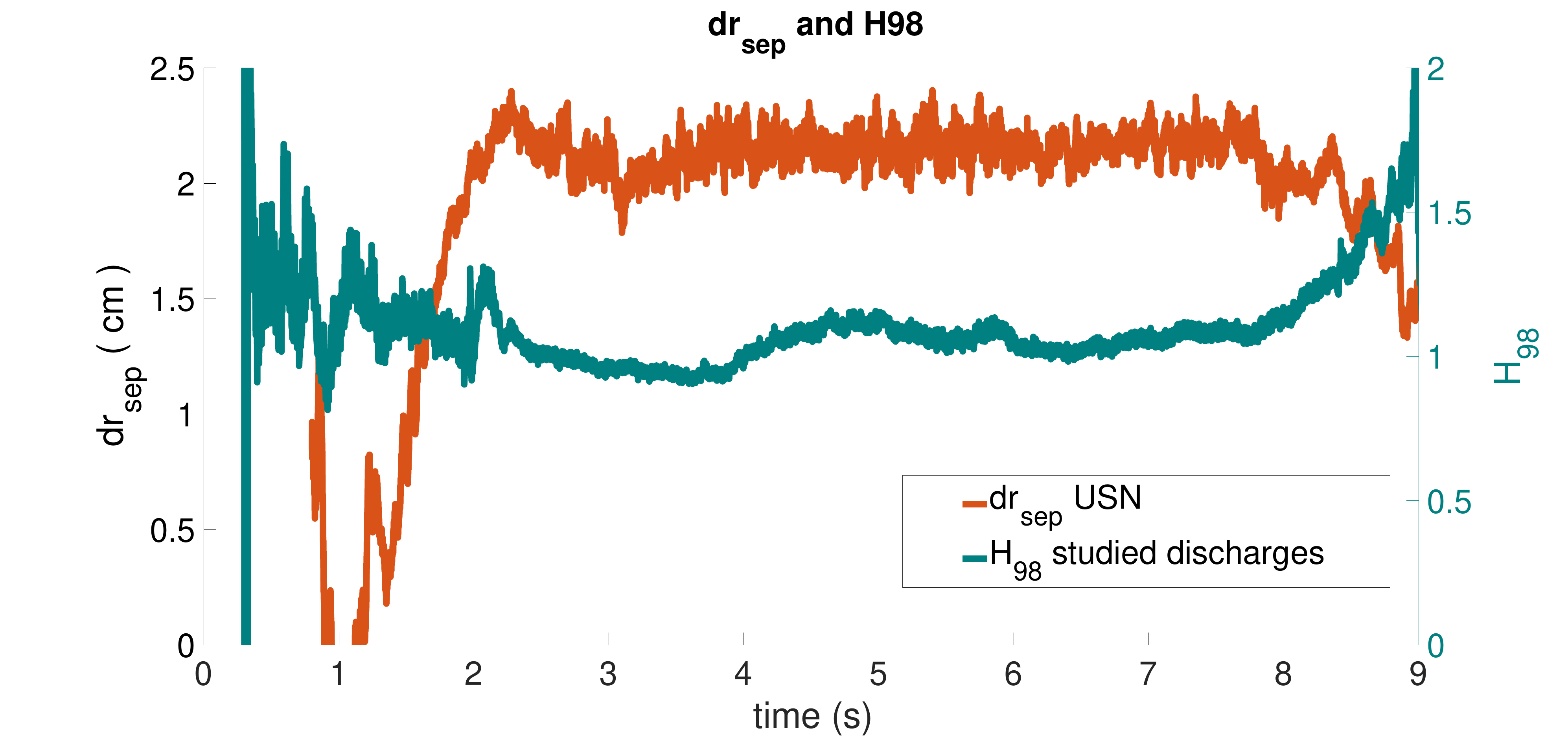}
		\caption{$\mathrm{dr_{sep}}$ during the performed discharges (left axis) and the H98 profile which was similar for all discharges (right axis).}
		\label{subfig:drs_H98}
	\end{subfigure}
	\begin{subfigure}{6cm}
		\centering
		\medskip
		\includegraphics[width=6cm]{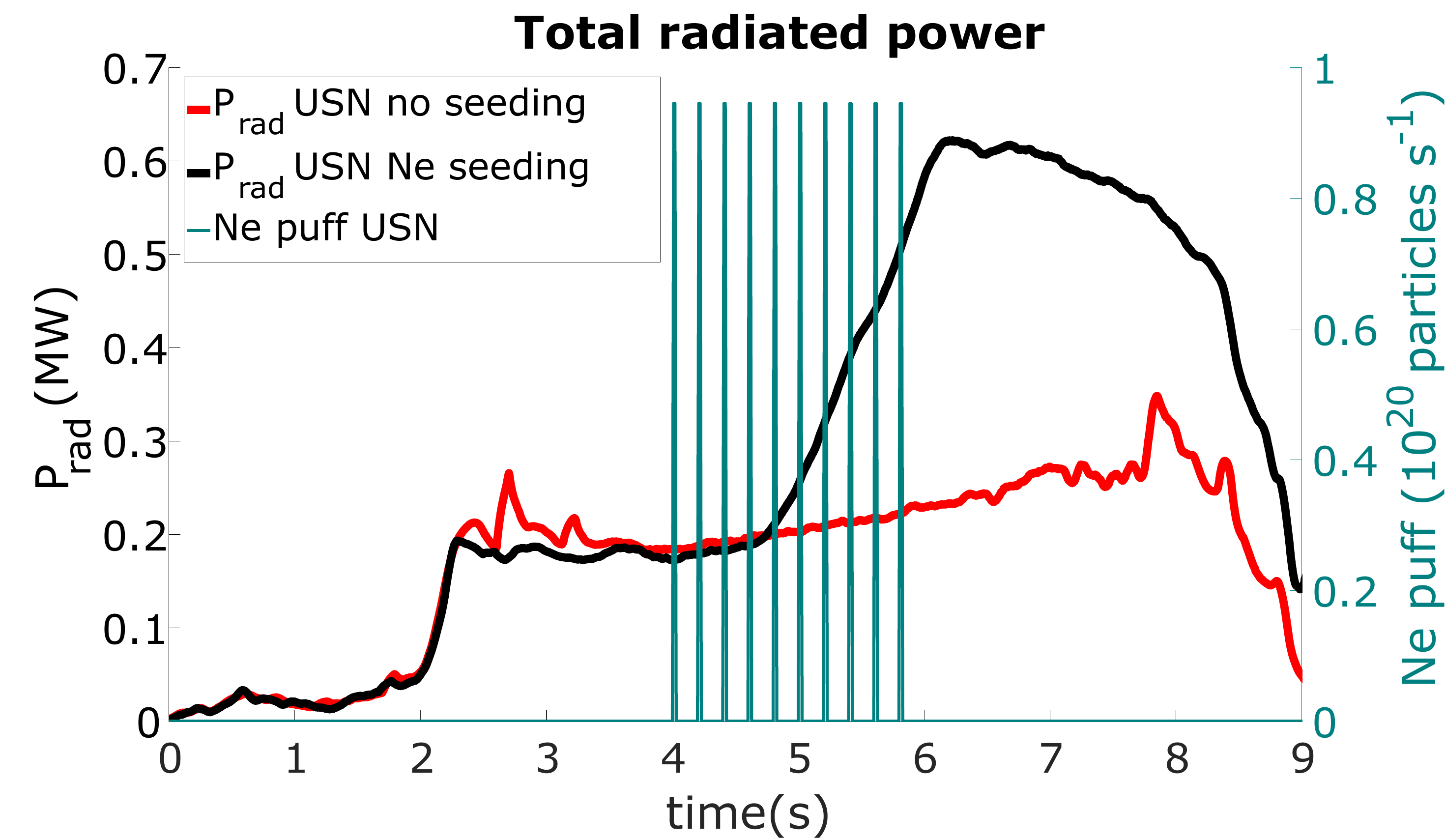}
		\caption{The core radiated power fraction measured with the AXUV system for the reference discharge (no seeding) and during Ne seeding. On the right axis the injected amount of Ne is displayed.}
		\label{subfig:AXUV}
	\end{subfigure}
	\begin{subfigure}{6cm}
		\centering
		\medskip
		\includegraphics[width=6cm]{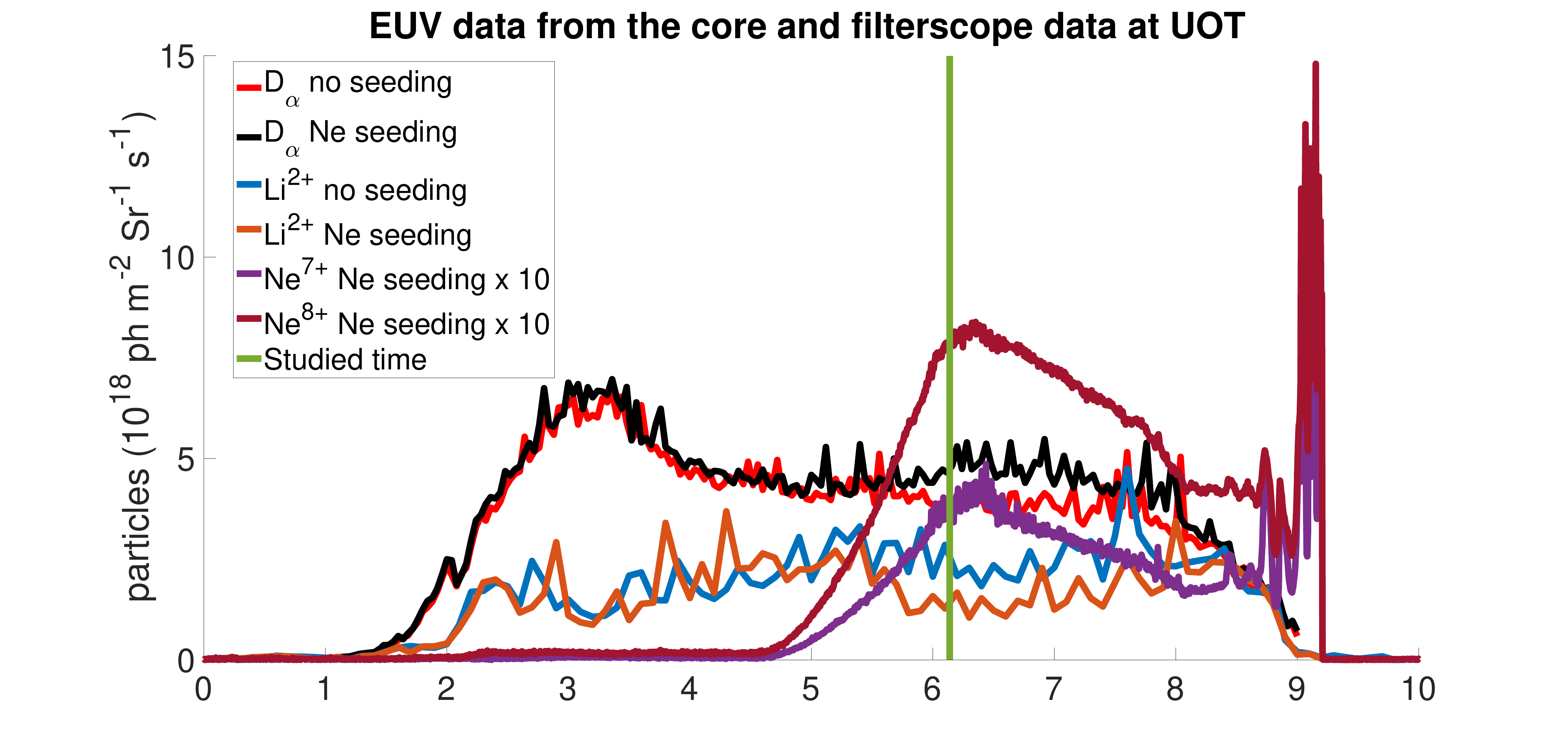}
		\caption{The line emission at the UOT measured with the filterscope for $\mathrm{D_{\alpha}}$ and $\mathrm{Li^{2+}}$ for the reference discharge (no seeding) and during Ne seeding. The line emission from the core measured with the EUV is shown for $\mathrm{Ne^{7+}}$ and $\mathrm{Ne^{8+}}$ during the neon seeded discharge. The investigated time slot in both experiments is indicated in green.}
		\label{subfig:filterscope}
	\end{subfigure}
	\caption{Overview of the main experimental parameters for the studied discharges in this paper.}
	\label{fig:overview_discharges}
\end{figure}

The performed analysis is based on two EAST discharges in upper single null (USN) configuration, performed during the campaign of 2019, and described in ref. \cite{boeyaert2021towards}. The two investigated discharges (87626 without seeding and 87628 with Ne seeding) are H-mode experiments to which a total heating power of 2.25 MW was applied. The magnetic field was 2.5 T and the plasma current 400 kA. Deuterium was injected at the OMP using a feedback loop to keep the electron density at the separatrix constant ($n_{e,sep} \sim 0.9 \cdot 10^{19} \mathrm{m^{-3}}$). Due to the used control system for the magnetic field, USN discharges in EAST are in fact disconnected double null (DDN) discharges with a large separation $\mathrm{dr_{sep}}$. In the studied experiments $\mathrm{dr_{sep} \sim 2.5 \, cm}$. In shot 87628 Ne seeding was added in a $50 \, \%$ Ne, $50 \, \%$ $\mathrm{D_2}$ mixture at the strike point in the upper outer target (UOT). 10 pulses of this Ne-$\mathrm{D_2}$ mixture were added and afterwards the plasma state (discharge 87628) is compared with the case in which no seeding was added (discharge 87626). The length of a Ne injection is in total 10 ms, from which 5 ms flattop. An overview of the main discharge parameters is given in figure \ref{fig:overview_discharges}.

\begin{figure}
	\centering
		\centering
		\medskip
		\includegraphics[width=6cm]{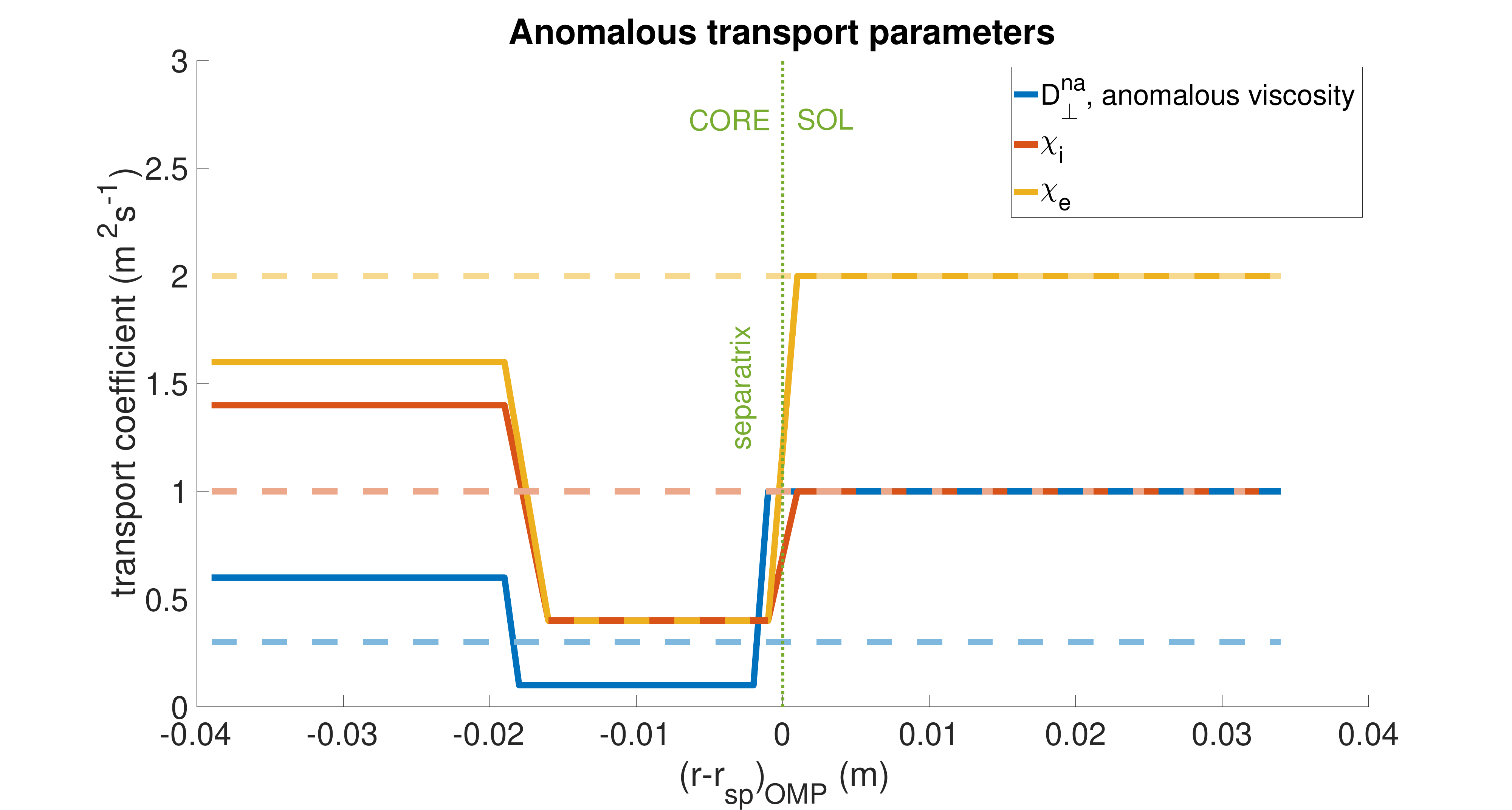}
	\caption{Overview of the transport coefficients used in the presented SOLPS-ITER simulations. The transport coefficients in the divertor and private flux regions are shown with dotted, more transparent curves.}
	\label{fig:transport}
\end{figure}

The setup for the SOLPS-ITER simulations is based on the one in refs. \cite{boeyaert2021towards,boeyaert2022numerical,boeyaert2024numerical}. As in these references, the 3.0.7 master version of the code is used. Although that the experiments are in fact carried out in a DDN configuration, the simulations use a USN setup and are limited to 2.5 cm outside the last closed flux surface. 
At the core boundary of the plasma grid, the density is imposed for the continuity equation of the deuterium ions: $n_i = 2.8 \cdot 10^{19} m^{-3}$ based on the OMP profiles of the Thomson Scattering (TS) system of EAST \cite{zang2011upgraded} discussed in section \ref{sec:up_downstream_profiles}. The simulated Ne puff is located at the strikeline location at the UOT. Figure \ref{subfig:AXUV} indicates that 10 puffs of 10 ms each (5 ms flattop) with a puffed Ne fraction of $10^{20}$ particles s$^{-1}$. As this is a pulsed injection and as SOLPS-ITER is a steady-state code, these puffs are added to come to a total injection of $\sim7.2$ $10^{18}$ particles s$^{-1}$. This assumes that all Ne has an effect at the same time. As the Ne is injected in a Ne-$\mathrm{D_2}$ mixture, a similar puff for $\mathrm{D_2}$ is added to the Ne-seeded simulation (SOLPS Ne 1). From the neon ions there is no exact measurement of each ion individually. As the TS profiles do not show a large increase in electron density at the core boundary of the grid, it is assumed that the lower charge states of neon are fully ionized and that their density is zero. For Ne$^{5+}$ - Ne$^{8+}$ small amounts in the range of $n_i = 5.0 \cdot 10^{14} m^{-3}$  to $n_i = 2.0 \cdot 10^{17} m^{-3}$  are imposed at the core boundary. Higher charge states are assumed not to be present. The deuterium level at the OMP is imposed through a feedback loop ($n_{e,sep} = 0.9 \cdot 10^{19} \mathrm{m^{-3}}$) and kept similar for both simulations. Such a feedback loop makes the overall simulation and especially the inclusion of drifts more challenging from a numerical viewpoint \cite{boeyaert2024numerical}. The input power to the simulation is imposed as a boundary condition for the ion and electron energy equations. It is assumed that power is equally spread over ions and electrons. As boundary condition, the input power in the experiment, shown in figure \ref{subfig:power} is decreased with the radiated power fraction given in figure \ref{subfig:AXUV}. For the purely deuterium simulation, this results in an input power of 2.05 MW. As deuterium is radiating further downstream, no edge radiation within the viewing cords of the AXUV system are expected. Ne, on the other hand, will also radiate around the OMP. Therefore, not the full 0.6 MW of figure \ref{subfig:AXUV} is subtracted from the 2.25 MW input power, but only 0.45 MW is assumed to be core radiation resulting in an input power of 1.8 MW for the simulation. Figure \ref{subfig:filterscope} indicates that also other species are radiating. As the goal of the presented research is to investigate the effect of neon, these species are not included in the SOLPS-ITER simulations. At the targets, Bohm-Chodura conditions are applied and a recycling coefficient of one is imposed for all ions. More details about the used boundary conditions can be found in refs. \cite{boeyaert2022numerical,boeyaert2024numerical}. 
Anomalous transport is imposed ad-hoc according to the profiles of figure \ref{fig:transport}. Similar anomalous transport is assumed for all species. 
Grid information can be found in refs. \cite{boeyaert2022numerical, boeyaert2024numerical}.

\begin{figure}
	\centering
		\centering
		\medskip
		\includegraphics[width=6cm]{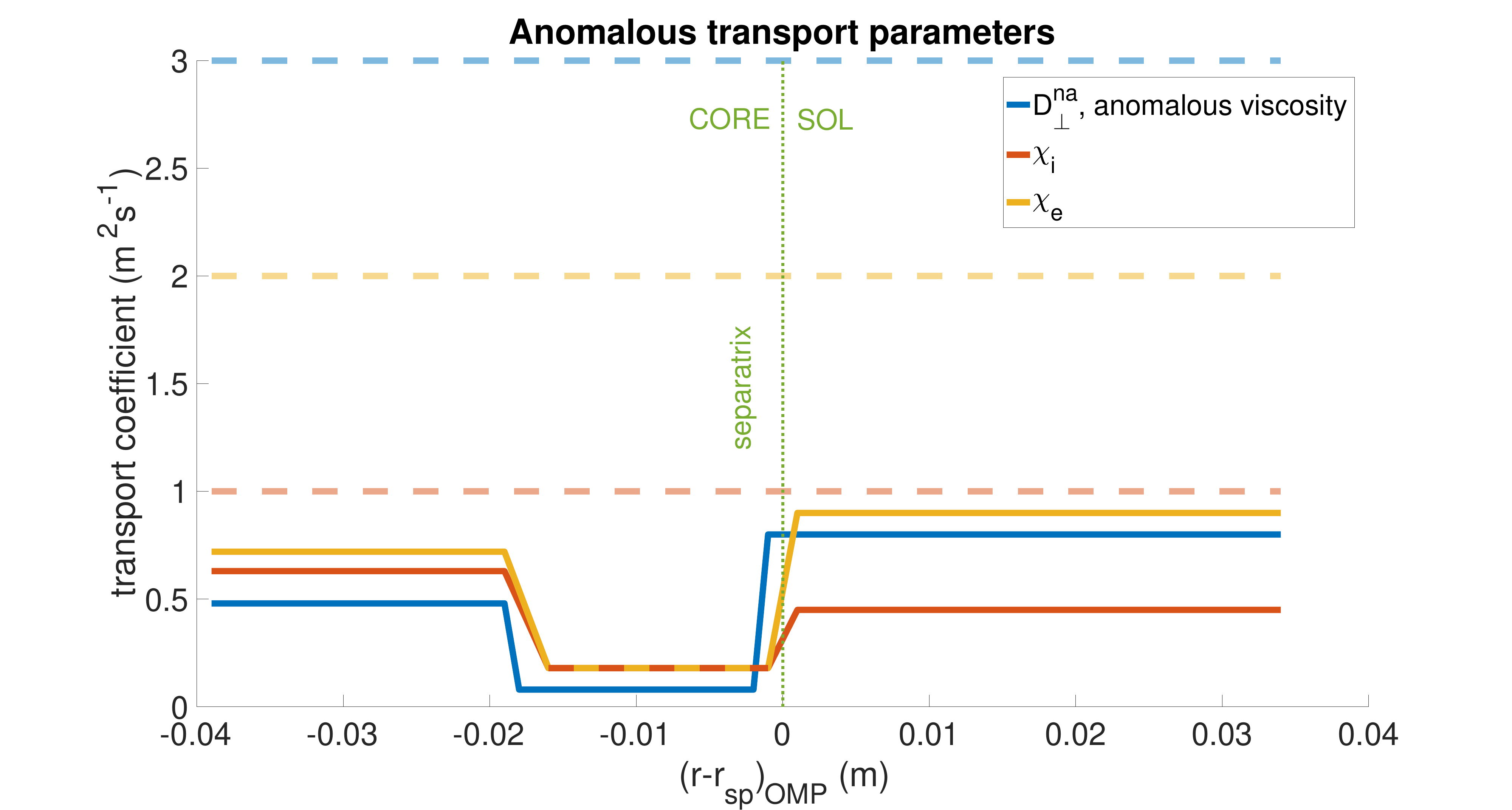}
	\caption{Overview of the transport coefficients used in the additional SOLPS-ITER simulation with neon impurities (SOLPS Ne 2). The transport coefficients in the divertor and private flux regions are shown with dotted, more transparent curves.}
	\label{fig:transport_2}
\end{figure}

In the performed analysis, also a second SOLPS-ITER simulation including Ne is studied (SOLPS Ne 2). For this later simulation, the focus is on better matching the up- and downstream profiles with the experimental observations. Therefore, the transport profiles of figure \ref{fig:transport} are replaced by the profiles in figure \ref{fig:transport_2}, the neon puff is increased with a factor 20, and also the upstream density is increased. As mentioned above, it is unlikely that this level of Ne can be achieved in an H-mode experiment, but this made it possible to achieve better downstream agreement between simulations and experiments as discussed in section \ref{sec:up_downstream_profiles}. 

As mentioned before, figures \ref{subfig:AXUV} and \ref{subfig:filterscope} show that not only neon is the radiating impurity. As deuterium cannot radiate at the temperatures expected in the core, other radiators have to be present. Figure \ref{subfig:filterscope} indicates the presence of lithium in the vicinity of the UOT. In fact, several impurity species are expected in EAST plasmas \cite{mao2017impacts}: the lower divertor, the shine-through area for the NBI, and the outboard guard limiters are from carbon, where the first wall is made from molybdenum. To reduce the impact of the carbon wall parts, but also to avoid hydrogen in a deuterium plasma, lithium wall conditioning is used to make long pulse operation possible. The experimental data show, however, that this results in at least carbon, lithium and tungsten in the main plasma. To account for these impurities, the artificial radiation switch in SOLPS-ITER is used. In the "SOLPS Ne 2" simulation, it is assumed that the concentration of this artificial radiator is 1 $\%$ in the plasma. The effective radiation is then determined by: $\mathrm{P_{rad,art.} = 0.01 \cdot n_a n_e L(T_e)}$ with $\mathrm{n_a}$ the deuterium density, $\mathrm{n_e}$ the electron density and $\mathrm{L_{T_e}}$ the radiative loss function which is built-in in SOLPS-ITER and based on a mixture of carbon-oxygen. To also account for this additional radiation in the purely deuterium simulations, an additional deuterium simulation with 1 $\%$ of this artificial radiator is performed.

This brings the total number of examined simulations to four: two with deuterium ("SOLPS $\mathrm{D_2}$ 1" and "SOLPS $\mathrm{D_2}$ 2") which are using the same simulation setup apart from the artificial radiation, one Ne simulation which has a similar SOLPS-ITER setup as the deuterium ones apart from the Ne-$\mathrm{D_2}$ injection (SOLPS Ne 1), and a further elaborated Ne simulation with artificial radiation, and several changes to the simulation setup to match the downstream profiles at the UOT (SOLPS Ne 2).

In SOLPS-ITER, the neutral kinetics are handled by the EIRENE code \cite{reiter2005eirene}. The included neutral reactions in the performed simulations are given in table \ref{tab:EIRENE_neutral_reactions}. To detach the plasma, some types of reactions will play a more important role than others, as discussed further in this paper. 
The reactions at the bottom of the table are only included in the neon simulations.


\begin{table}[h!]
	\centering
	\begin{tabular}{c | c |c}
		Reaction & Reaction type & Reference\\
		\hline
		D + D $\rightarrow$ D$_2^+$ + e & elastic collision & AMMONX H.2 R-H-H \\
		D + D$_2$ $\rightarrow$ D$_3^+$ + e & elastic collision &AMMONX H.2 R-H-H2\\
		D + e $\rightarrow$ D$^+$ + 2e & ionization & AMJUEL H.4,10 2.1.5 \\
		D + D$^+$ $\rightarrow$ D$^+$ + D & charge exchange & HYDHEL H.1,3 3.1.8\\
		D$_2$ + D $\rightarrow$ 3 D & elastic collision & AMMONX H.2 R-H2-H\\
		D$_2$ + D$_2$ $\rightarrow$ D$_2$ + D + D & elastic collision & AMMONX H.2 R-H2-H2\\
		e + D$_2$ $\rightarrow$ 2e + D$_2^+$ & ionization & AMJUEL H.4 2.2.9\\
		e + D$_2$ $\rightarrow$ e + D + D & dissociation &AMJUEL H.4 2.2.5g\\
		e + D$_2$ $\rightarrow$ 2e + D + D$^+$ & ionizing dissociation & AMJUEL H.4 2.2.10\\
		D$_2$ + D$^+$ $\rightarrow$ D$^+$ + D$_2$ & elastic collision & AMJUEL H.0,1,3 0.3T\\
		D$_2$ + D$^+$ $\rightarrow$ D$_2^+$ + D & charge exchange & AMJUEL H.2 3.2.3\\
		e + D$_2^+$ $\rightarrow$ e + D + D$^+$ & dissociation & AMJUEL H.4 2.2.12\\
		e + D$_2^+$ $\rightarrow$ 2e + D$^+$ + D$^+$ & ionizing dissociation & AMJUEL H.4 2.2.11\\
		e + D$_2^+$ $\rightarrow$ D + D & recombining dissociation & AMJUEL H.4,8 2.2.14\\		
		D$^+$ + 2e $\rightarrow$ D + $h \nu$& recombination & AMJUEL H.4,10 2.1.8\\
		\hline
		e + Ne$^{\mathrm{(x-1)+}}$ $\rightarrow$ e + Ne$^{\mathrm{x+}}$ + e & ionization & AMJUEL H.2 2.10B0 \\
		Ne$\mathrm{^{x+}}$ + e $\rightarrow$ Ne$\mathrm{^{(x-1)+}}$ & recombination &ADAS H.4,10 acd96,prb96\\
	\end{tabular}
	\caption{The included neutral reactions in the EIRENE mode for deuterium and neon.}
	\label{tab:EIRENE_neutral_reactions}
\end{table}

\section{Upstream and downstream profiles}
\label{sec:up_downstream_profiles}

Key parameters to study plasma edge transport in both simulations and experiments are the upstream plasma conditions at the OMP, and the downstream ones  at the inner and outer targets (UIT and UOT) \cite{stangeby2000plasma, pitcher1997experimental}. Therefore, the plasma quantities at these locations are examined for the performed experiments and simulations before studying the transport mechanisms themselves.

As mentioned in the previous section, the upstream data are measured with the TS system. For the downstream experimental data, the divertor Langmuir probes (DivLP) from EAST are used \cite{xu2016upgrade}. The installed DivLP in the upper divertor are triple probes. This means that the electron temperature ($T_e$), density ($n_e$), heat flux ($q_t$), and ion saturation current ($j_s$) can be determined based on the voltage at two pins and the current through the third pin. They are calculated using the following formulas:
\begin{equation}
	j_s = \frac{I_s}{A_{pr}}
	\label{eq:j_s}
\end{equation}
\begin{equation}
	T_e = \frac{V_{pb} - V_f}{log(2)} 
	\label{eq:T_e}
\end{equation}
\begin{equation}
	n_e = \frac{I_s}{eA_{pr}\sqrt{\frac{T_e+T_i}{m_i}}}
	\label{eq:n_e}
\end{equation}
\begin{equation}
	q_{||} = \gamma \frac{j_s T_e}{e},
	q_t = q_{||} sin(\theta)
	\label{eq:q}
\end{equation}
$I_s$ is the current measured on one probe tip, $V_{pb}$ and$V_f$ are the voltages measured on the other two tips. $A_{pr}$ is the surface area of the probe tip, and $\gamma$ the sheath heat transmission coefficient which is assumed to be seven \cite{stangeby2000plasma}. For the electron density and the heat flux, it is assumed that $T_e = T_i$. To enable later an easier comparison between simulations and experiments, the presented $q_t$ profiles from SOLPS-ITER use the same $q_t$ approximation as in equation \ref{eq:q}. The advantage of triple probes is that it gives a quick qualitative measurement \cite{chen1965instantaneous}. 
As EAST is used for long pulse operation, the probe tip area is sometimes damaged meaning that the exact probe tip surface area is not known making an exact measurement difficult. The measurement of $j_s$ is the most reliable as it depends on the current measurement but does not rely on the voltage measurements. The only difficulty is its dependence on the probe tip area. Measurements at locations from which it is believed the probe tip is damaged are not shown in the figures throughout this paper. Temperature measurements are the second most reliable as they depend on the two voltage measurements but not on the current measurement and probe tip area, where heat flux and density measurements depend on all measured values. In case voltages do not change when the plasma conditions vary significantly, or when they always have a zero output, the data are not shown in the figures throughout this paper.

As no ELM-filtering is available for the DivLP data, the median of five data points, 50 ms apart from each other around the studied time in the discharge is taken to avoid the influence of ELMs on the measured profiles. Error bars of 20 $\%$ are are assumed for all DivLP profiles \cite{xu2016upgrade}. They, however, do not count for problems with the voltage measurements or damaged probe tips. Similar ELM-filtering for the TS data is not possible as only few measurements per discharge are available.

Also on the simulations, error bars will be present due to numerical errors, modeling errors, etc. In ref. \cite{boeyaert2022numerical} the optimal numerical parameters are determined to minimize the numerical errors in simulations without drifts. Drifts cause numerical peaks in the simulation results making a dedicated error estimation impossible. However, by using the numerical parameters of ref. \cite{boeyaert2022numerical} it is aimed to minimize the numerical errors.

\begin{figure}
 	\centering
 	\medskip
 	\includegraphics[width=8cm]{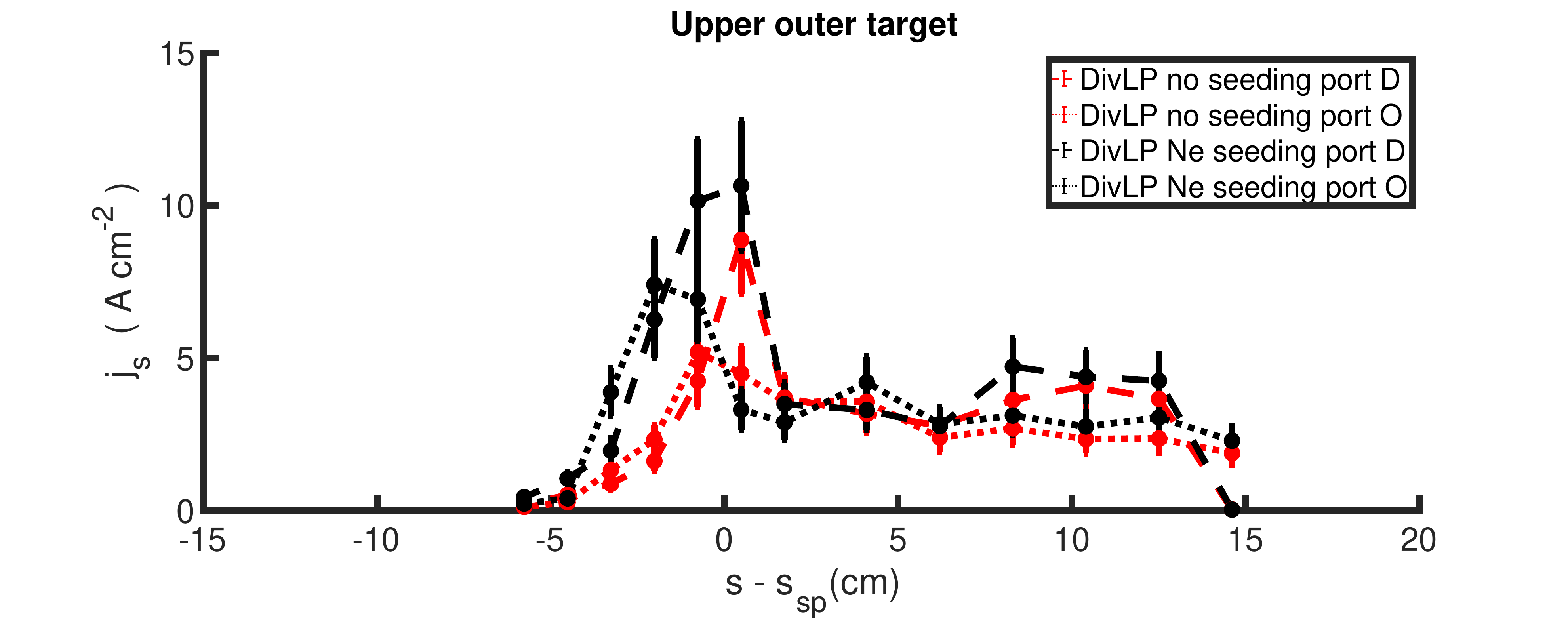}
 	\caption{The DivLP data for the ion saturation current at the upper outer target. Data are taken at two toroidal locations (port D and port O).}
 	\label{fig:DivLP_UOT_js}
 \end{figure}

 In the performed analysis toroidal symmetry is assumed. Figure \ref{fig:DivLP_UOT_js}, however, shows that an asymmetry between measurements at different ports is present. As the presented work aims to investigate the effect of Ne seeding on the plasma transport, rather than having an exact comparison between experiments and simulations, simulations are only compared to the experimental target data at port D. Port D was chosen as the initial comparison with the simulations in ref. \cite{boeyaert2021towards} was performed for this port.
 
\begin{figure}
	\centering
	\begin{subfigure}{6cm}
		\centering
		\medskip
		\includegraphics[width=6cm]{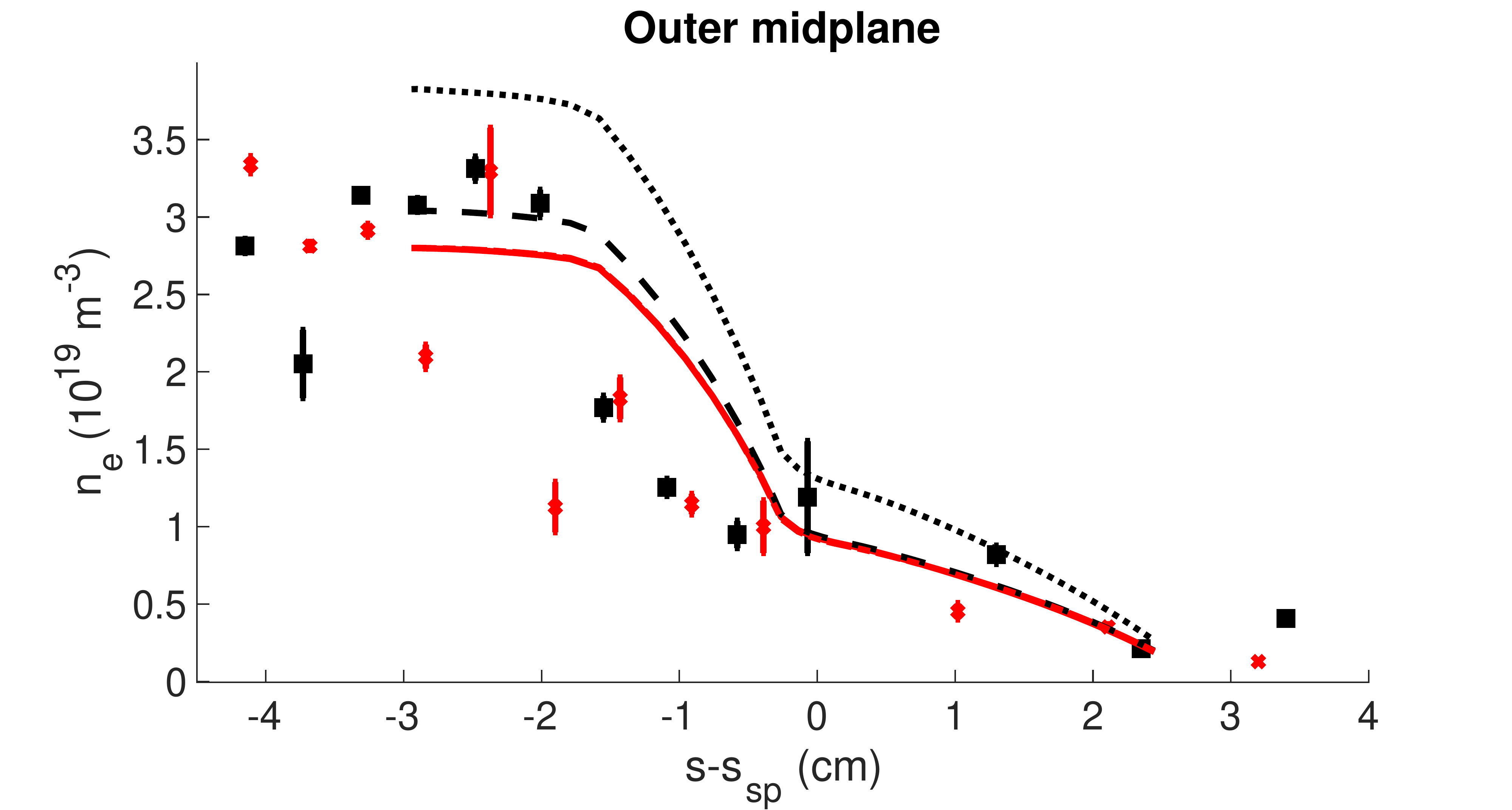}
		\caption{The electron density profiles measured with the TS system and from the SOLPS-ITER simulations at the OMP.}
		\label{subfig:ne_OMP}
	\end{subfigure}
	\begin{subfigure}{6cm}
		\centering
		\medskip
		\includegraphics[width=6cm]{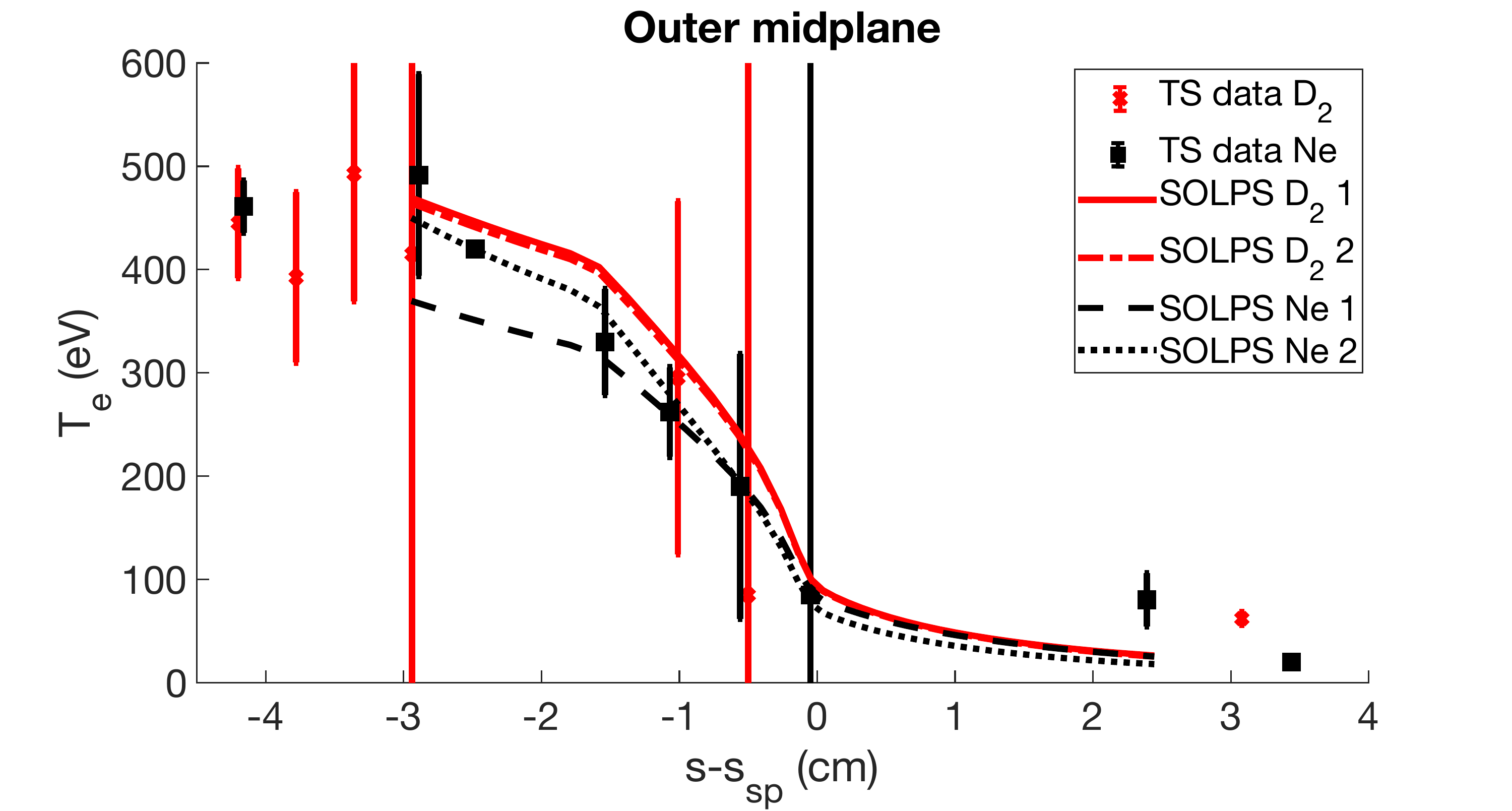}
		\caption{The electron temperature profiles measured with the TS system and from the SOLPS-ITER simulations at the OMP.}
		\label{subfig:Te_OMP}
	\end{subfigure}
	\caption{The upstream profiles at the OMP for electron density and temperature during the purely deuterium and neon seeded discharges and simulations.}
	\label{fig:TS}
\end{figure}

The two-point model shows that the upstream conditions at the OMP have a strong influence on the downstream quantities \cite{stangeby2000plasma}. The upstream OMP profiles from the TS system have the advantage that $\mathrm{n_e}$ and $\mathrm{T_e}$ profiles are obtained at exactly the same location. The TS system is installed at port L of the EAST device. Figure \ref{fig:TS} shows a limited influence of the neon seeding on the experimental OMP profiles: the density is slightly increased and the temperature decreased. Based on these experimental profiles, the experimental decay lengths for temperature and density are estimated to be: $\lambda_{T_e} \approx \mathrm{4.5 \, mm}$ and $\lambda_{n_e} \approx \mathrm{2.0 \, cm}$. As the SOLPS-ITER grid only extends slightly further into the SOL than $\lambda_{n_e}$, it has been verified in ref. \cite{boeyaert2024numerical} that maximum $\sim 25 \, \%$ of the power is leaving the simulation domain through the grid boundary closest to the first wall. 

\begin{figure}
	\centering
	\begin{subfigure}{8cm}
		\centering
		\medskip
		\includegraphics[width=8cm]{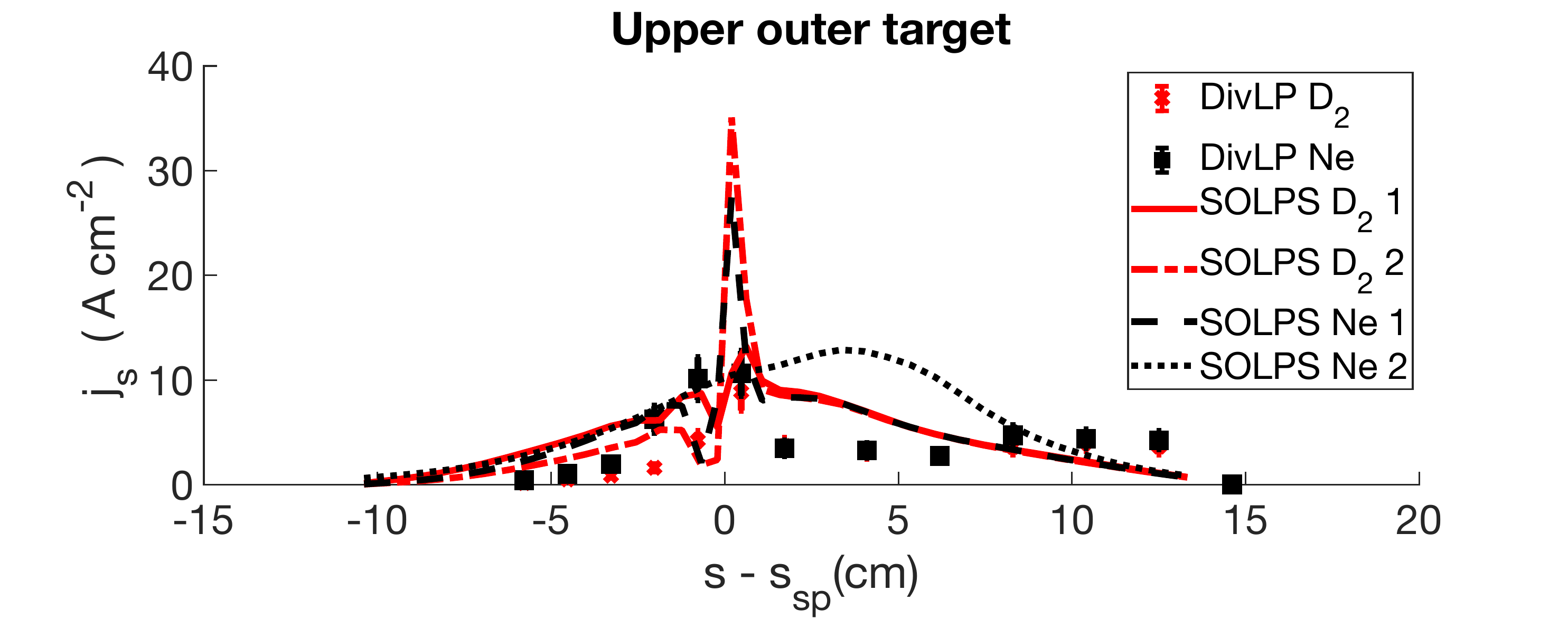}
		\caption{}
		\label{subfig:js_UOT}
	\end{subfigure}
	\begin{subfigure}{8cm}
		\centering
		\medskip
		\includegraphics[width=8cm]{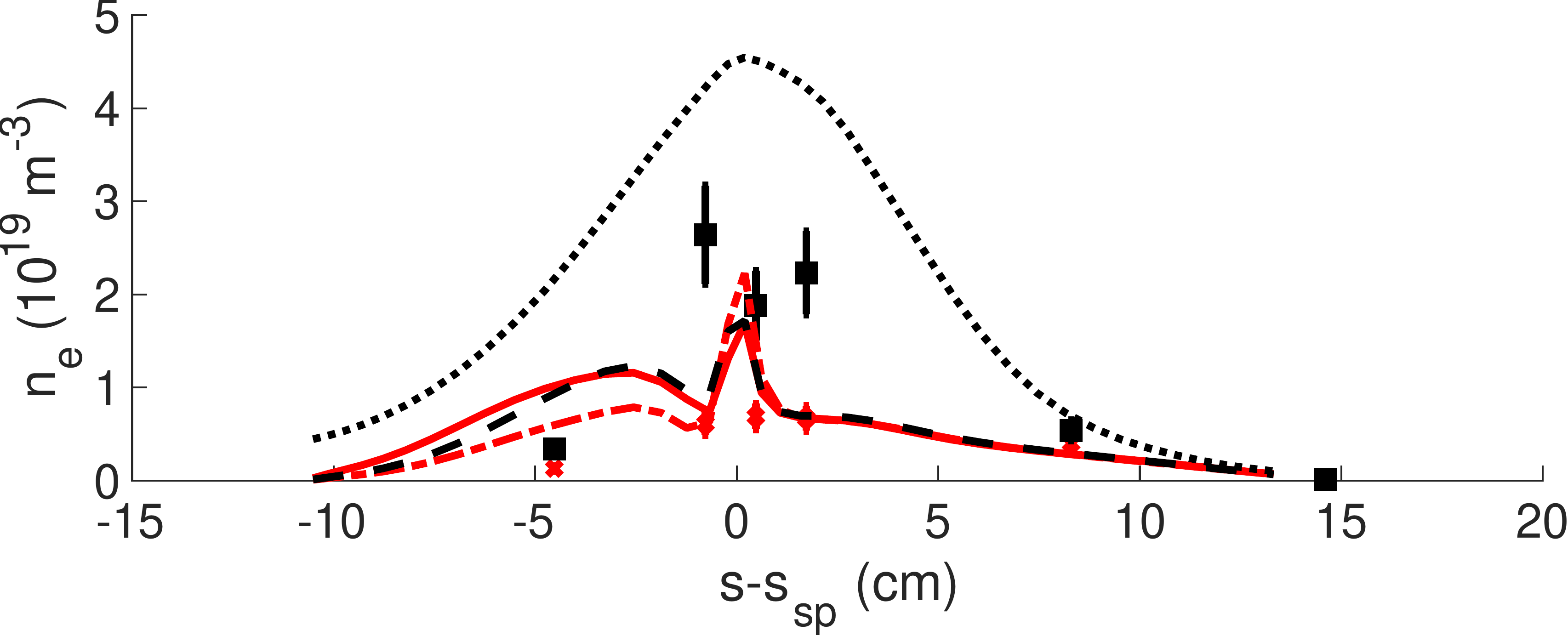}
		\caption{}
		\label{subfig:ne_UOT}
	\end{subfigure}
	\begin{subfigure}{8cm}
		\centering
		\medskip
		\includegraphics[width=8cm]{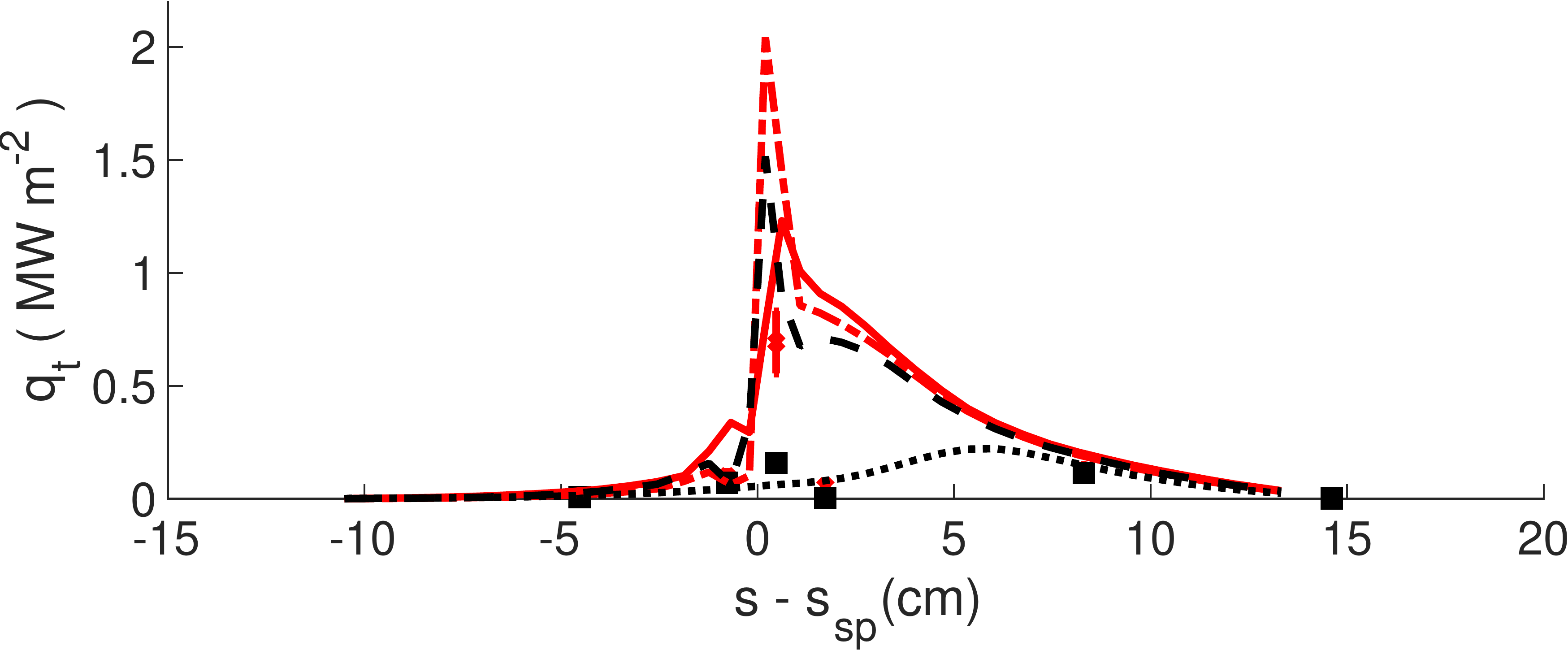}
		\caption{}
		\label{subfig:qt_UOT}
	\end{subfigure}
	\begin{subfigure}{8cm}
		\centering
		\medskip
		\includegraphics[width=8cm]{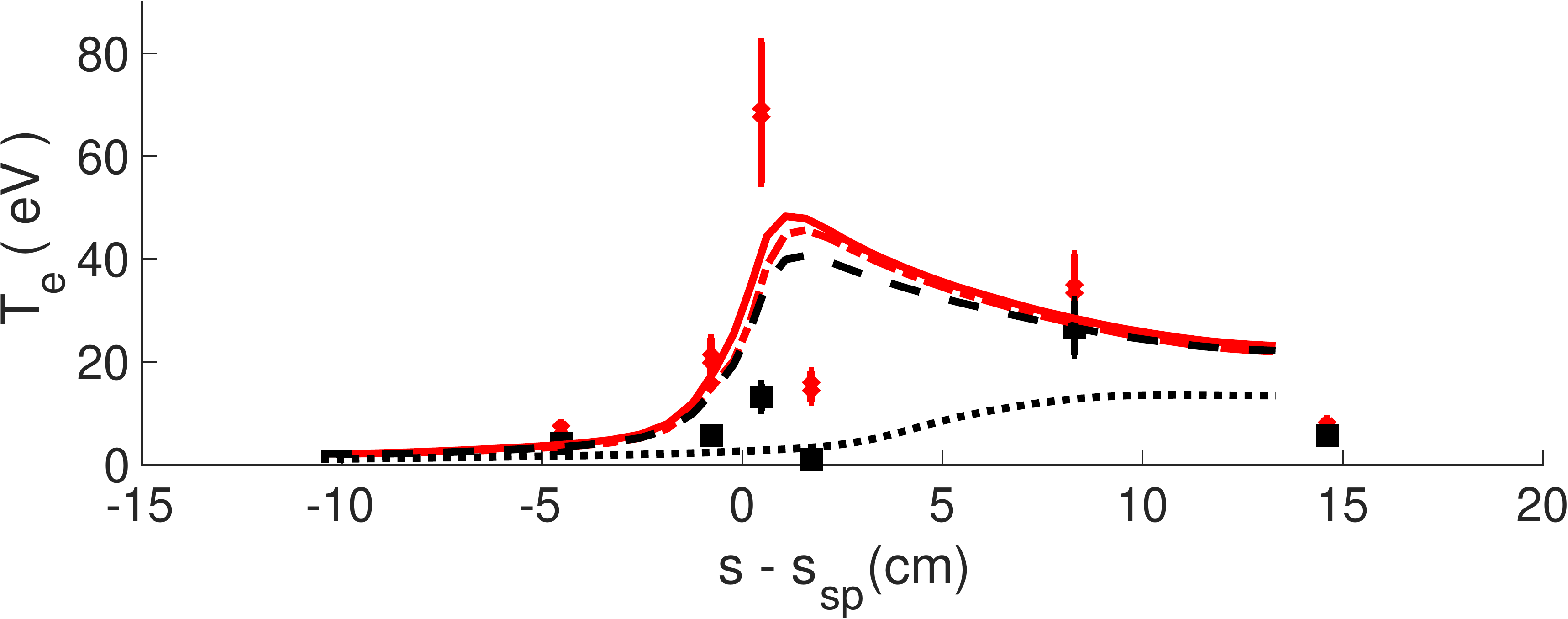}
		\caption{}
		\label{subfig:Te_UOT}
	\end{subfigure}
	\caption{The profiles of the ion saturation current (a), density (b), heat flux (c) and temperature (d) at the UOT during the purely deuterium and the neon seeded discharge.}
	\label{fig:DivLP_SOLPS_UOT}
\end{figure}

The purely deuterium simulations (SOLPS D$_2$ 1 and SOLPS D$_2$ 2) have similar upstream profiles as the TS data from the experiment as can be seen in figure \ref{fig:TS}. It only appears that the increase in density happens further inside the separatrix in the experiment than in the simulation. Figure \ref{fig:DivLP_SOLPS_UOT} shows the profiles at the UOT from the simulations and DivLP. This figure shows that, apart from the spikes around the separatrix due to the inclusion of drifts (see section \ref{sec:drifts}), the peak values agree reasonable between simulations and experiment. The included artificial radiation in simulation "SOLPS $\mathrm{D_2}$ 2" lowers slightly the temperature at the UOT, but the spike in the $j_s$, $n_e$ and as a result also in the $q_t$ profiles right outside the separatrix, is largely increased. 

The effect of only Ne transport can be seen by comparing simulations "SOLPS D$_2$ 1" and "SOLPS Ne 1". The only differences between these simulations are the Ne-D$_2$ injection at the UOT, and the decreased input power into the SOL. Figure \ref{fig:TS} shows that the reduced input power lowers the electron temperature at the OMP close to the core boundary of the grid. The TS data suggest that a such large decrease does not take place in the experiments. Also the simulated temperature at the OMP separatrix slightly decreases. The profiles at the UOT, on the other hand, show that the influence of Ne is larger in the experiment, than in the simulation. The trends for $j_s$, $n_e$ and $T_e$ are nevertheless similar, where this is difficult to determine for $q_t$ due to spikes in the profile around the separatrix. The effect of adding the injected Ne-D$_2$ mixture to the simulation on the target profiles is similar (but not the same) as adding artificial radiation. This indicates that in both simulations the changes are determined by an additional radiated power fraction (see section \ref{sec:radiated_power}) and the increased peak in the $j_s$ profile of simulation "SOLPS Ne 1" is not determined by the additional Ne-D$_2$ injection at this location as the ionization profiles which are analyzed in section \ref{sec:drifts}, do not show an increased ionization in the vicinity of the UOT.

Simulation "SOLPS Ne 2" manages to match better the experimental data at the UOT. This suggests that the amount of Ne in the device is larger than the injected amount (which is possible as other Ne experiments took place before discharge 87628, but it is unlikely there was 20 times as much Ne present), or that the anomalous transport profiles have changed. Therefore, also the transport physics of simulation "SOLPS Ne 2" is analyzed in the next sections.

Based on the experimental profiles of figures \ref{fig:TS} and \ref{fig:DivLP_SOLPS_UOT}, the one dimensional particle balance ($\mathrm{2n_tT_t = f_{mom}n_uT_u}$) with $\mathrm{f_{mom}}$ the momentum loss factor in the flux tube, is $f_{mom} \approx 0.88$ and $f_{mom} \approx 0.65$ in the neon seeded experiment. This indicates a larger momentum loss because of the neon seeding which is expected due to the decreased temperatures at the target. In fact, the decrease in maximum $T_e$ at the UOT means that conduction becomes more important in comparison with the purely deuterium experiment as can be concluded from the temperature-dependent edge physical processes of figure \ref{fig:detachment_physics}.

\begin{figure}
	\centering
	\begin{subfigure}{8cm}
		\centering
		\medskip
		\includegraphics[width=8cm]{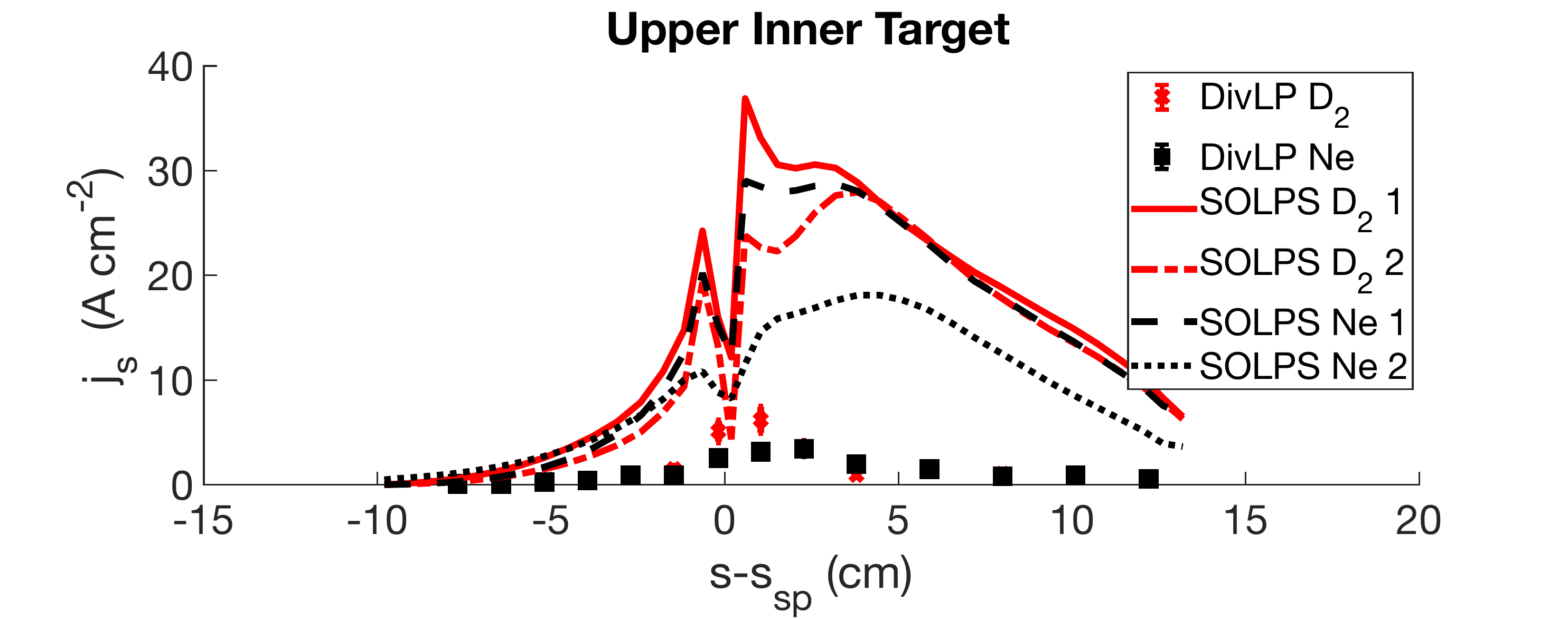}
		\caption{}
		\label{subfig:js_UIT}
	\end{subfigure}
	\begin{subfigure}{8cm}
		\centering
		\medskip
		\includegraphics[width=8cm]{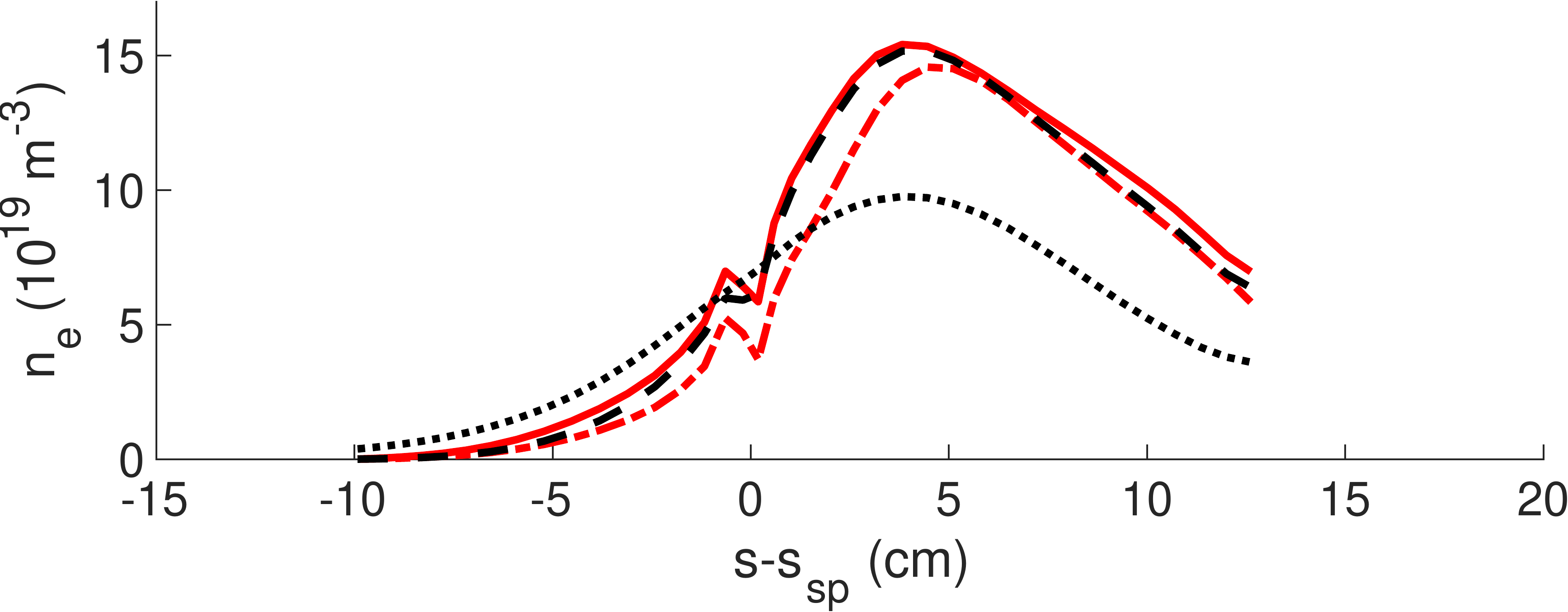}
		\caption{}
		\label{subfig:ne_UIT}
	\end{subfigure}
	\begin{subfigure}{8cm}
		\centering
		\medskip
		\includegraphics[width=8cm]{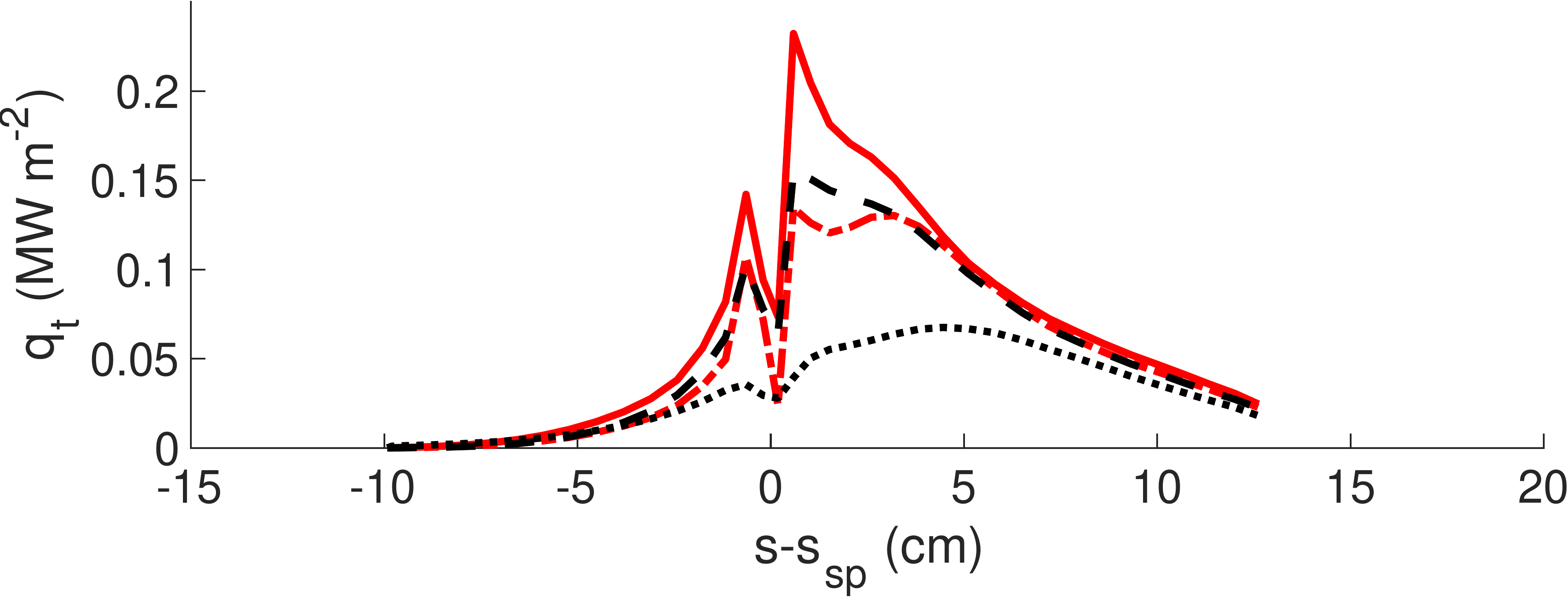}
		\caption{}
		\label{subfig:qt_UIT}
	\end{subfigure}
	\begin{subfigure}{8cm}
		\centering
		\medskip
		\includegraphics[width=8cm]{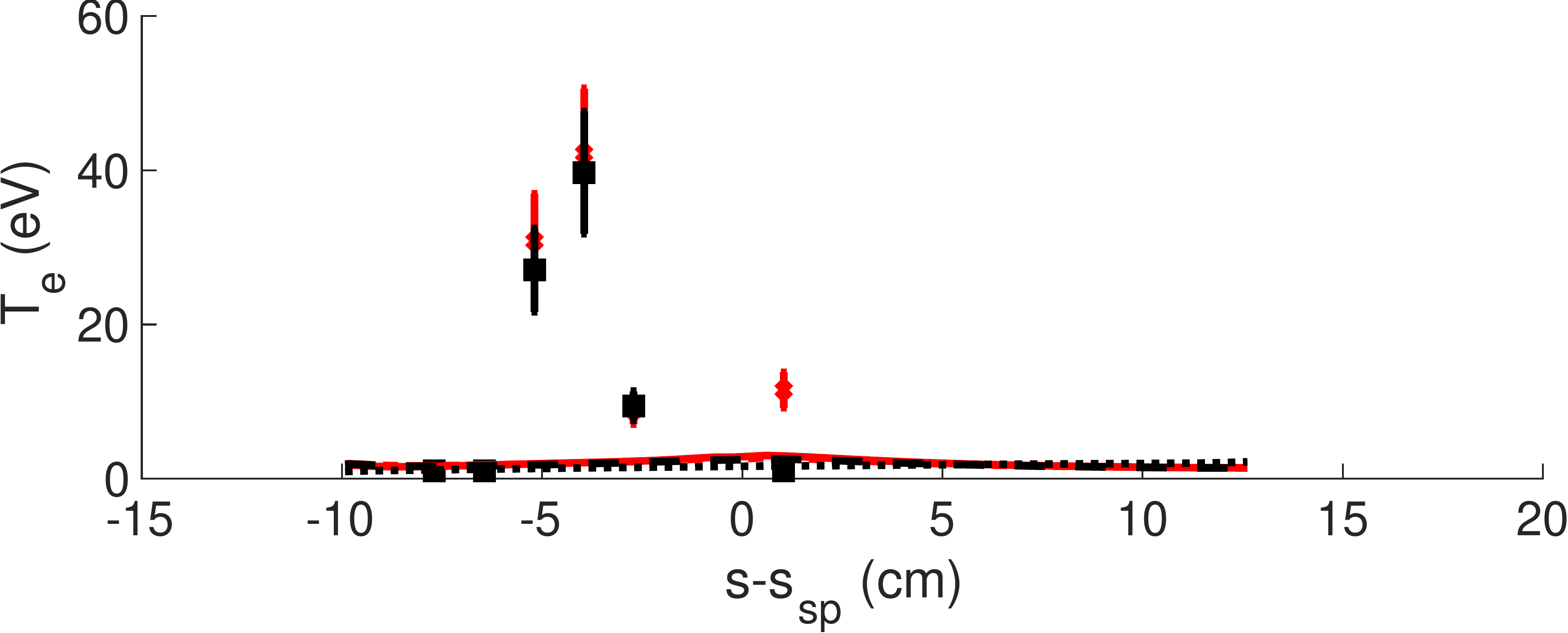}
		\caption{}
		\label{subfig:Te_UIT}
	\end{subfigure}
	\caption{The profiles of the ion saturation current (a), density (b), heat flux (c) and temperature (d) at the UIT during the purely deuterium and the neon seeded discharge.}
	\label{fig:DivLP_SOLPS_UIT}
\end{figure}

The data at the UIT of figure \ref{fig:DivLP_SOLPS_UIT}, on the other hand, disagree for all cases between simulations and experiments. Especially the temperature at the inner target is extremely low in the simulations.

For this disagreement between simulations and experiments at the UIT a couple of remarks should be taken into account. First, the quality of the DivLP data at the UIT was not as good as the quality at the UOT making a general overview more challenging. Secondly, it is known that the anomalous transport at the inboard side can differ from the one at the outboard side \cite{reimold2014experimental} which is not taken into account in the performed analysis. Simulations were performed in which the anomalous diffusion in the entire divertor region was raised. This increased indeed the maximum temperature at the UIT, but decreased it at the UOT. However, a large increase in anomalous transport is needed to cover the temperature difference. A more in-depth study of the difference in anomalous transport between UIT and UOT is left for future work.


In the following sections first the effect of drifts on the observed profiles is examined with the SOLPS-ITER simulations and later the effects of the neon on the radiation and the plasma-neutral interactions are discussed.

\section{Effect of drifts}
\label{sec:drifts}

\begin{figure}
	\centering
	\begin{subfigure}{6cm}
		\centering
		\medskip
		\includegraphics[width=6cm]{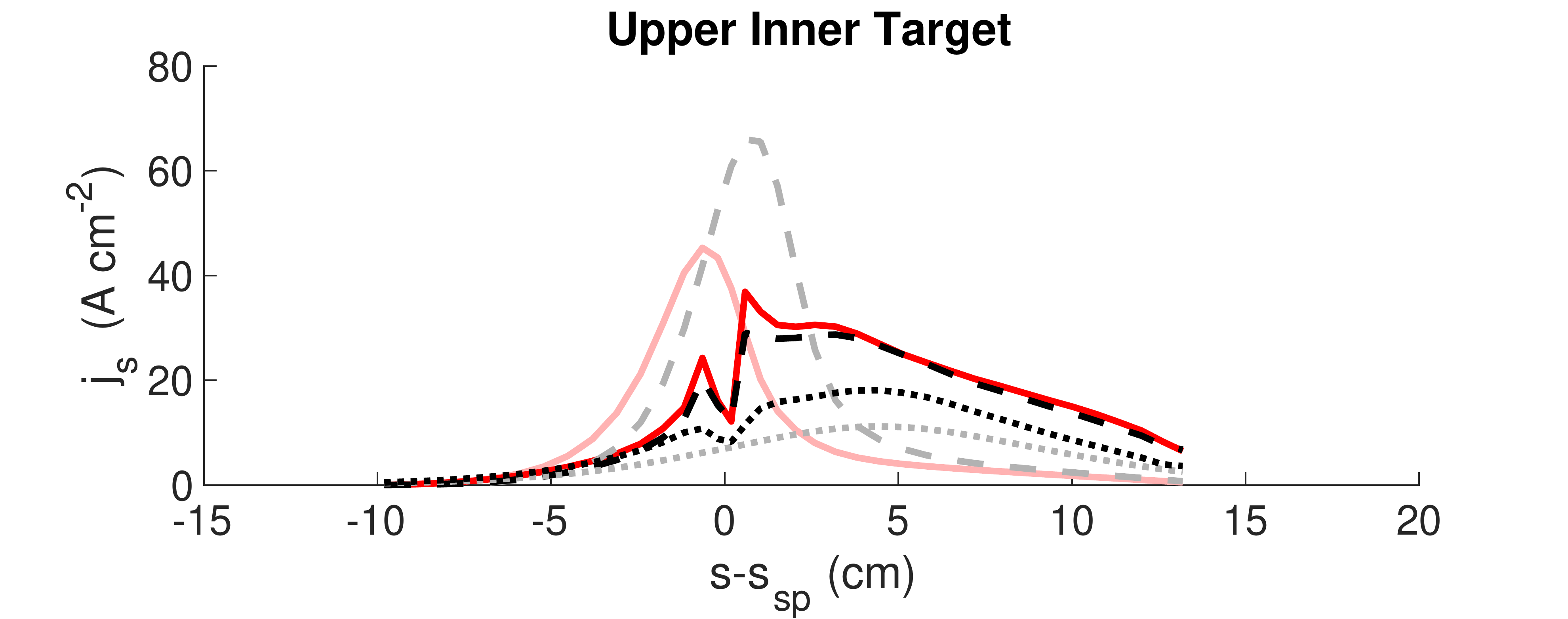}
		\caption{}
		\label{subfig:js_UIT_drifts}
	\end{subfigure}
	\begin{subfigure}{6cm}
		\centering
		\medskip
		\includegraphics[width=6cm]{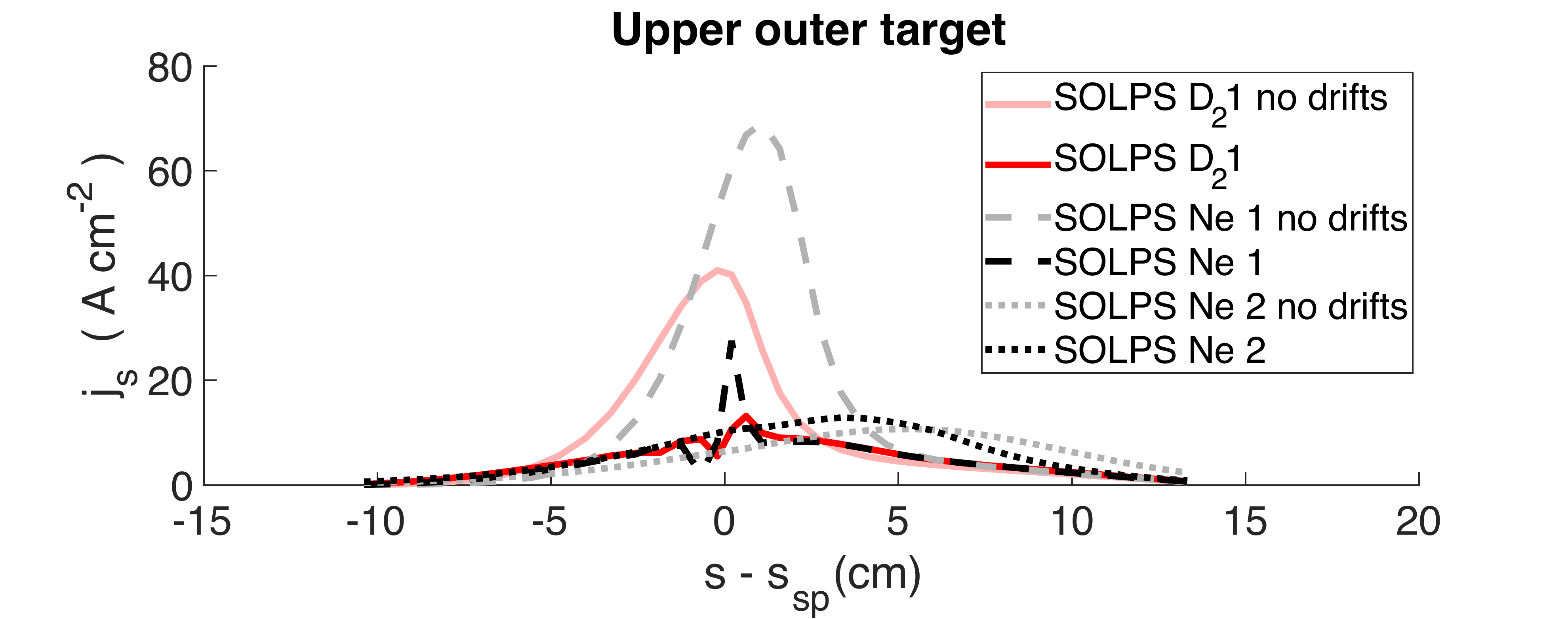}
		\caption{}
		\label{subfig:js_UOT_drifts}
	\end{subfigure}
	\begin{subfigure}{6cm}
		\centering
		\medskip
		\includegraphics[width=6cm]{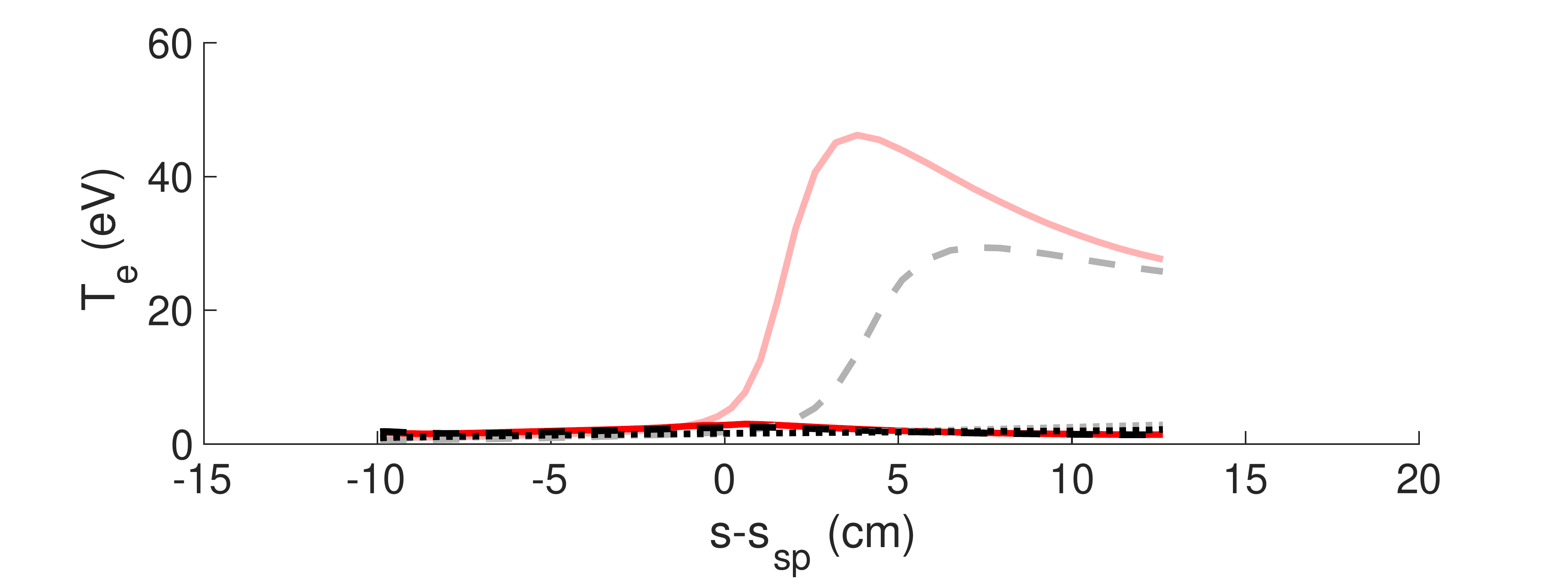}
		\caption{}
		\label{subfig:Te_UIT_drifts}
	\end{subfigure}
	\begin{subfigure}{6cm}
		\centering
		\medskip
		\includegraphics[width=6cm]{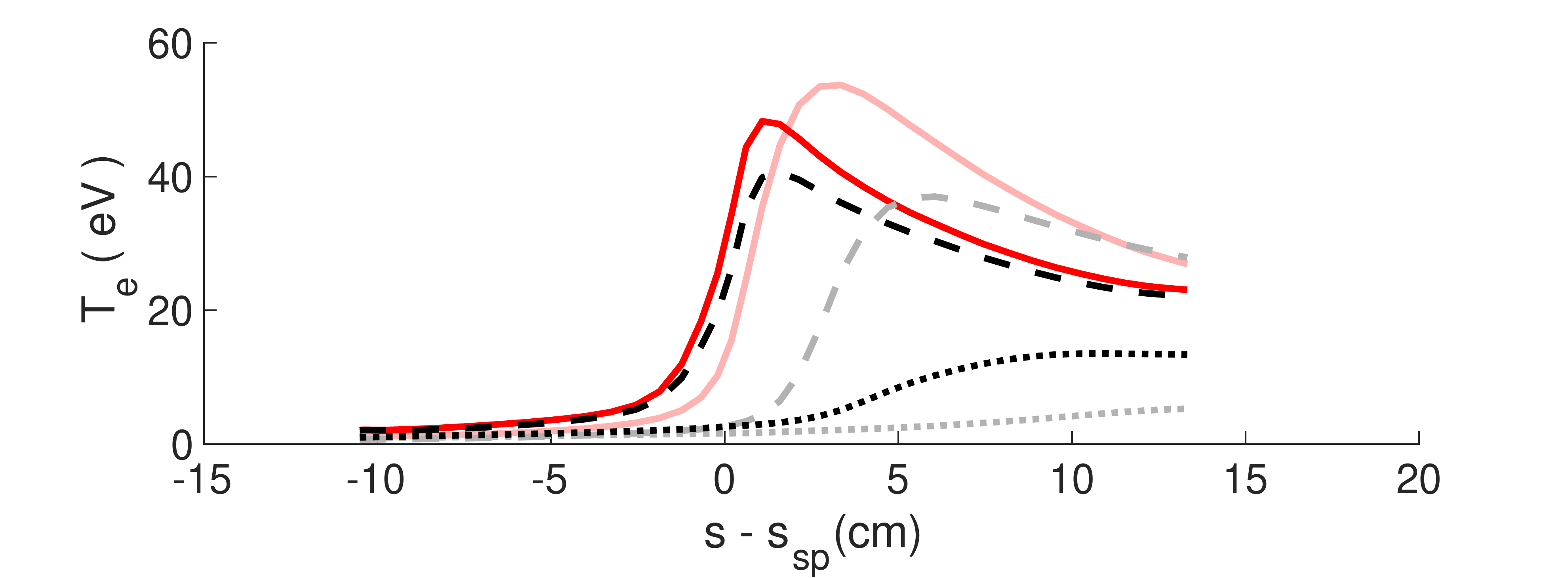}
		\caption{}
		\label{subfig:Te_UOT_drifts}
	\end{subfigure}
	\caption{The influence of drift flows on the electron temperature and ion saturation current profiles at the divertor targets.}
	\label{fig:targets_drifts}
\end{figure}

The analysis of the drifts effects is mainly performed for the "SOLPS $\mathrm{D_2}$ 1", "SOLPS Ne 1", and "SOLPS Ne 2" simulations. As the "SOLPS $\mathrm{D_2}$ 2" simulation has properties of all of them, it is not analyzed in detail.
Figure \ref{fig:targets_drifts} shows the influence of drift flows in the SOLPS-ITER simulations for $j_s$ and $T_e$. The upstream profiles at the OMP are only slightly influenced by drift flows. Figures \ref{fig:DivLP_SOLPS_UOT} and \ref{fig:DivLP_SOLPS_UIT} indicate a shift in peak location of the $j_s$ and $T_e$ profiles between the inner and outer targets. Where this shift is small for $j_s$, $T_e$ undergoes a large shift.

\begin{figure}
	\centering
	\begin{subfigure}{8cm}
		\centering
		\medskip
		\includegraphics[width=8cm]{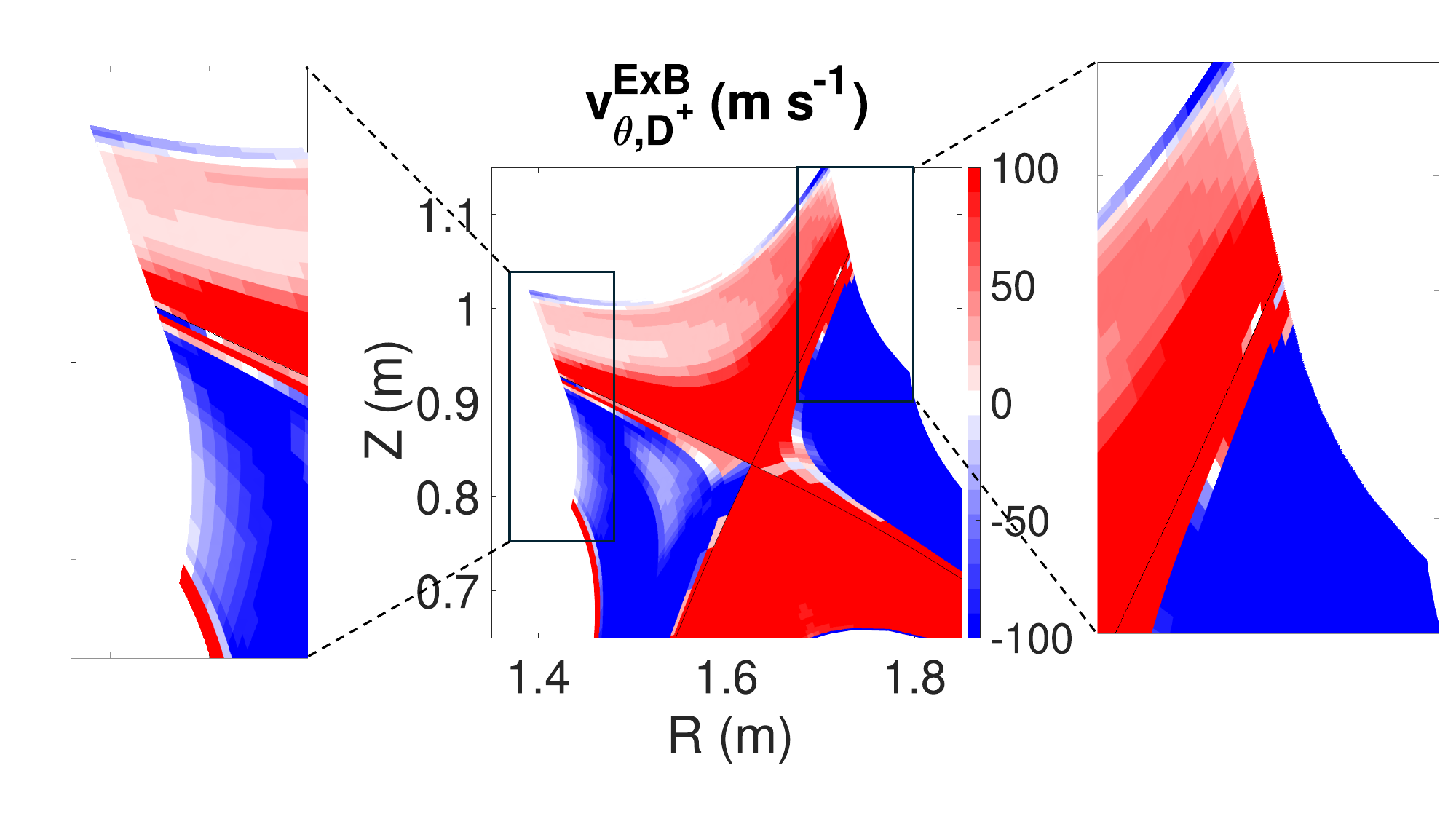}
		\caption{}
		\label{subfig:v_ExB_D_D2}
	\end{subfigure}
	\begin{subfigure}{8cm}
		\centering
		\medskip
		\includegraphics[width=8cm]{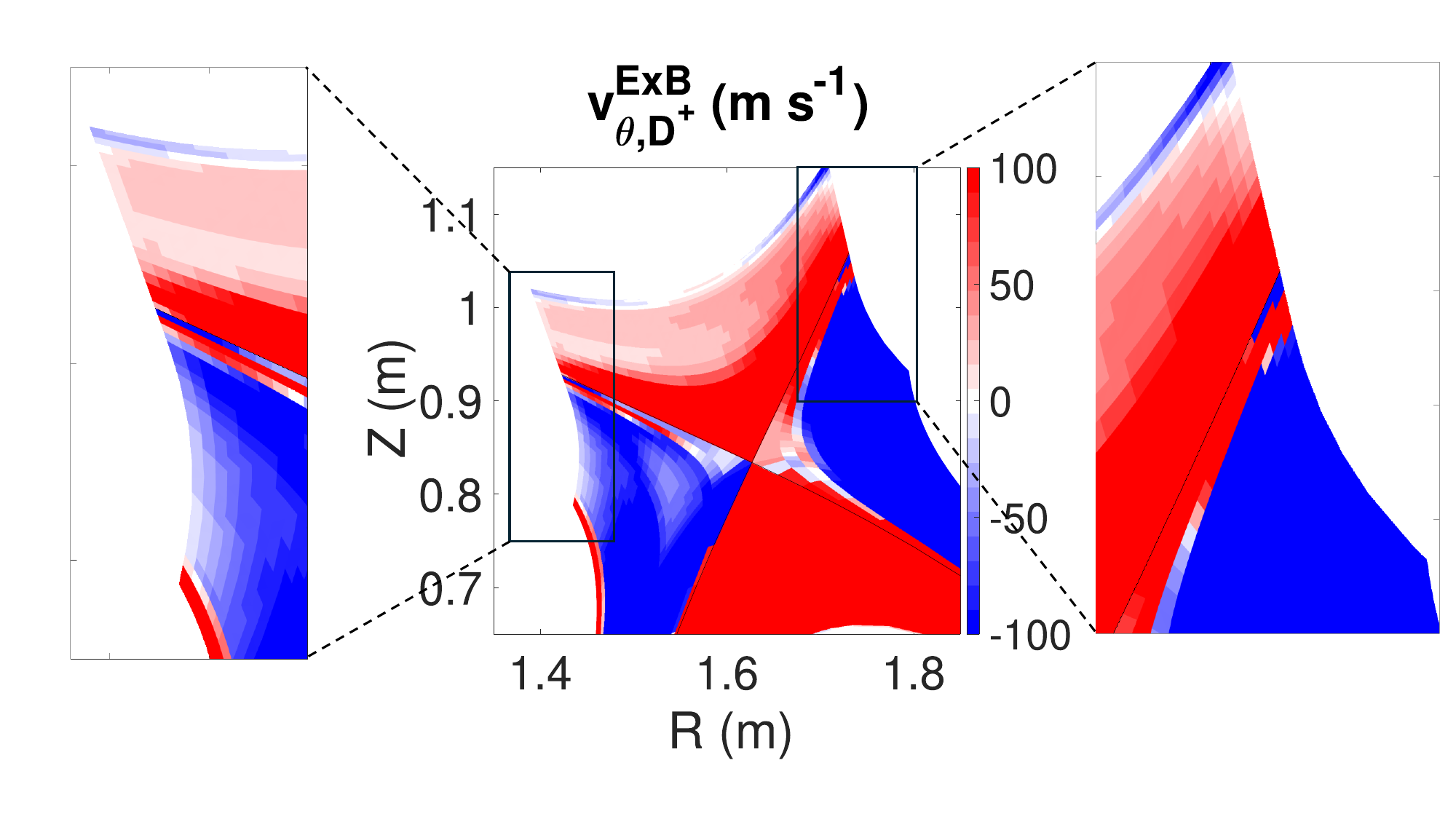}
		\caption{}
		\label{subfig:v_ExB_D_Ne}
	\end{subfigure}
	\begin{subfigure}{8cm}
		\centering
		\medskip
		\includegraphics[width=8cm]{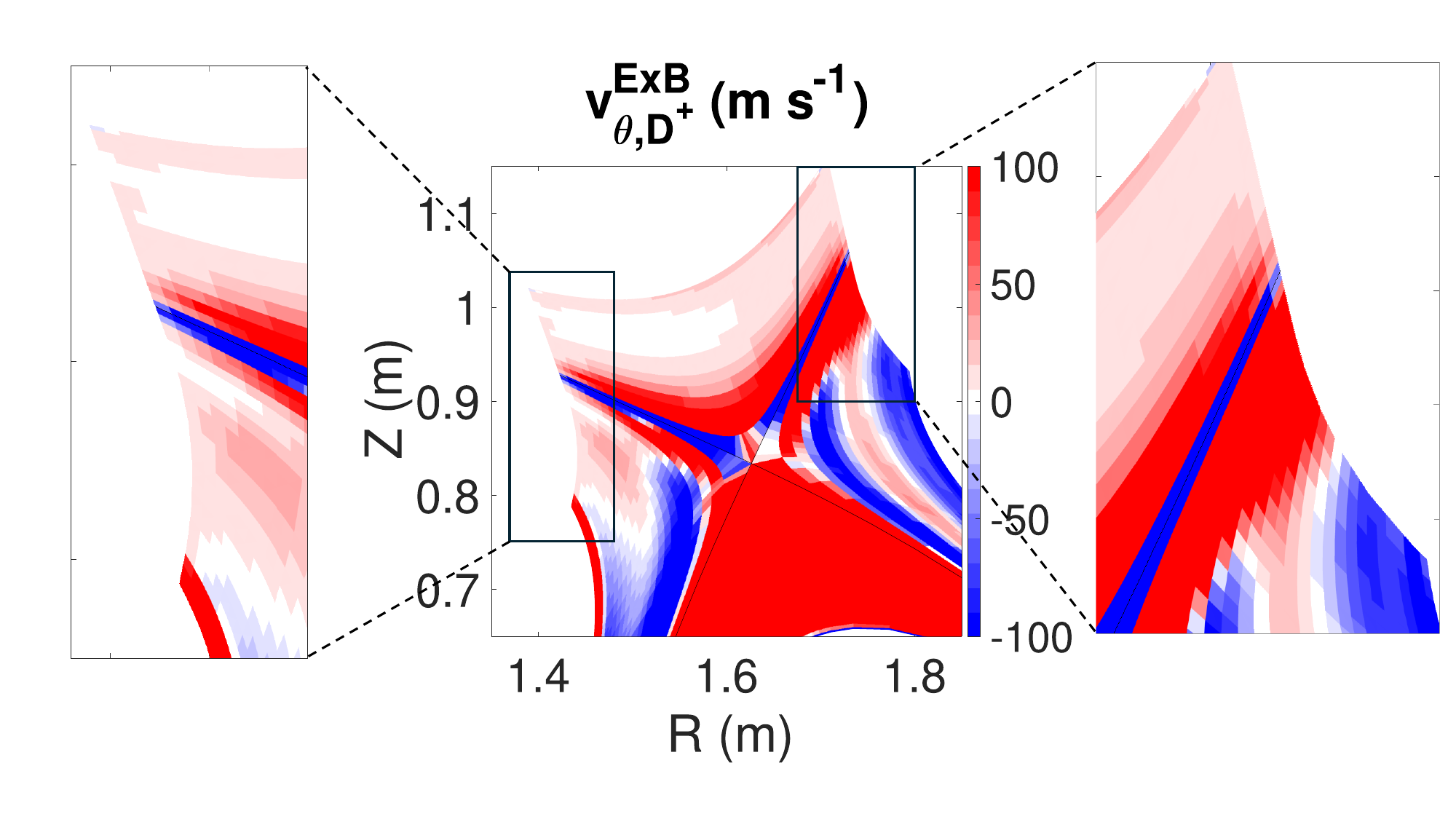}
		\caption{}
		\label{subfig:v_ExB_D_Ne2}
	\end{subfigure}
	\caption{The ExB drift velocity of D$^+$ in the poloidal direction shown for the  "SOLPS D$_2$ 1" (a), "SOLPS Ne 1" (b) and "SOLPS Ne 2" (c) simulations in the vicinity of the divertor. The separatrix in all plots is indicated in black.}
	\label{fig:v_ExB_D}
\end{figure}

As studied in ref. \cite{reimold2017high} for ASDEX Upgrade, this shift in peak of the target profiles is caused by the ExB drift flows. These flows cause additional particle sources in the far SOL. Figure \ref{fig:v_ExB_D} shows the ExB drift velocities for the different simulations. Due to the definition of the computational grid, a positive velocity implies that the plasma particles are moving towards the UIT, while a negative one indicates they are moving towards the UOT. Where the ExB drift velocities in the private flux region (PFR) stay limited (especially further away from the separatrix), they indeed have a larger impact in the SOL. 

For simulations "SOLPS D$_2$ 1" and "SOLPS Ne 1", there appears a double peak in the $j_s$ profile due to drifts. Part of these oscillations in plasma profiles around the separatrix can be explained by numerical effects (see ref. \cite{boeyaert2024numerical}). However, a double peaking is also observed in other experimental and computational EAST studies \cite{jia2022role} as well as in DIII-D, TCV, ASDEX Upgrade... \cite{colandrea2024investigation}. For the performed simulations, the profile of the ExB drift velocities in the vicinity of the divertor targets explain this double peaking. Figure \ref{fig:v_ExB_D} indicates small regions around the UIT and UOT separatrix where the direction of the ExB drift velocity is changed. These regions agree with the location of a peak (UOT) or the location of a valley in between the two peaks (UIT). The different behavior between UOT and UIT corresponds with the additional drift flow in the SOL which transports particles from the inboard to the outboard side as also observed in ref. \cite{jia2022role}. It explains why the double peak is observed at the inner target and two valleys are present at the outer target. Figure \ref{subfig:v_ExB_D_Ne} shows that the effect is larger at the UOT in the "SOLPS Ne 1" simulation in comparison with the "SOLPS D$_2$ 1" one. In the vicinity of the inner target, figure \ref{subfig:v_ExB_D_Ne} indicates that the region of reversed ExB velocity penetrates deeper into the plasma when neon is added to the simulation. The effect on the particle flux (figure \ref{subfig:js_UIT_drifts}) stays however limited. 

Ref. \cite{kaveeva2021solps} points out that this double peaking vanishes when the transport in the PFR and divertor region is increased. In simulations "SOLPS Ne 2", the anomalous diffusion in the divertor region, is three times larger than in the other simulations, and as can be seen on figure \ref{fig:targets_drifts}, this simulation does not show a double peaking. Figure \ref{subfig:v_ExB_D_Ne2} confirms that here the influence of the ExB ion velocity is smaller than in the other simulations.

\begin{figure}
	\centering
	\begin{subfigure}{6cm}
		\centering
		\medskip
		\includegraphics[width=6cm]{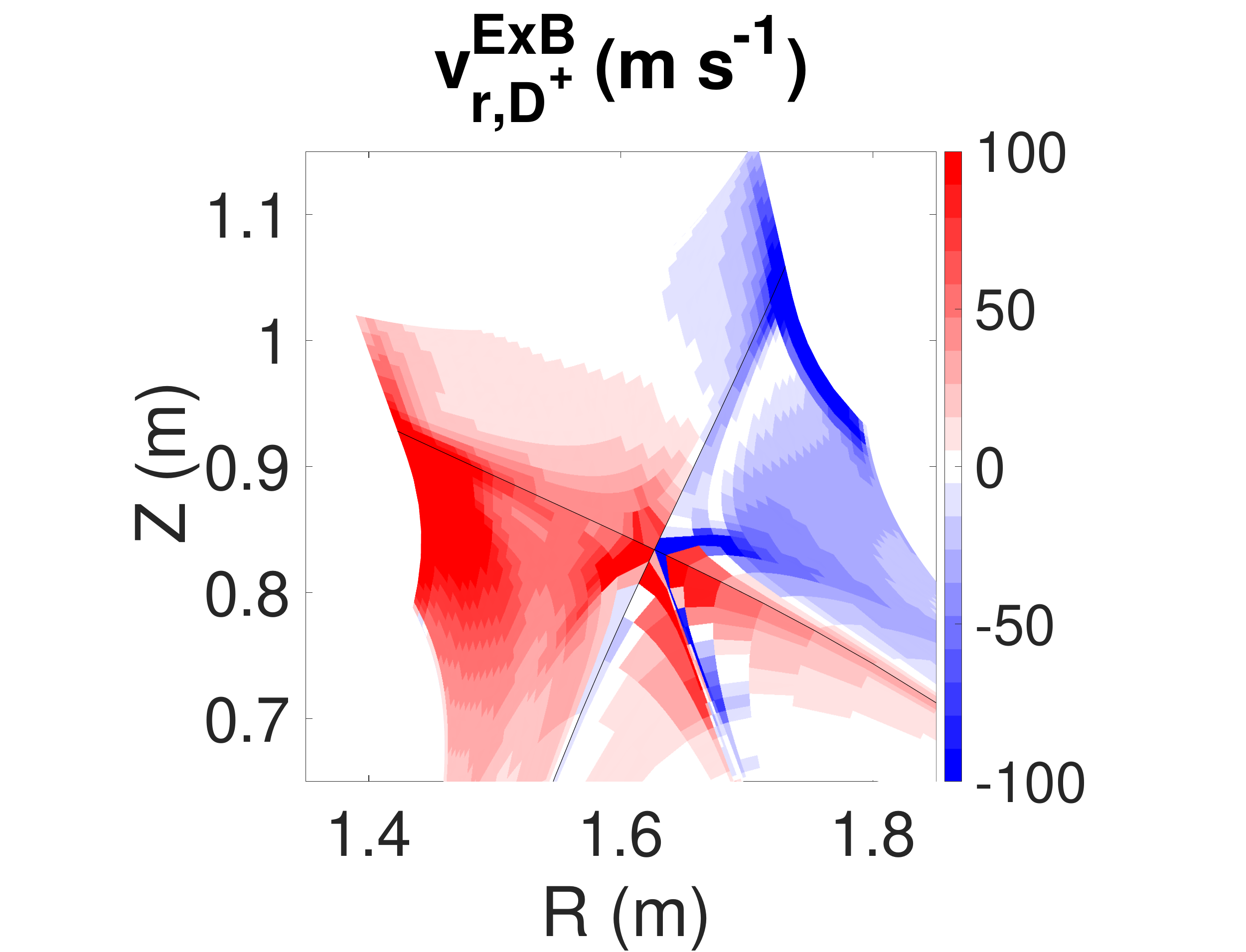}
		\caption{}
		\label{subfig:v_ExB_r_D_D2}
	\end{subfigure}
	\begin{subfigure}{6cm}
		\centering
		\medskip
		\includegraphics[width=6cm]{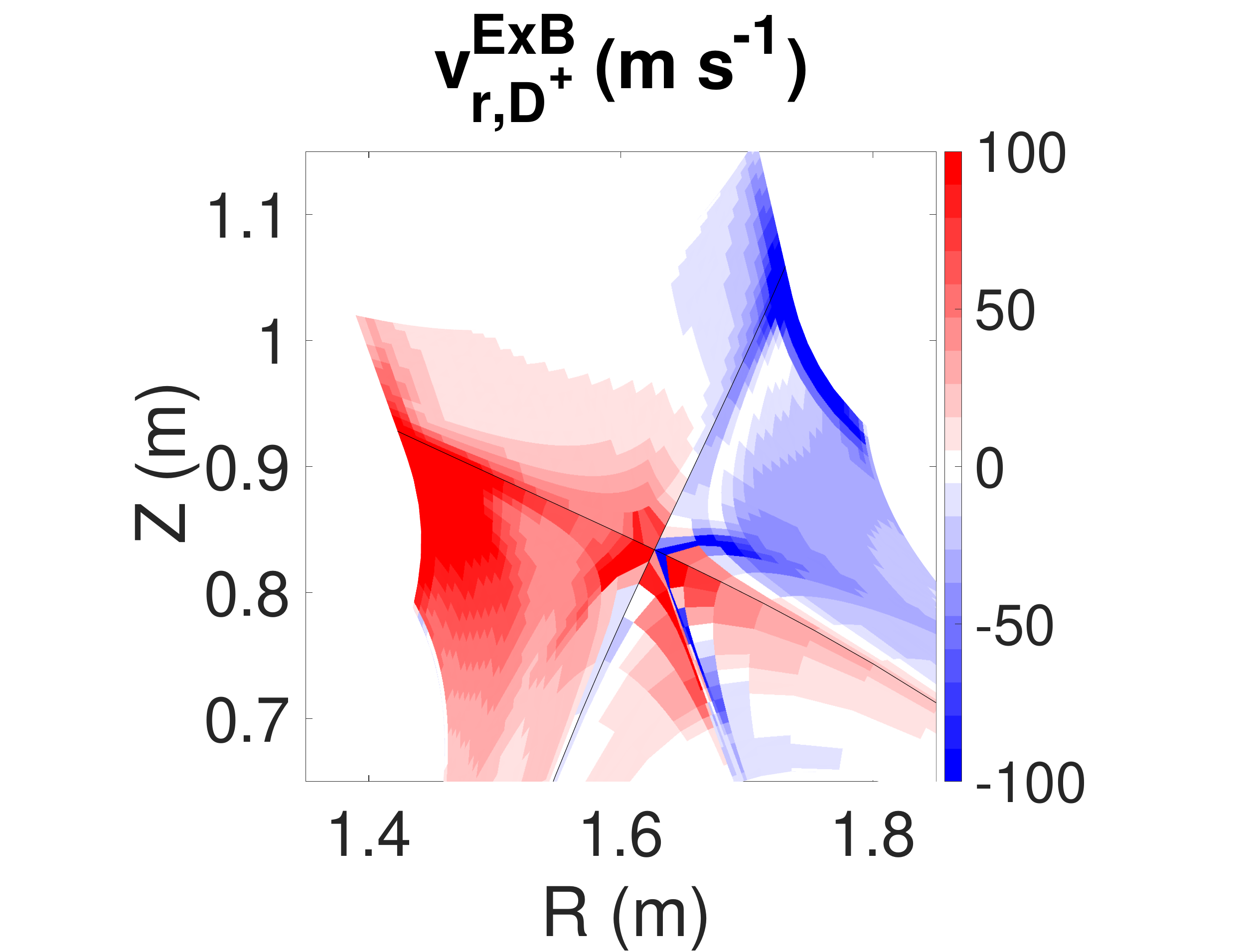}
		\caption{}
		\label{subfig:v_ExB_r_D_Ne}
	\end{subfigure}
	\begin{subfigure}{6cm}
		\centering
		\medskip
		\includegraphics[width=6cm]{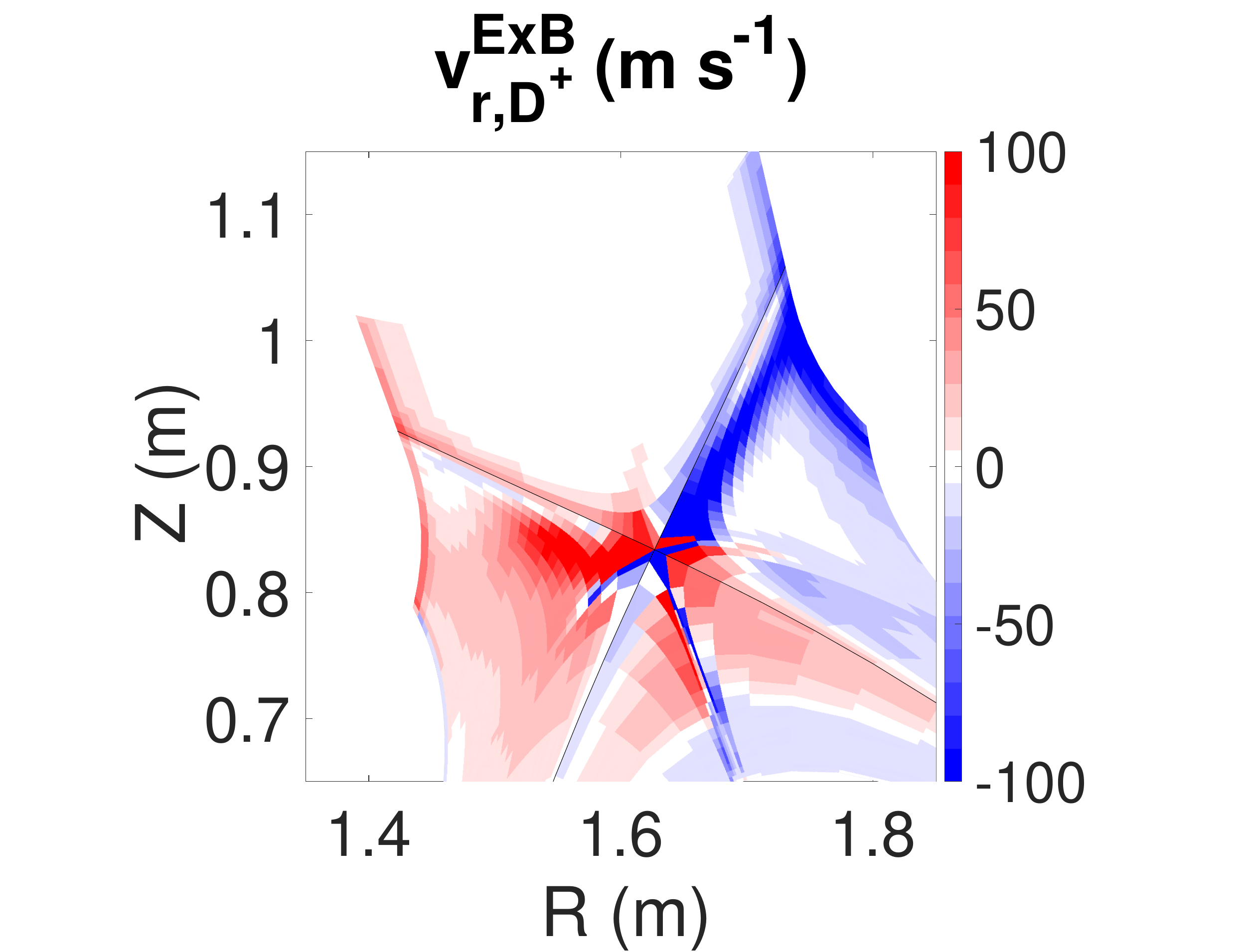}
		\caption{}
		\label{subfig:v_ExB_r_D_Ne2}
	\end{subfigure}
	\caption{The ExB drift velocity of D$^+$ in the radial direction shown for the  "SOLPS D$_2$ 1" (a), "SOLPS Ne 1" (b) and "SOLPS Ne 2" (c) simulations in the vicinity of the divertor. The separatrix in all plots is indicated in black.}
	\label{fig:v_ExB_r_D}
\end{figure}

\begin{figure}
	\centering
	\begin{subfigure}{6cm}
		\centering
		\medskip
		\includegraphics[clip,width=6cm]{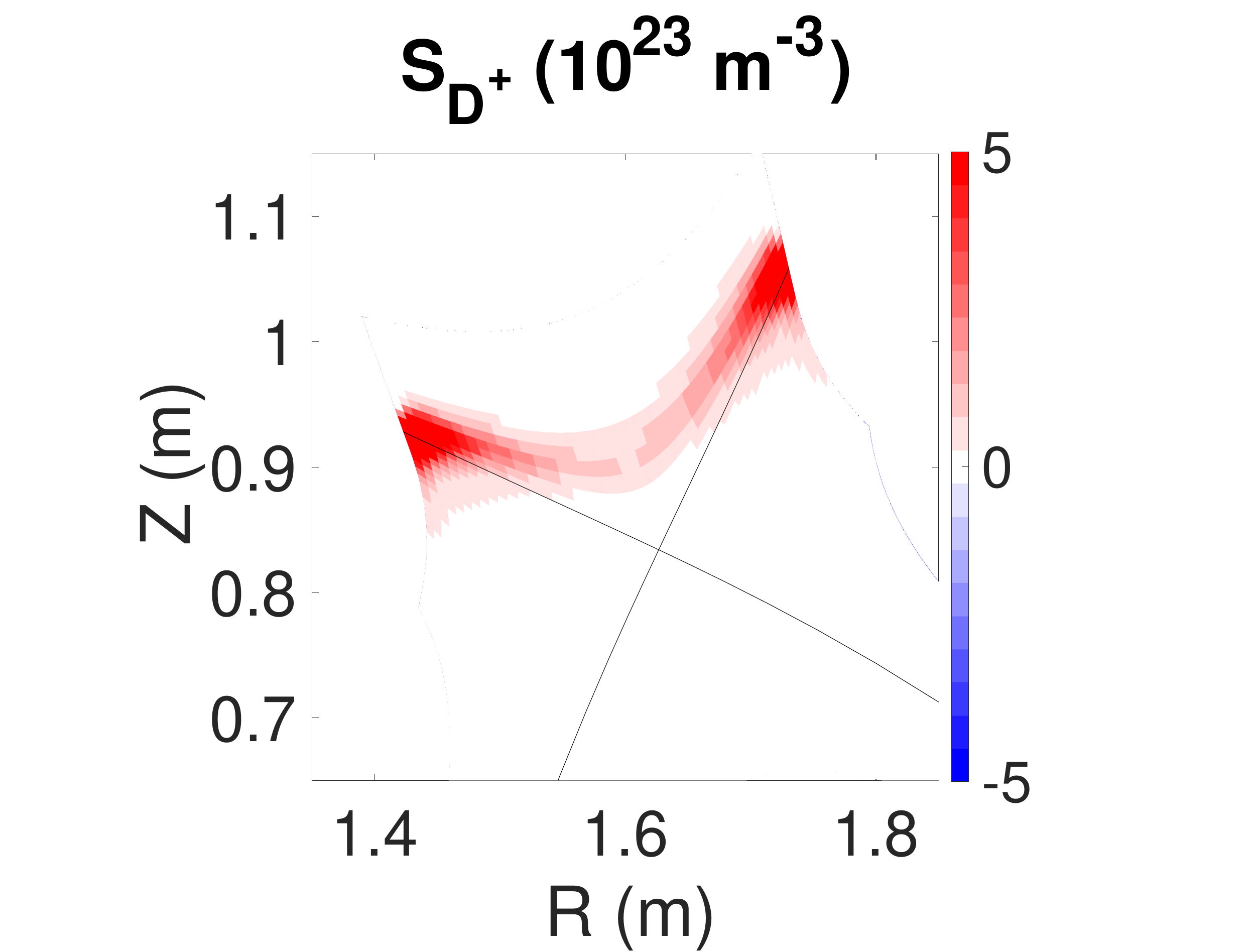}
		\caption{}
		\label{subfig:ionization_D_D2_detail_no_drifts}
	\end{subfigure}
	\begin{subfigure}{6cm}
		\centering
		\medskip
		\includegraphics[clip,width=6cm]{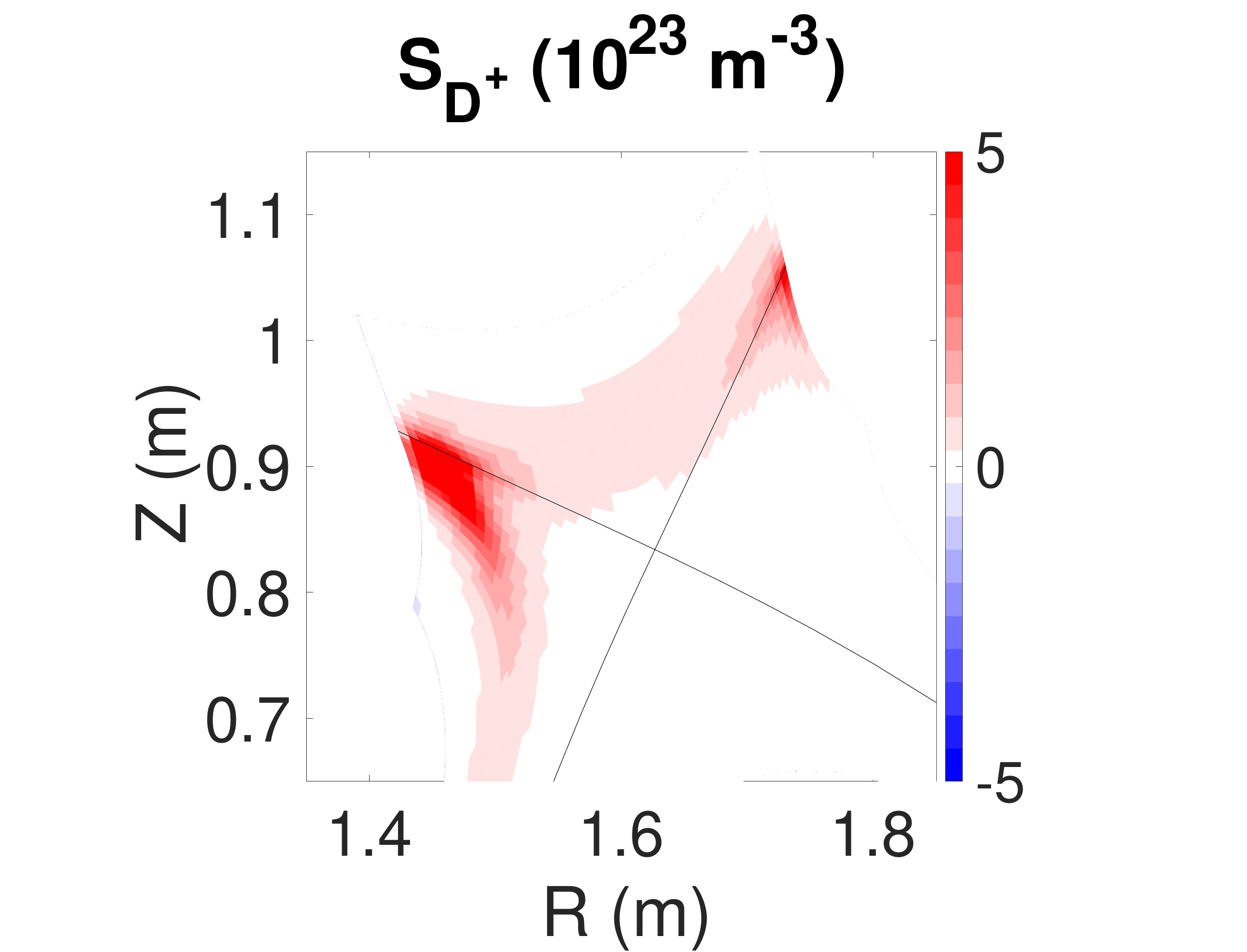}
		\caption{}
		\label{subfig:ionization_D_D2_detail}
	\end{subfigure}
		\begin{subfigure}{6cm}
		\centering
		\medskip
		\includegraphics[clip,width=6cm]{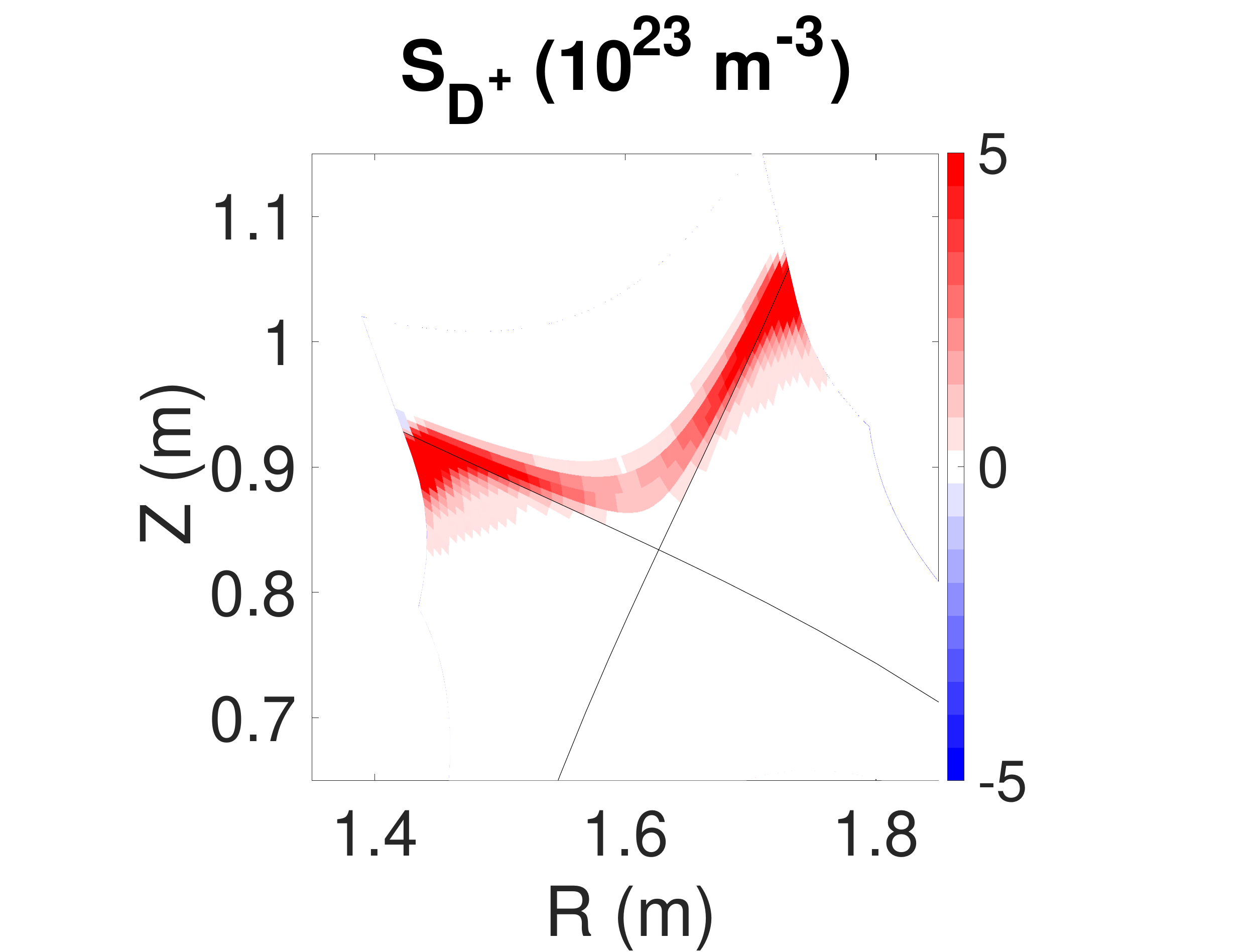}
		\caption{}
		\label{subfig:ionization_D_Ne_detail_no_drifts}
	\end{subfigure}
	\begin{subfigure}{6cm}
		\centering
		\medskip
		\includegraphics[clip,width=6cm]{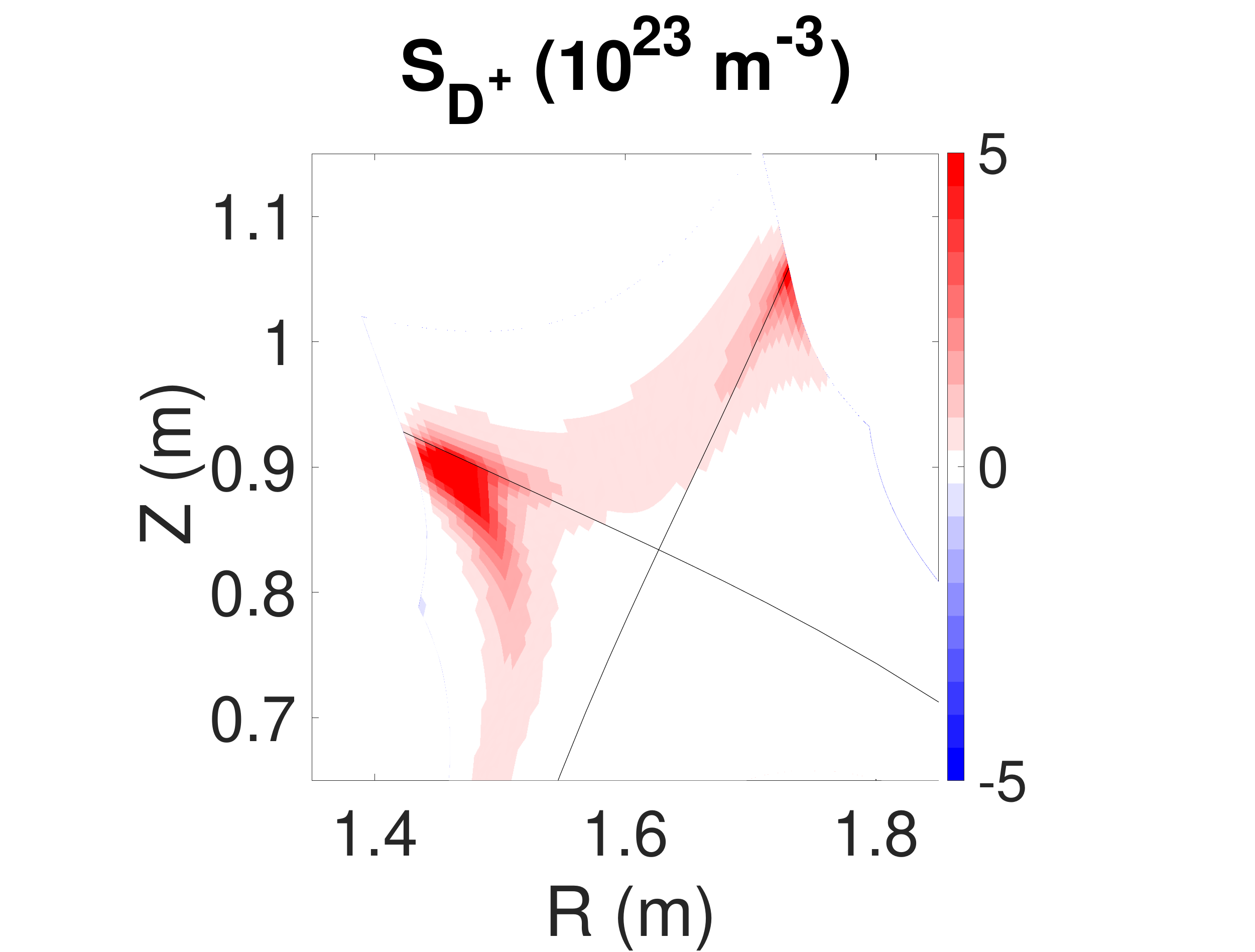}
		\caption{}
		\label{subfig:ionization_D_Ne_detail}
	\end{subfigure}
	\caption{The ionization source for D$^+$ in the vicinity of the divertor region in the "SOLPS D$_2$ 1" and "SOLPS Ne 1" simulations without (a,c) and with (b,d) drifts.}
	\label{fig:ionization_detail_drifts}
\end{figure}

Figure \ref{fig:v_ExB_r_D} shows for all the simulated cases that the radial ExB drift velocity changes sign between UIT and UOT. In this figure, positive means towards the SOL, where negative means towards the PFR. 
This explains the shift in maximum plasma quantities at the UOT from further into the SOL towards the separatrix. As drifts do not only transport particles, but also momentum and energy, their influence is not only limited to the $n_e$ profiles, but also the $T_e$ profiles are influenced. Due to the double peaking/the valley the observed shift is less pronounced for the $j_s$ profiles than for the $T_e$ ones. The shift in peaked $T_e$ shifts the location where more energy is present as well as the ionization location of D$^+$. At the UIT the ionization location shifts further into the SOL, where at the UOT this shifts towards the separatrix with its peak value just into the SOL as can be seen in figure \ref{fig:ionization_detail_drifts}. As an additional D$_2$ puff is added in the "SOLPS Ne 1" simulation, this effect is more pronounced over there in comparison with the "SOLPS D$_2$ 1" simulation. This change in ionization location explains the shift in peak value of $j_s$ at the targets. At the UOT the peak shifts closer towards the separatrix for the Ne-seeded simulations where no clear shift is observed for the purely deuterium simulation. At the UIT, the shift in peak for $j_s$ is less pronounced, but if it is observed, it moves away from the separatrix into the SOL. At the inner divertor, due to the low $T_e$ at the UIT itself, the ionization profiles mainly change further away from the target towards the divertor entrance due to drift effects. Over there, a clear shift of the ionization into the SOL region is observed which explains the larger $j_s$ UIT values in the SOL region (right of the $j_s$ peak in figure \ref{subfig:js_UIT_drifts}) in the drift simulations.

The redistribution of the particle due to the ExB drifts, causes in ASDEX Upgrade and JET, a high density region in the inner divertor  \cite{reimold2017high}. The relative importance of drifts can be quantified by the ratio of the parallel particle flux to the outer divertor and the ExB flux towards the PFR part of the inner divertor \cite{sytova2019comparing}. Table \ref{tab:Gamma_ExB_comparison} gives this ratio for the performed simulations. This indicates that for the "SOLPS $\mathrm{D_2}$ 1" and "SOLPS $\mathrm{D_2}$ 2" simulations around 17 $\%$ of the flux arriving at the UOT is redistributed to the UIT due to the ExB drifts. The inclusion of neon in the simulation reduces the effect of drifts. This can also be seen in figure \ref{fig:targets_drifts} where the differences between non-drift and drift simulations are reduced for the "SOLPS Ne 1" simulation. The importance of drifts is nearly vanishing in simulation "SOLPS Ne 2" due to the increased anomalous transport profiles in the PFR and divertor region for that simulation. 

\begin{table}
	\centering
		\begin{tabular}{c | c } 
			& $\mathrm{\frac{\Gamma_{ExB}^{UIT, PFR}}{\Gamma_{||}^{UOT}}}$ \\
			\hline
			SOLPS $\mathrm{D_2}$ 1  & 0.1627 \\ 
			SOLPS $\mathrm{D_2}$ 2 & 0.1760 \\
			SOLPS Ne 1 & 0.1175 \\
			SOLPS Ne 2 & $\sim 0$\\
		\end{tabular}
		\caption{The relative importance of the ExB flows.}
		\label{tab:Gamma_ExB_comparison}
\end{table}

\begin{figure}
	\centering
	\begin{subfigure}{6cm}
		\centering
		\medskip
		\includegraphics[width=7cm]{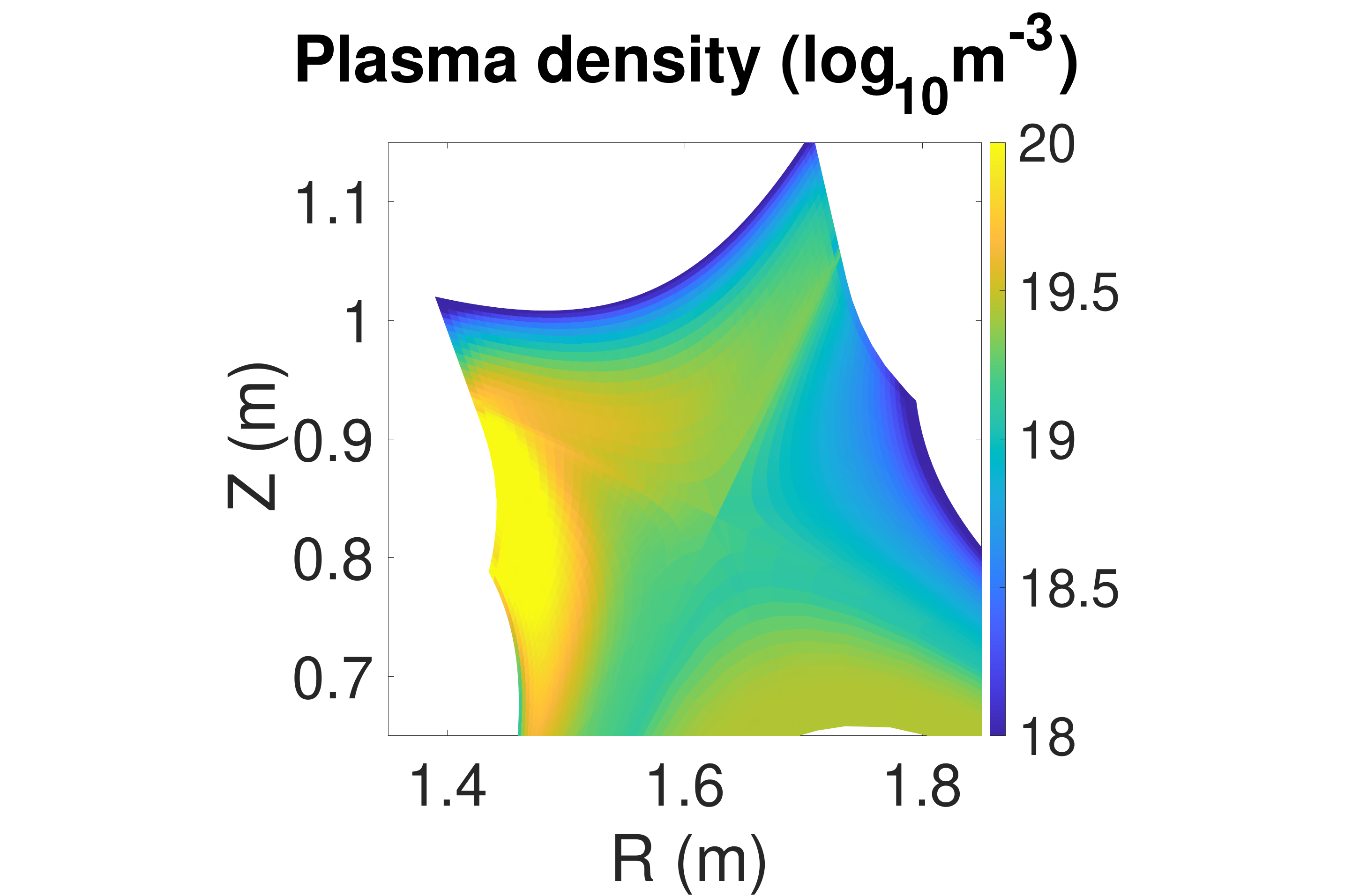}
		\caption{}
		\label{subfig:ne_div_D2_2D}
	\end{subfigure}
	\begin{subfigure}{6cm}
		\centering
		\medskip
		\includegraphics[width=7cm]{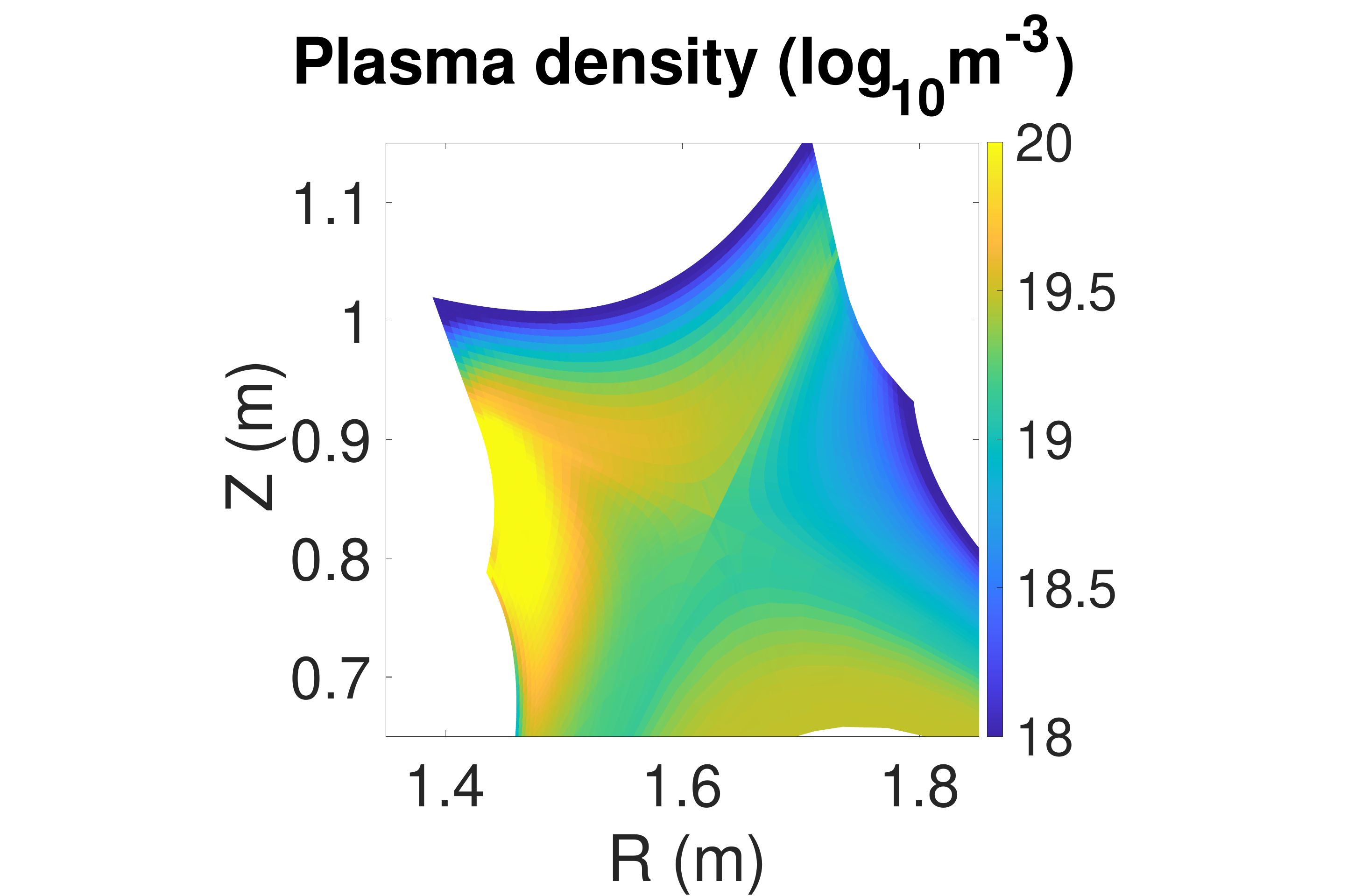}
		\caption{}
		\label{subfig:ne_div_Ne1_2D}
	\end{subfigure}
	\begin{subfigure}{6cm}
		\centering
		\medskip
		\includegraphics[width=6cm]{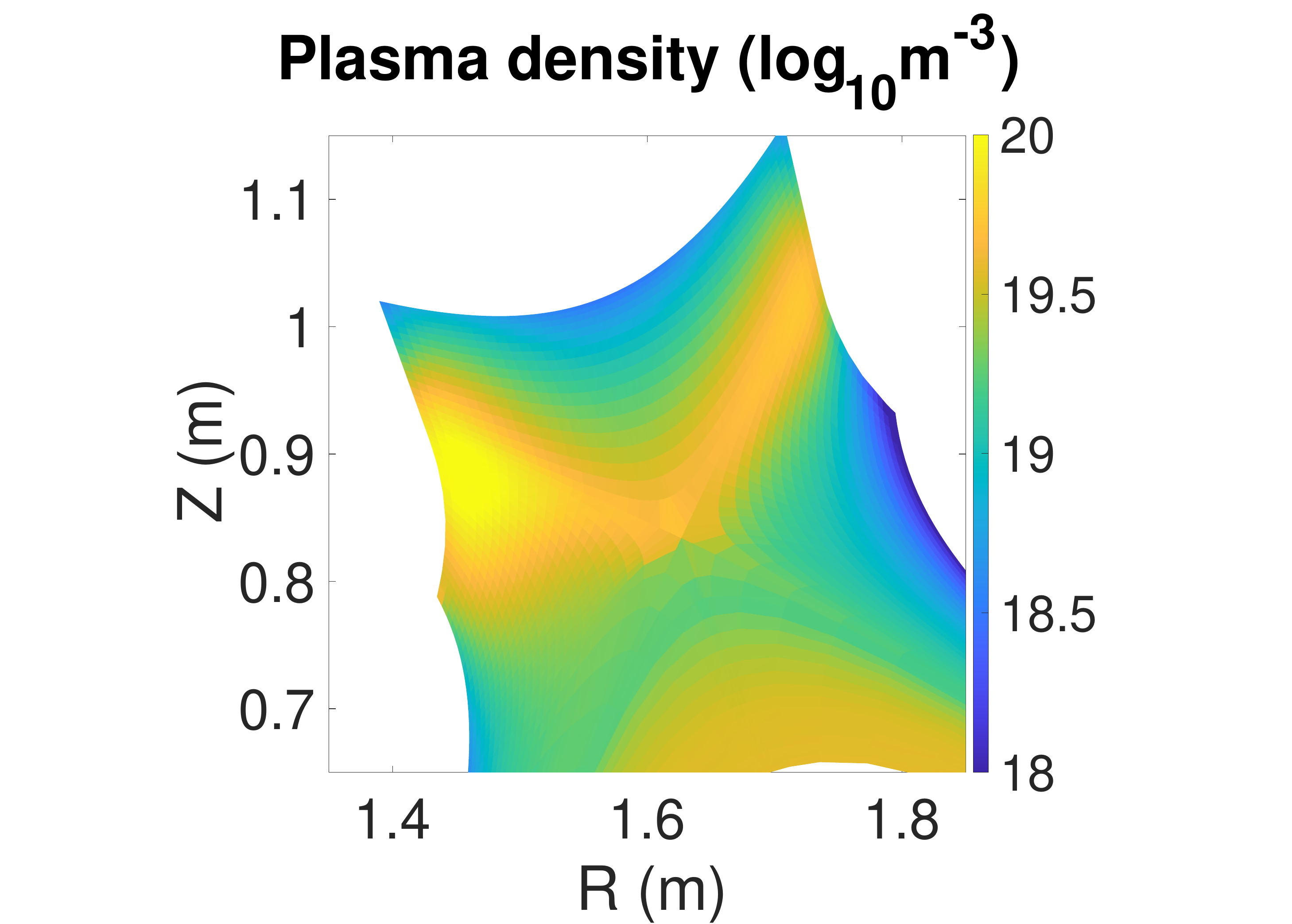}
		\caption{}
		\label{subfig:ne_div_Ne_2D}
	\end{subfigure}
	\caption{The simulated 2D density profiles in the vicinity of the divertor region for the SOLPS D$_2$ 1 (a), SOLPS Ne 1 (b) and SOLPS Ne 2 (c) simulations.}
	\label{fig:ne_div_2D}
\end{figure}

\begin{figure}
	\centering
	\medskip
	\includegraphics[width=8cm]{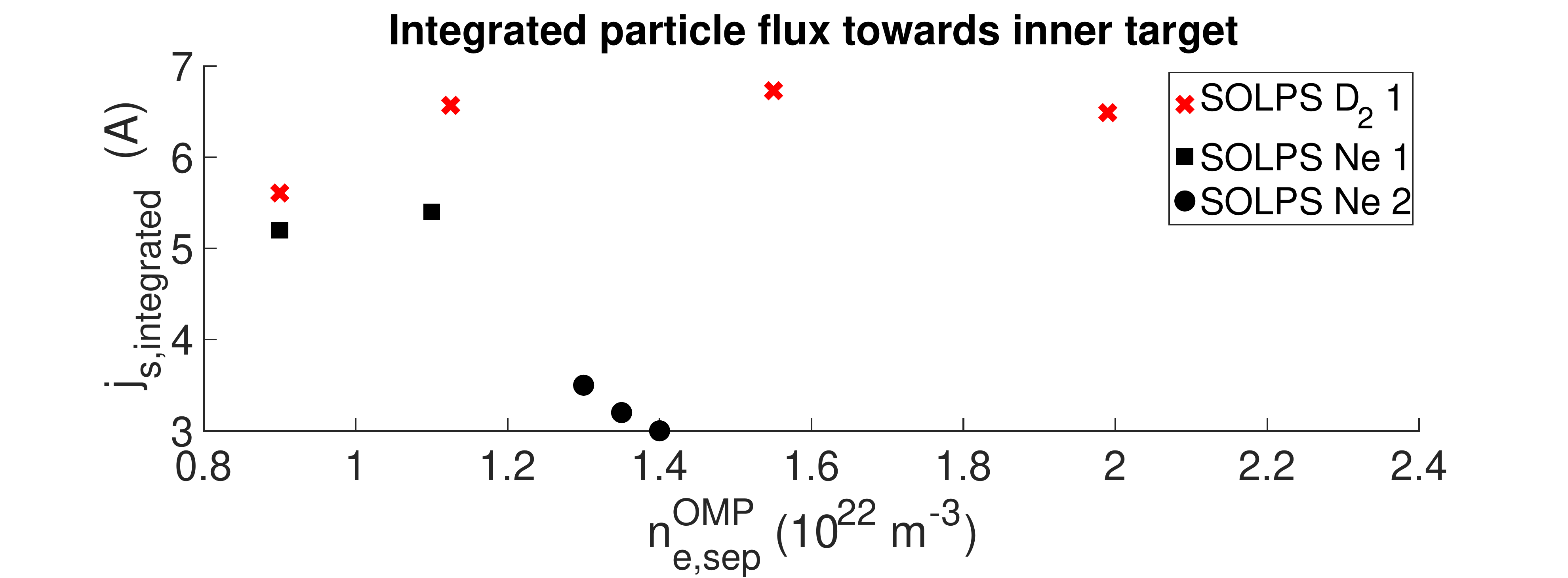}
	\caption{A density scan to study the degree of detachment for the different simulation setups. The original simulation is the most left point on the plot for all setups.}
	\label{fig:rollover_densityscan}
\end{figure}

This redistribution causes that the inner divertor becomes detached where the outer one stays attached, and is called the high field side high density. From figure \ref{fig:ne_div_2D} it is clear that a higher density is present at the inboard side for all the performed EAST simulations. The density scan in figure \ref{fig:rollover_densityscan} shows that the "SOLPS D$_2$ 1" and "SOLPS Ne 1" simulations are still attached and have not yet reached the rollover despite the low temperatures at the UIT. The "SOLPS Ne 2" simulation, on the other hand, has a decreasing particle flux when the density is further increased, so is detached at the UIT. A similar analysis shows that also the UOT in this last simulation is detached. The other simulations are attached at the UOT. As detachment is going together with a larger density, it is not surprising that the later simulation has the highest density profiles at the UIT (see figure \ref{subfig:ne_UOT}). As for the ASDEX Upgrade simulations from ref. \cite{reimold2014experimental,reimold2017high}, drifts are required to reproduce the higher density at the inner target. Drifts in fact increase the ionization sources at the inboard side as can be seen in figure \ref{fig:ionization_detail_drifts} for the "SOLPS D$_2$ 1" simulation.


\begin{figure}
\centering
\begin{subfigure}{6cm}
	\centering
	\medskip
	\includegraphics[trim={6.5cm 0 6.5cm 0},width=6cm]{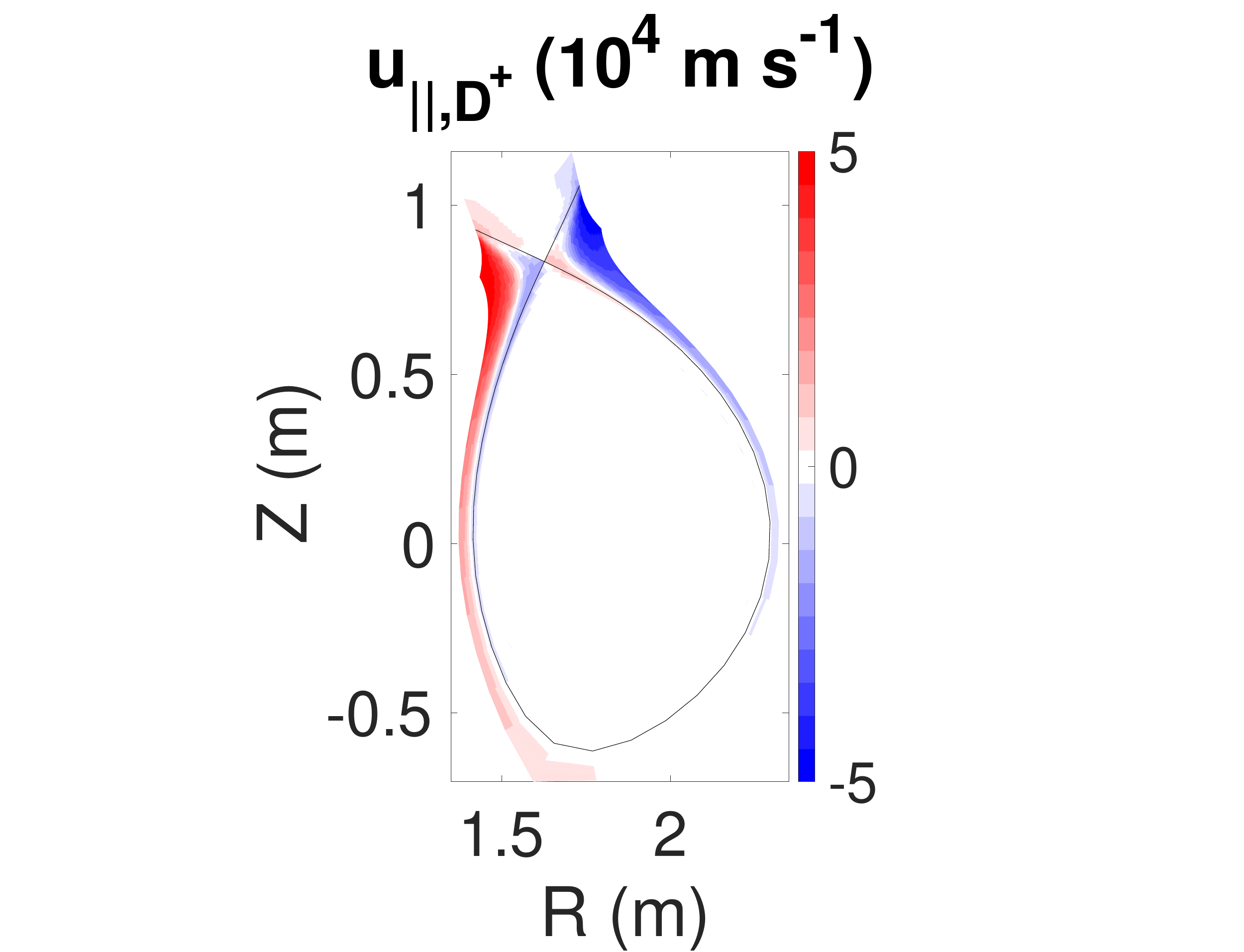}
	\caption{}
	\label{subfig:ua_D_D2_no_drifts}
\end{subfigure}
\begin{subfigure}{6cm}
	\centering
	\medskip
	\includegraphics[trim={6.5cm 0 6.5cm 0},width=6cm]{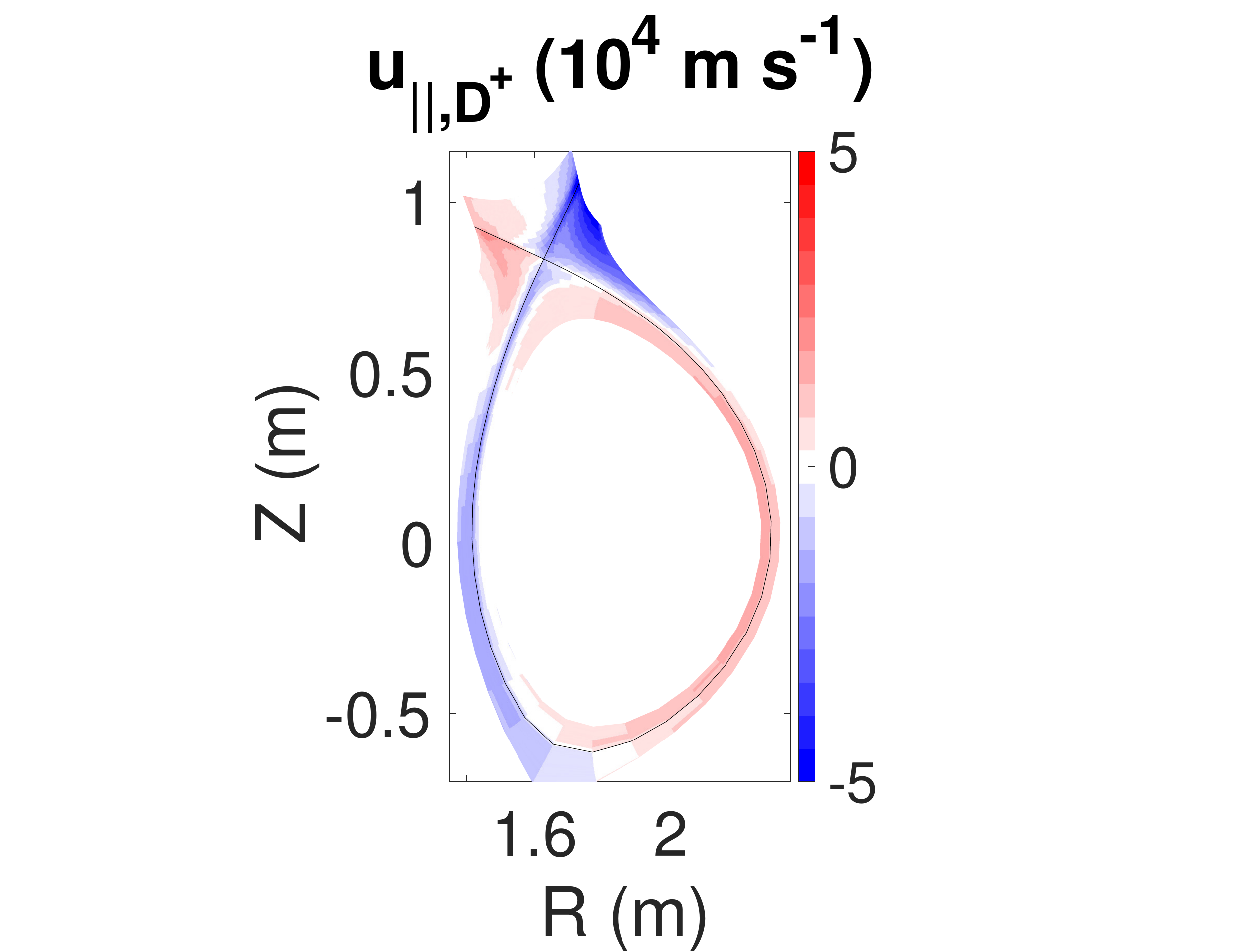}
	\caption{}
	\label{subfig:u_D_D2}
\end{subfigure}
\begin{subfigure}{6cm}
		\centering
		\medskip
		\includegraphics[trim={6.5cm 0 6.5cm 0},width=6cm]{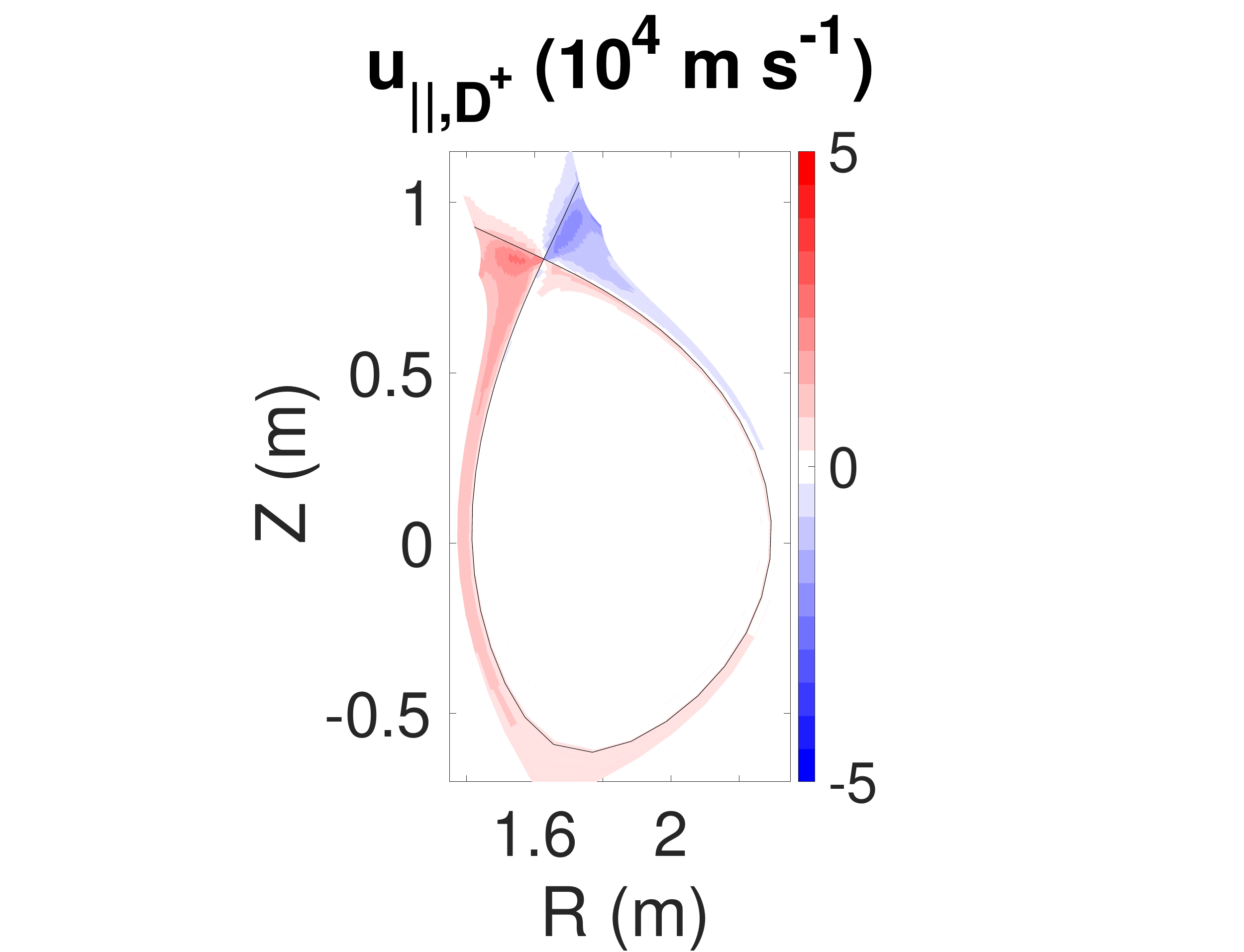}
		\caption{}
		\label{subfig:u_D_Ne2_no_drifts}
\end{subfigure}
\begin{subfigure}{6cm}
		\centering
		\medskip
		\includegraphics[trim={6.5cm 0 6.5cm 0},width=6cm]{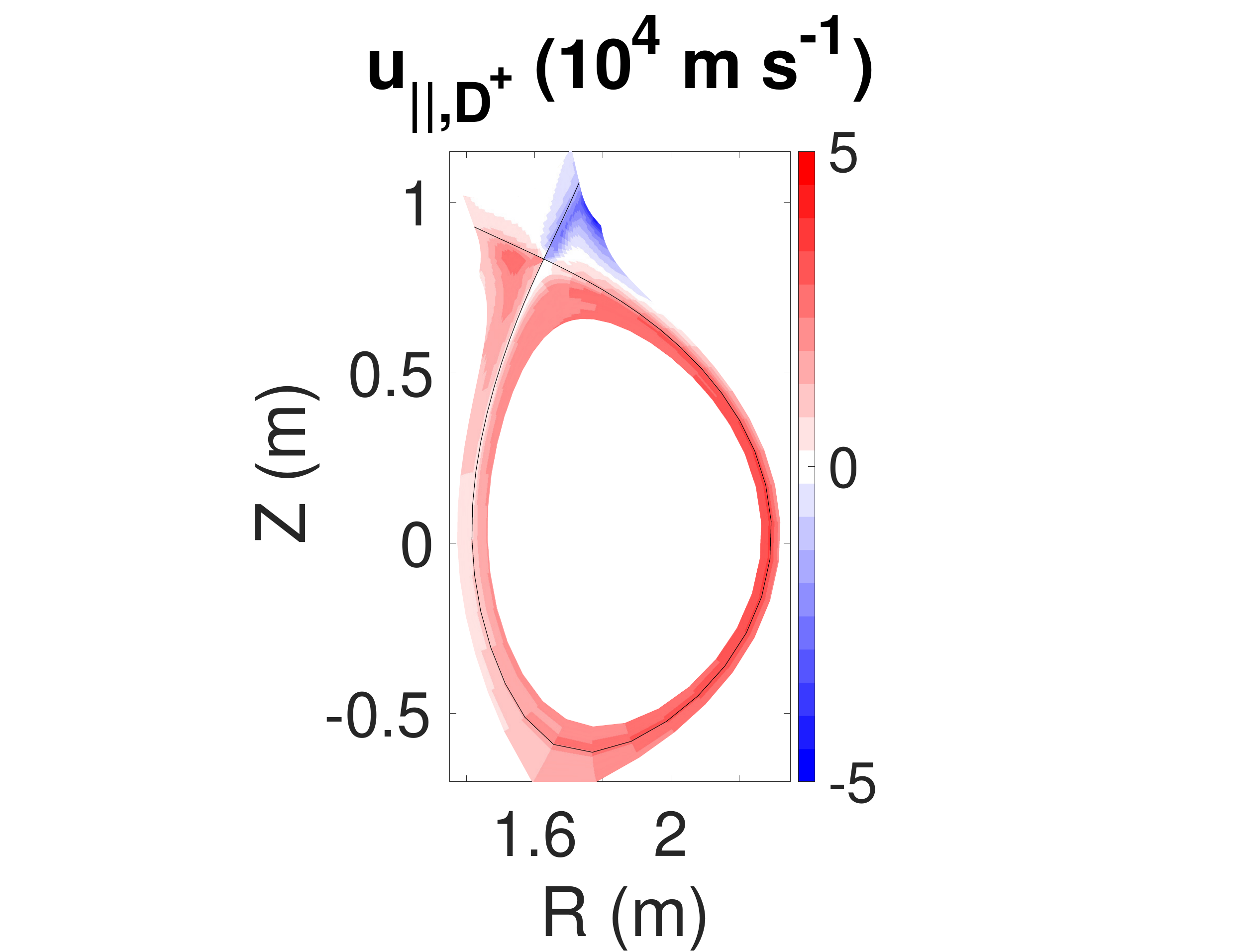}
		\caption{}
		\label{subfig:u_D_Ne2}
\end{subfigure}
\caption{The parallel velocity of D$^+$ for the different simulations: "SOLPS D$_2$ 1" without (a) and with (b) drifts, and "SOLPS Ne 2" without (c) and with (d) drifts. The separatrix in all figures is indicated in black.}
\label{fig:u_D}
\end{figure}

\begin{figure}
	\centering
	\begin{subfigure}{6cm}
		\centering
		\medskip
		\includegraphics[trim={6cm 0 6cm 0},width=6cm]{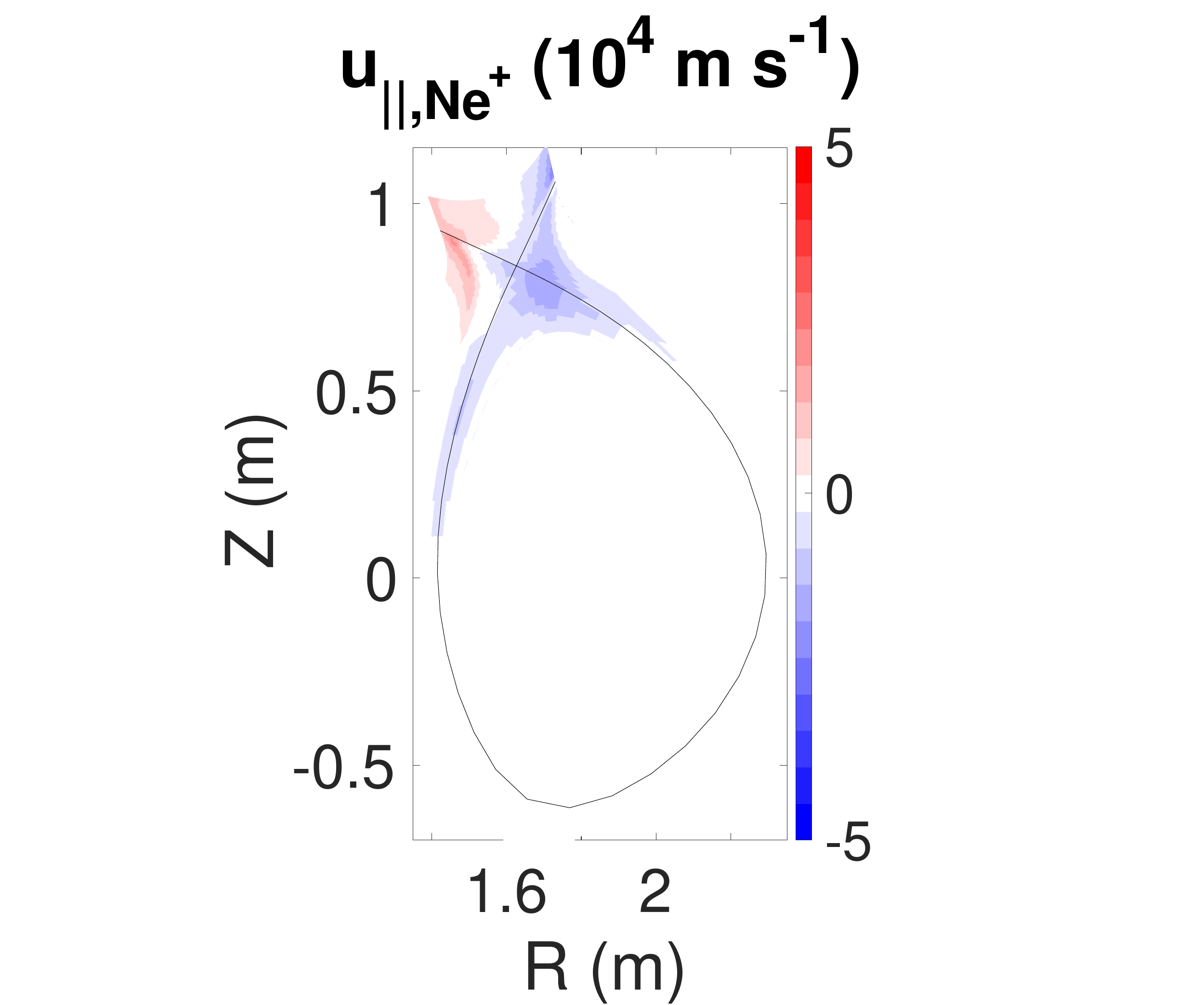}
		\caption{}
		\label{subfig:u_Ne1_Ne1}
	\end{subfigure}
	\begin{subfigure}{6cm}
		\centering
		\medskip
		\includegraphics[trim={6cm 0 6cm 0},width=6cm]{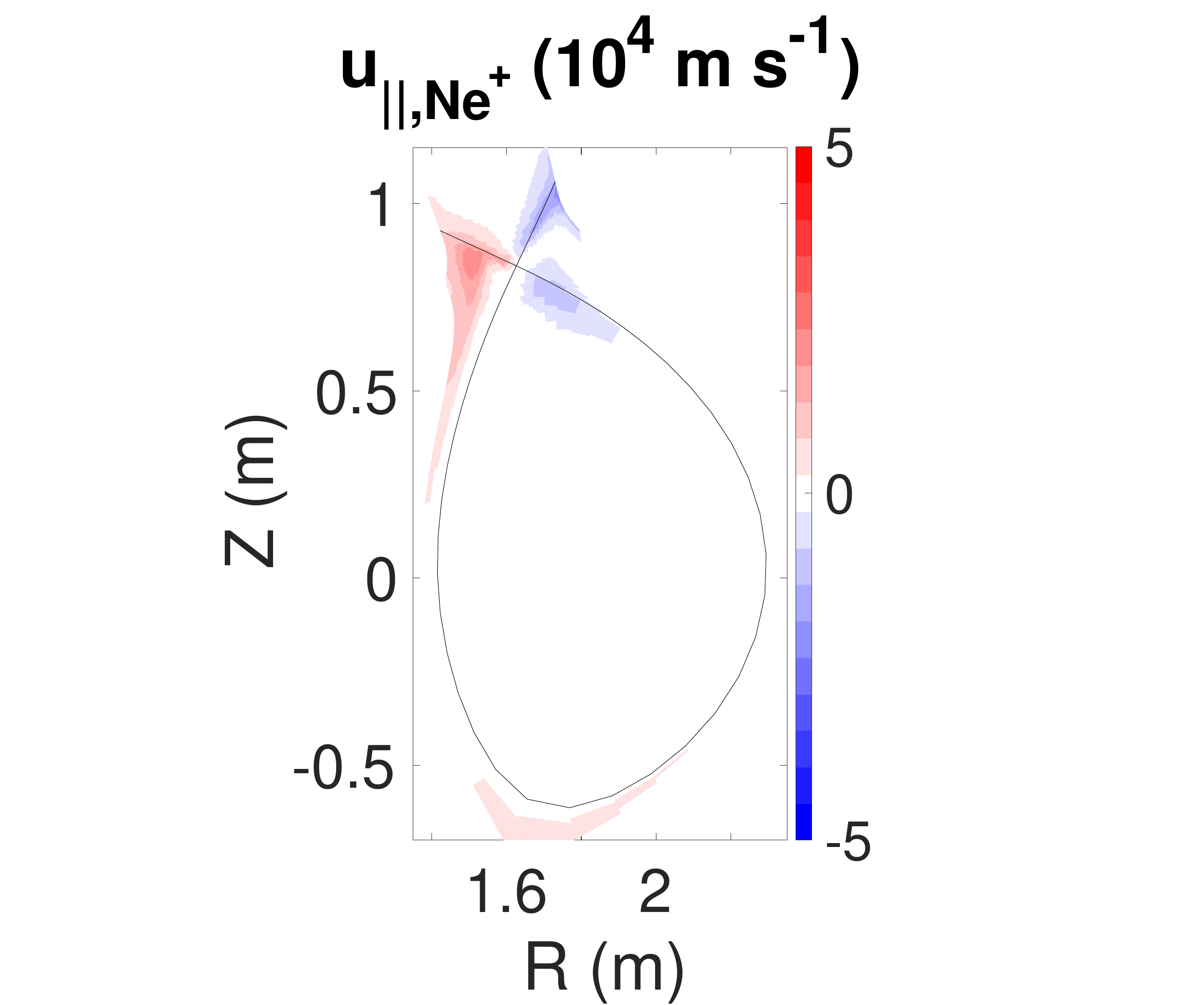}
		\caption{}
		\label{subfig:u_Ne1_Ne2}
	\end{subfigure}
	\caption{The parallel velocity of Ne$^{+}$ for the different simulations: "SOLPS Ne 1" (a), and "SOLPS Ne 2" (b). The separatrix in all figures is indicated in black.}
	\label{fig:u_Ne1}
\end{figure}

\begin{figure}
	\centering
	\begin{subfigure}{6cm}
		\centering
		\medskip
		\includegraphics[trim={6cm 0 6cm 0},width=6cm]{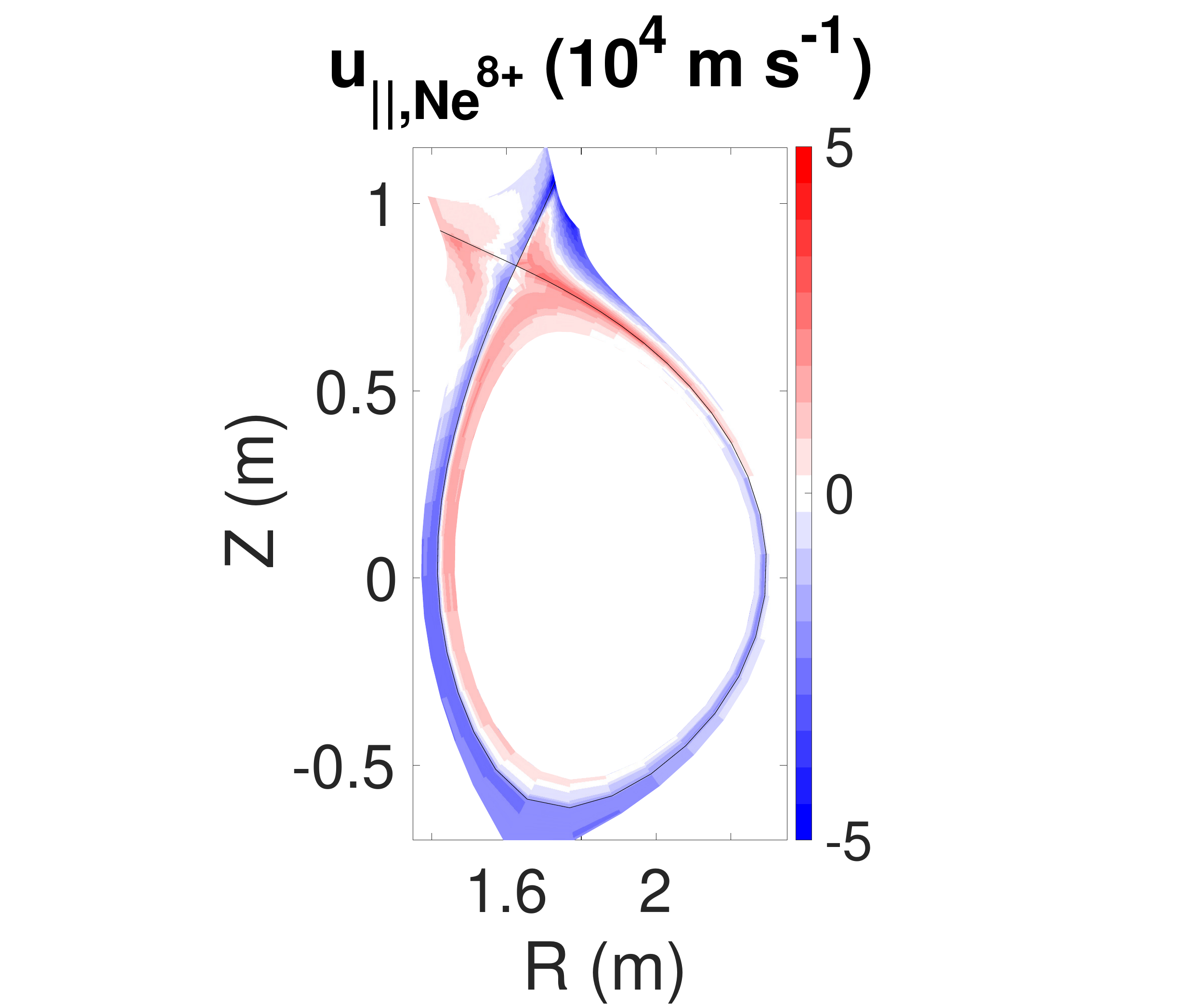}
		\caption{}
		\label{subfig:u_Ne8_Ne1}
	\end{subfigure}
	\begin{subfigure}{6cm}
		\centering
		\medskip
		\includegraphics[trim={6cm 0 6cm 0},width=6cm]{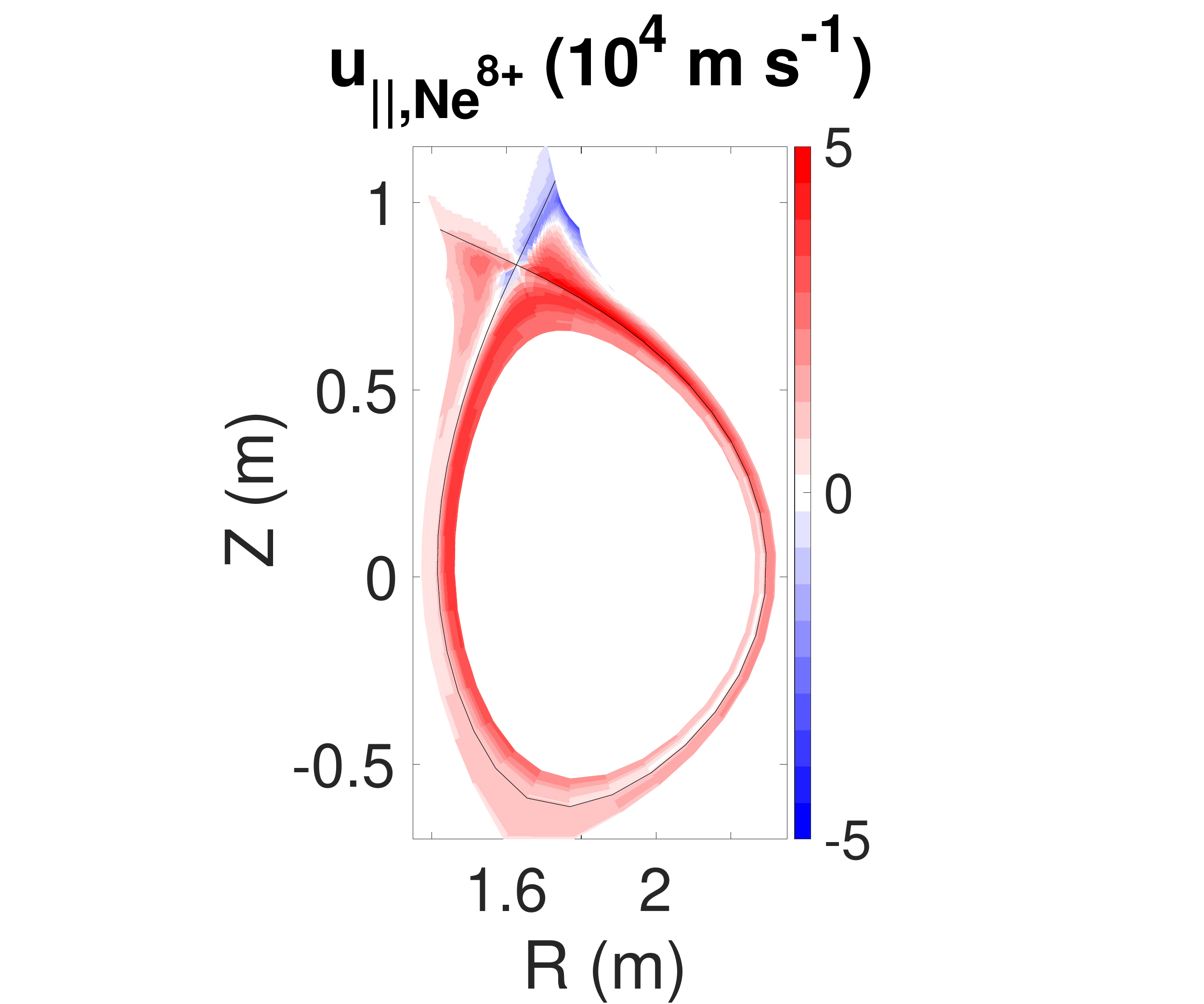}
		\caption{}
		\label{subfig:u_Ne8_Ne2}
	\end{subfigure}
	\caption{The parallel velocity of Ne$^{8+}$ for the different simulations: "SOLPS Ne 1" (a), and "SOLPS Ne 2" (b). The separatrix in all figures is indicated in black.}
	\label{fig:u_Ne8}
\end{figure}

Another way to study the redistribution from particles between the inner and outer target due to drift effects, is the analysis of the flow direction and flow reversal. In figures \ref{fig:u_D} the evolution of the parallel velocity for D$^+$ is studied for the "SOLPS D$_2$ 1" and "SOLPS Ne 2" simulations. the velocity profile for the "SOLPS Ne1" simulation is very similar to the one of "SOLPS D$_2$ 1". Similar figures for Ne$^+$ and Ne$^{8+}$ in the presence of drifts are shown in figures \ref{fig:u_Ne1} and \ref{fig:u_Ne8}. When the velocity is zero, the studied ion is not moving in the parallel direction and a stagnation point appears. The same conventions for the flow directions as for the ExB velocities are used.

Figure \ref{fig:u_D} shows that in the absence of drifts, the parallel velocity pattern is nearly symmetric and particles move towards the closest target. Drift effects change this picture completely. B $\nabla$B flows influence the velocity pattern mainly in the main SOL where the earlier discussed ExB flows influence the parallel velocity pattern mainly in the PFR and divertor region. Figure \ref{fig:u_D} shows that in the detached simulation "SOLPS Ne 2", there is the largest shift in movement towards the inner target. Due to the magnetic curvature, most particles will escape from the core towards the SOL around the position of the OMP. Figure \ref{subfig:u_D_Ne2} indicates that most of these particles flow towards the inner target in the "SOLPS Ne 2" simulation, where this influence is smaller in the absence of drifts.

In cases with impurities, the stagnation point of the impurity ions, in combination with the ionization location will also indicate if there is retention of the impurity in the divertor region, or if there is more leakage towards the SOL and to the core \cite{senichenkov2019mechanisms}. The mechanism behind this, is a balance of drifts and friction with the main ions. The OMP-to-target impurity flux ratio is determined by the fraction of impurity neutrals ionized beyond the stagnation point, with these impurities being carried upstream by friction with the main ions as described in ref. \cite{senichenkov2019mechanisms}. As drifts strongly influence the poloidal velocity profile of the main ions (see figure \ref{fig:u_D}), they also influence the one of the impurities. Figures \ref{fig:u_Ne1} and \ref{fig:u_Ne8} indicate that the asymmetry of the impurity velocities increases when the charge state is higher.

\begin{figure}
	\centering
		\begin{subfigure}{6cm}
		\centering
		\medskip
		\includegraphics[trim={8cm 0.5cm 8cm 0.5cm},clip,height=6cm]{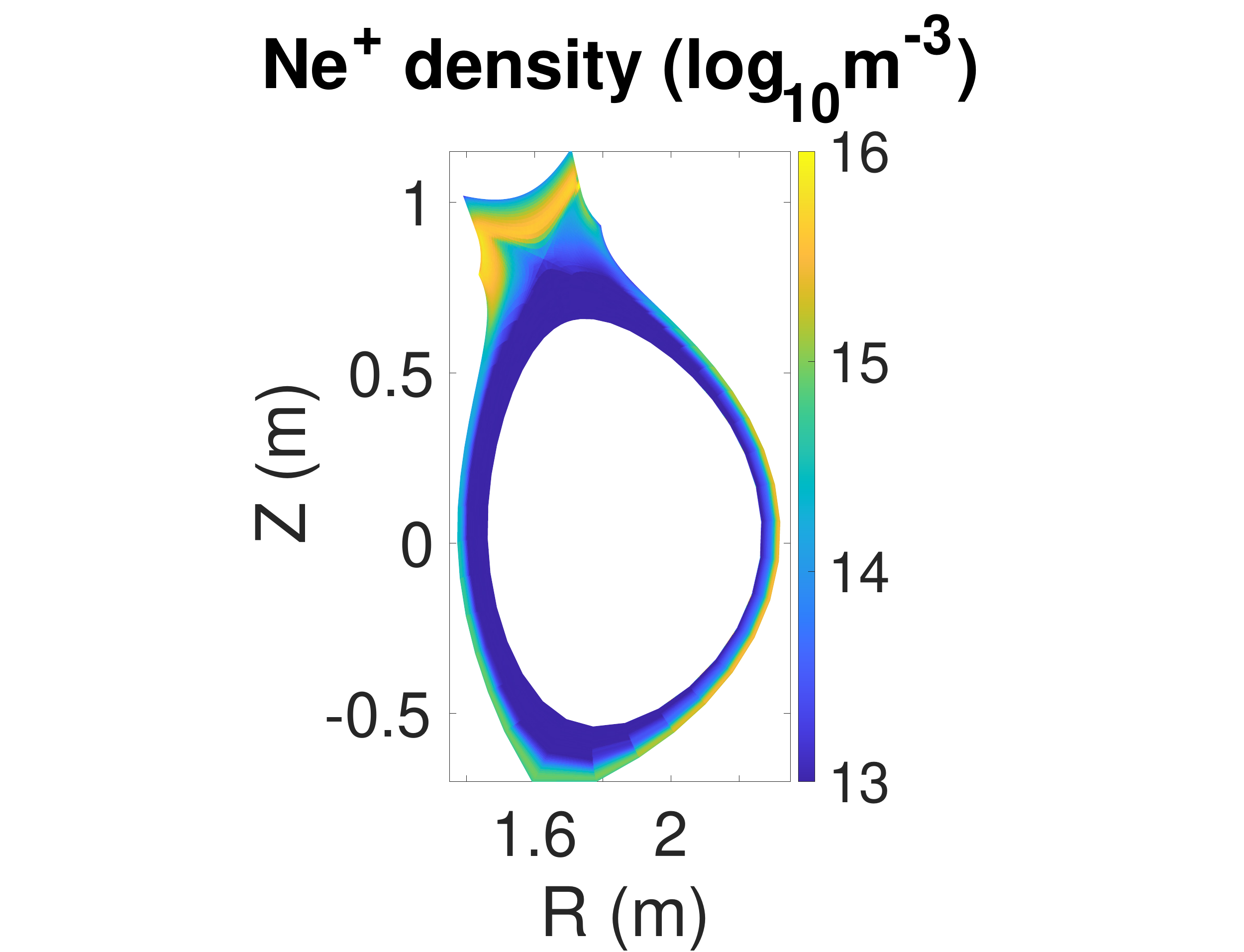}
		\caption{}
		\label{subfig:density_Ne+_Ne1_2D}
	\end{subfigure}
	\begin{subfigure}{6cm}
		\centering
		\medskip
		\includegraphics[trim={8cm 0.5cm 8cm 0.5cm},clip,height=6cm]{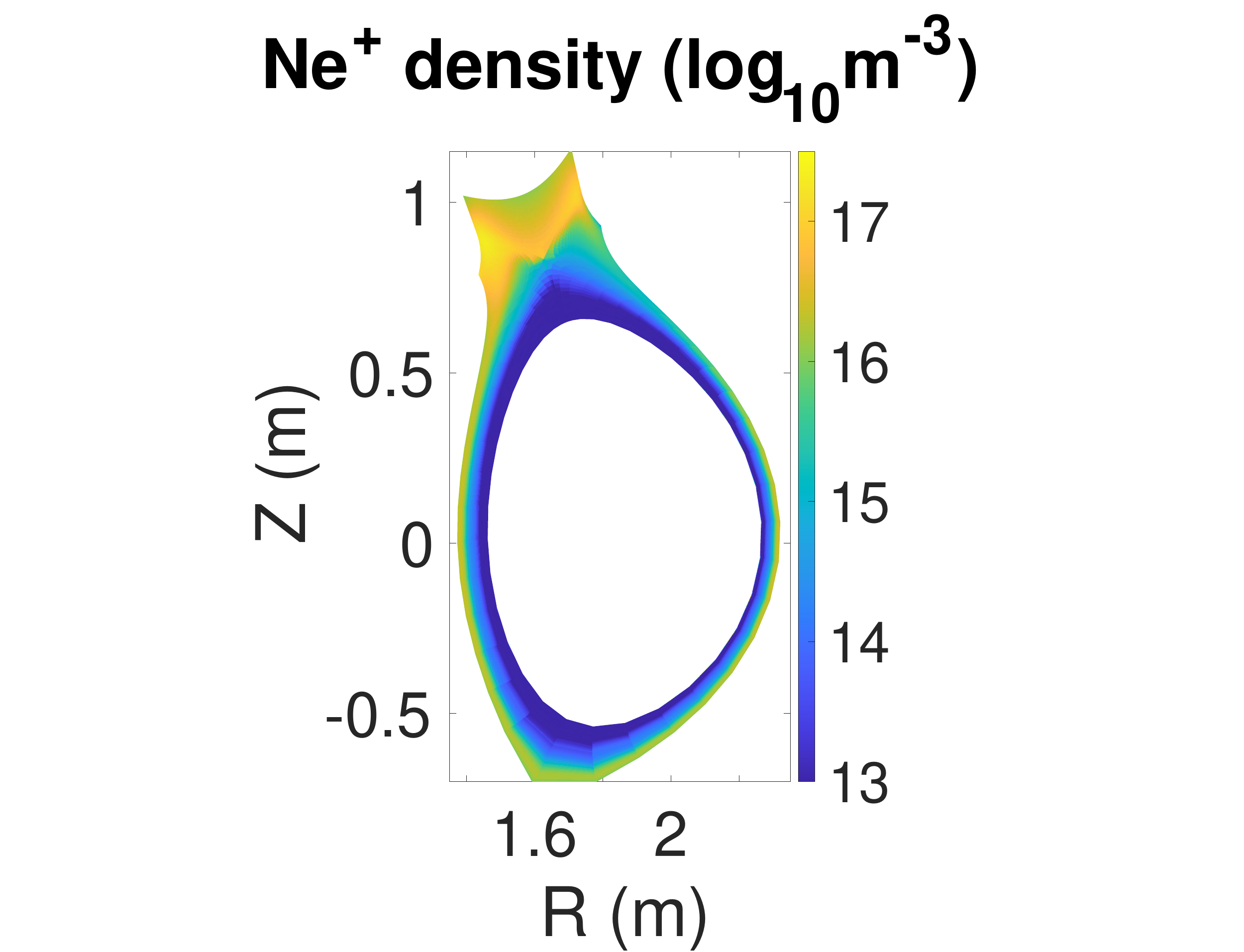}
		\caption{}
		\label{subfig:density_Ne+_Ne2_2D}
	\end{subfigure}
	\begin{subfigure}{6cm}
		\centering
		\medskip
		\includegraphics[trim={8cm 0.5cm 8cm 0.5cm},clip,height=6cm]{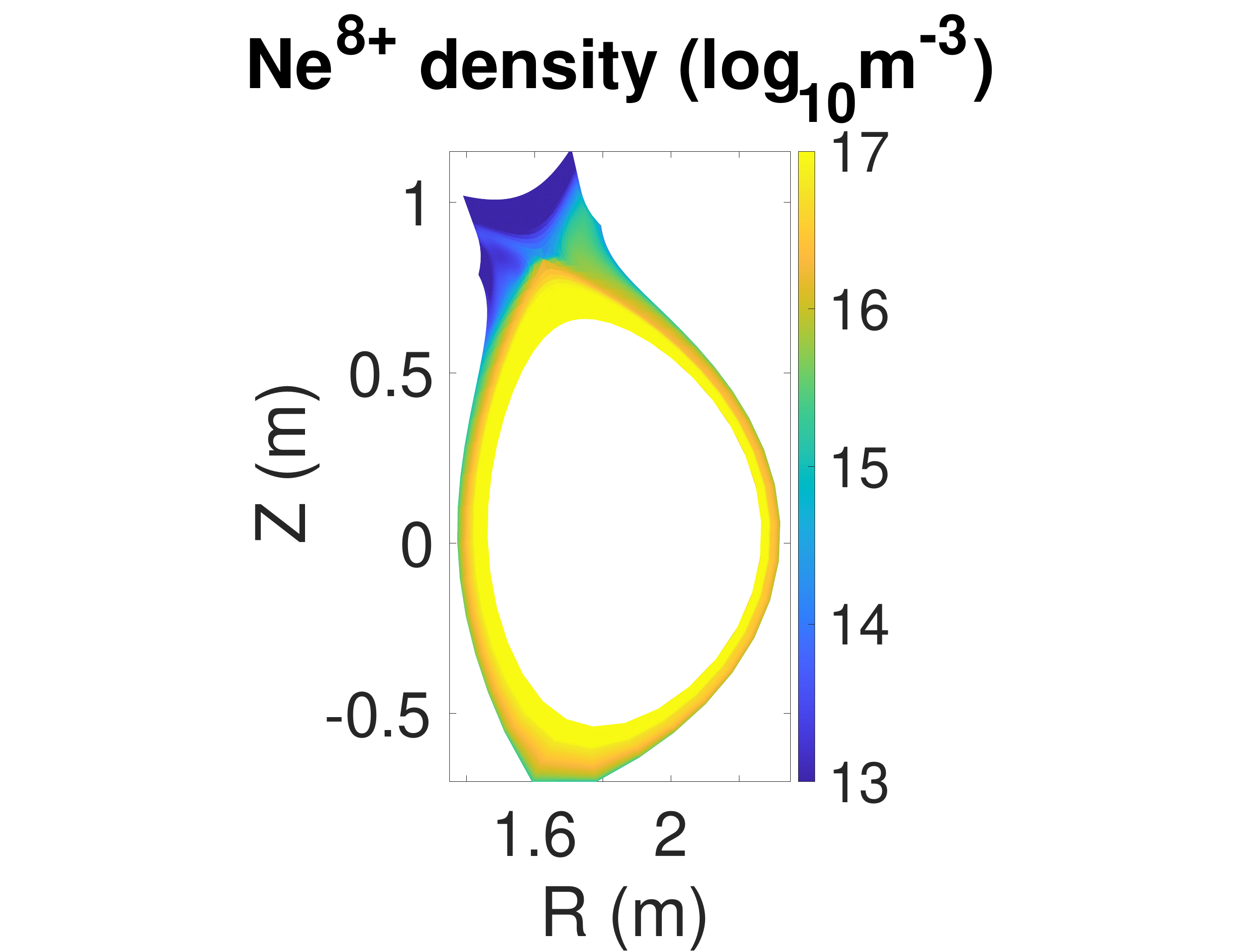}
		\caption{}
		\label{subfig:density_Ne8+_Ne1_2D}
	\end{subfigure}
	\begin{subfigure}{6cm}
		\centering
		\medskip
		\includegraphics[trim={8cm 0.5cm 8cm 0.5cm},clip,height=6cm]{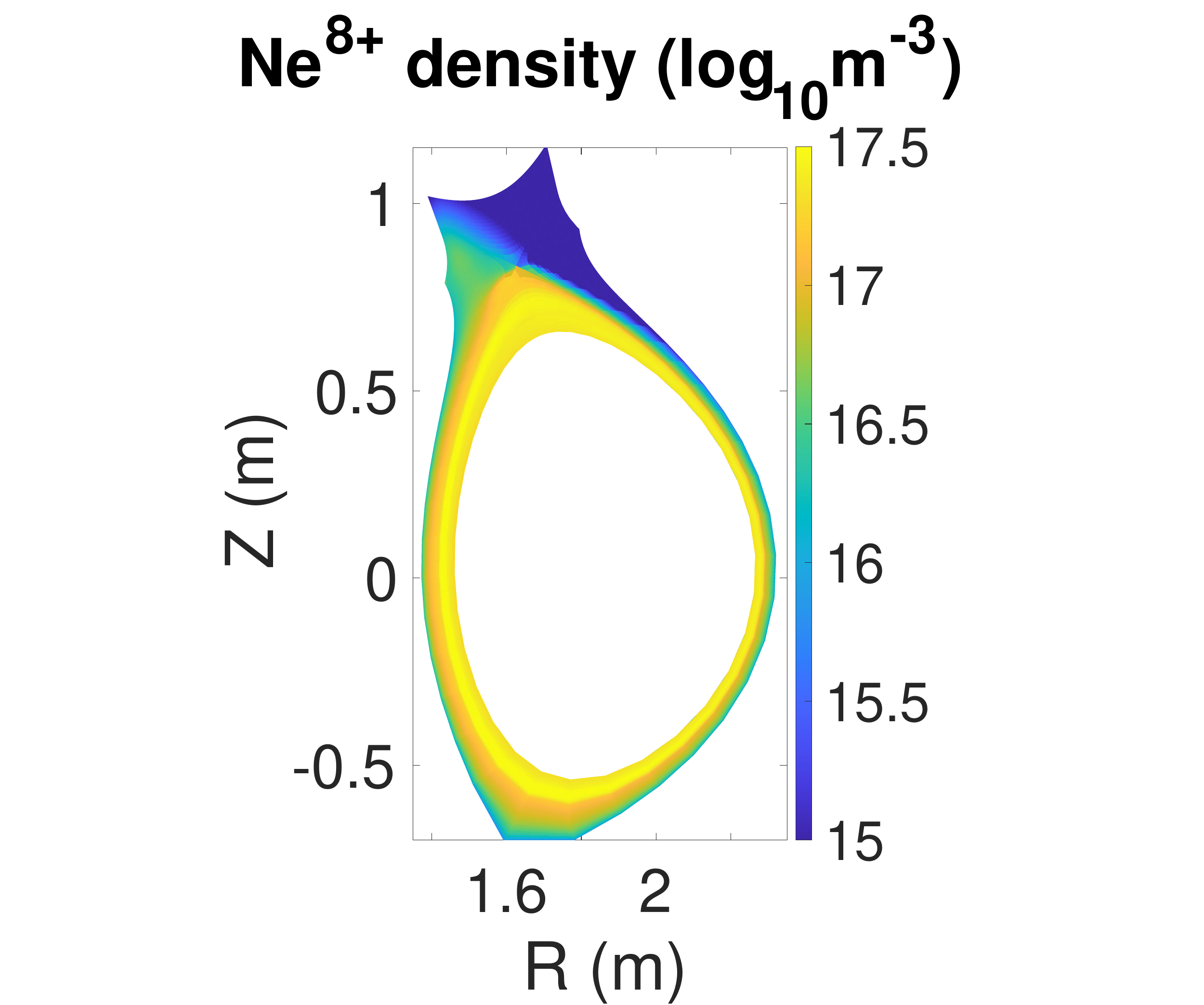}
		\caption{}
		\label{subfig:density_Ne8+_Ne2_2D}
	\end{subfigure}
	\caption{The densities for Ne$^+$ and Ne$^{8+}$ simulated with SOLPS-ITER for "SOLPS Ne 1" (a,c) and "SOLPS Ne 2" (b,d).}
	\label{fig:density_ion_Ne_2D}
\end{figure}

\begin{figure}
	\centering
	\begin{subfigure}{6cm}
		\centering
		\medskip
		\includegraphics[clip,height=6cm]{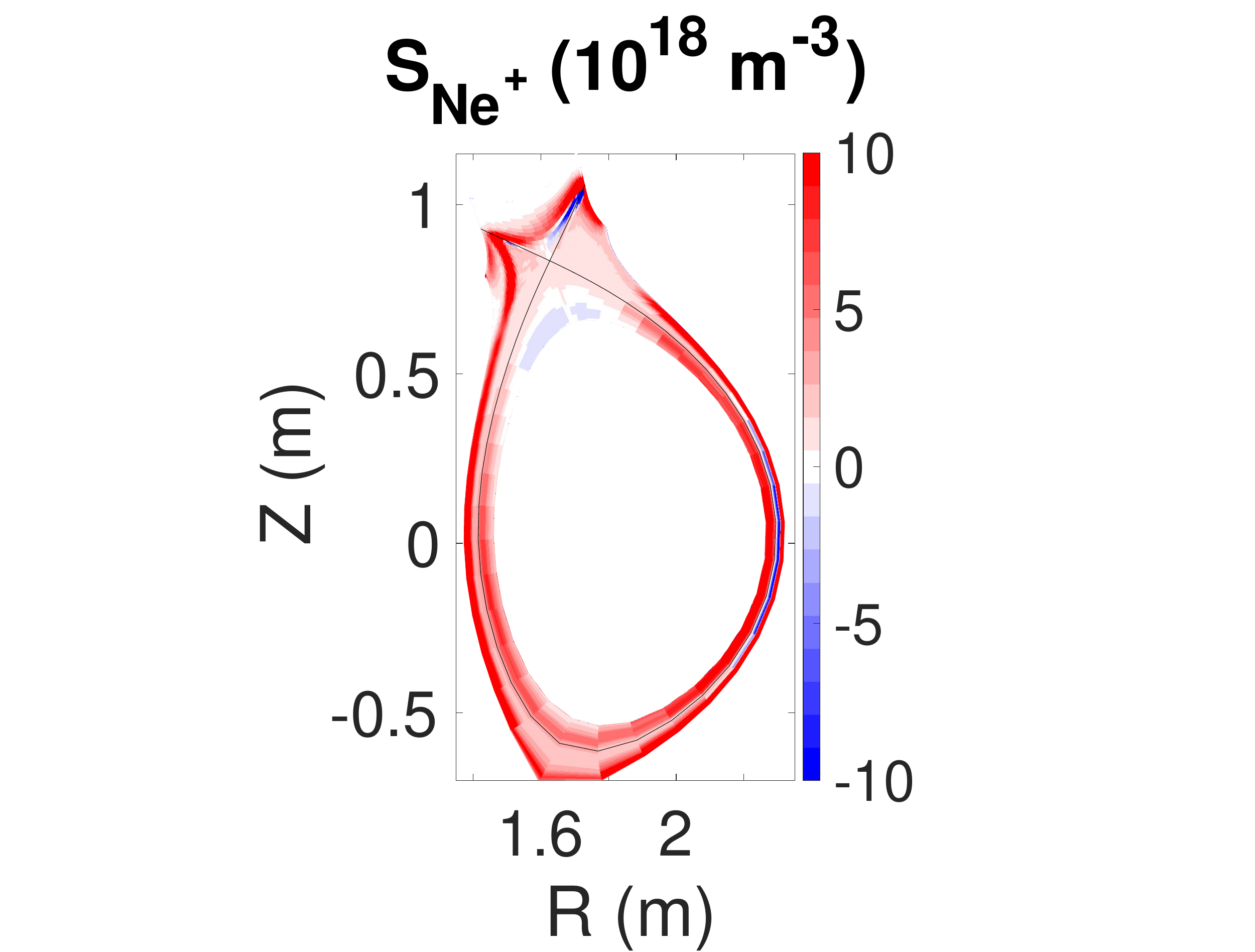}
		\caption{}
		\label{subfig:ionization_Ne+_Ne1_2D}
	\end{subfigure}
	\begin{subfigure}{6cm}
		\centering
		\medskip
		\includegraphics[,clip,height=6cm]{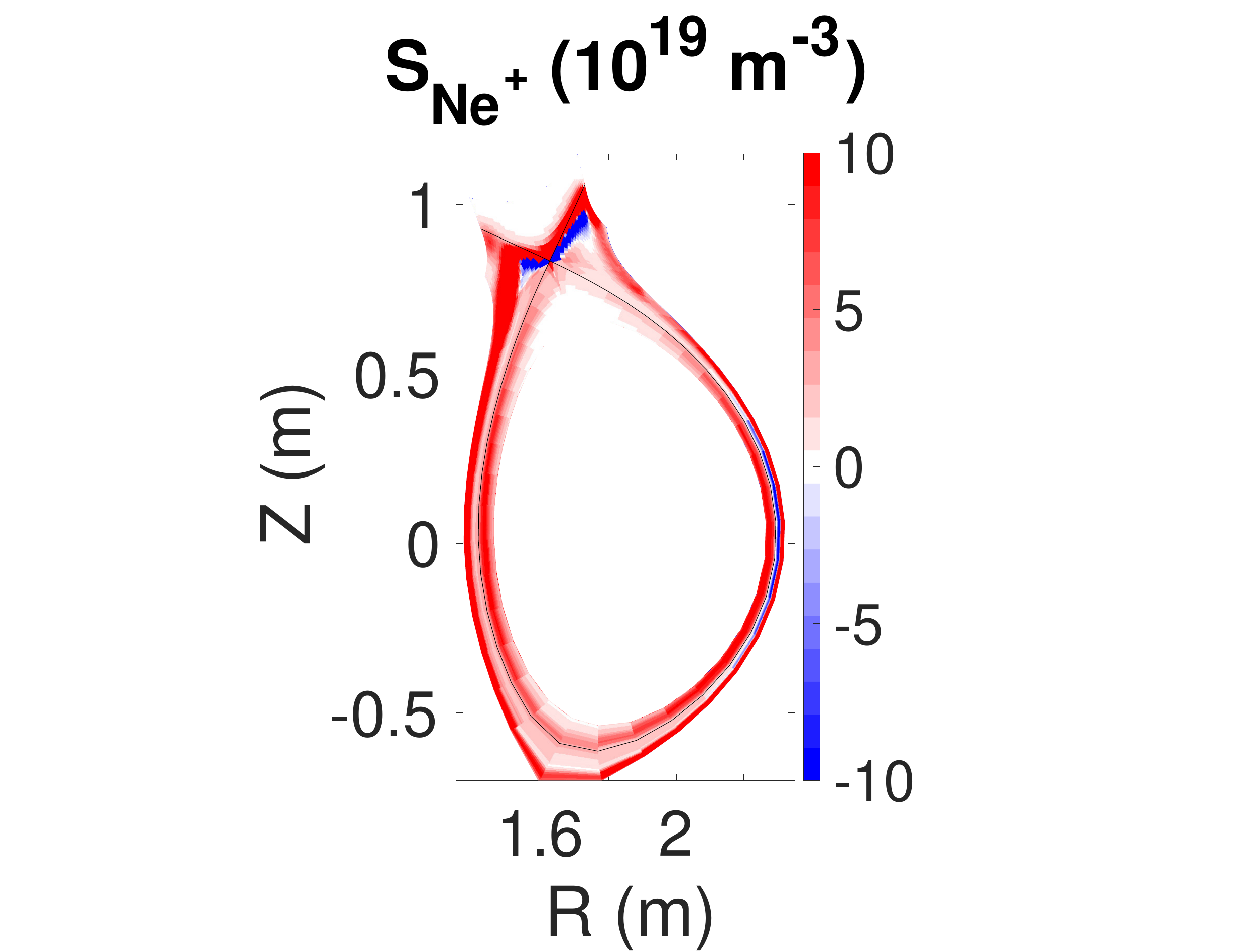}
		\caption{}
		\label{subfig:ionization_Ne+_Ne2_2D}
	\end{subfigure}
		\begin{subfigure}{6cm}
		\centering
		\medskip
		\includegraphics[clip,height=6cm]{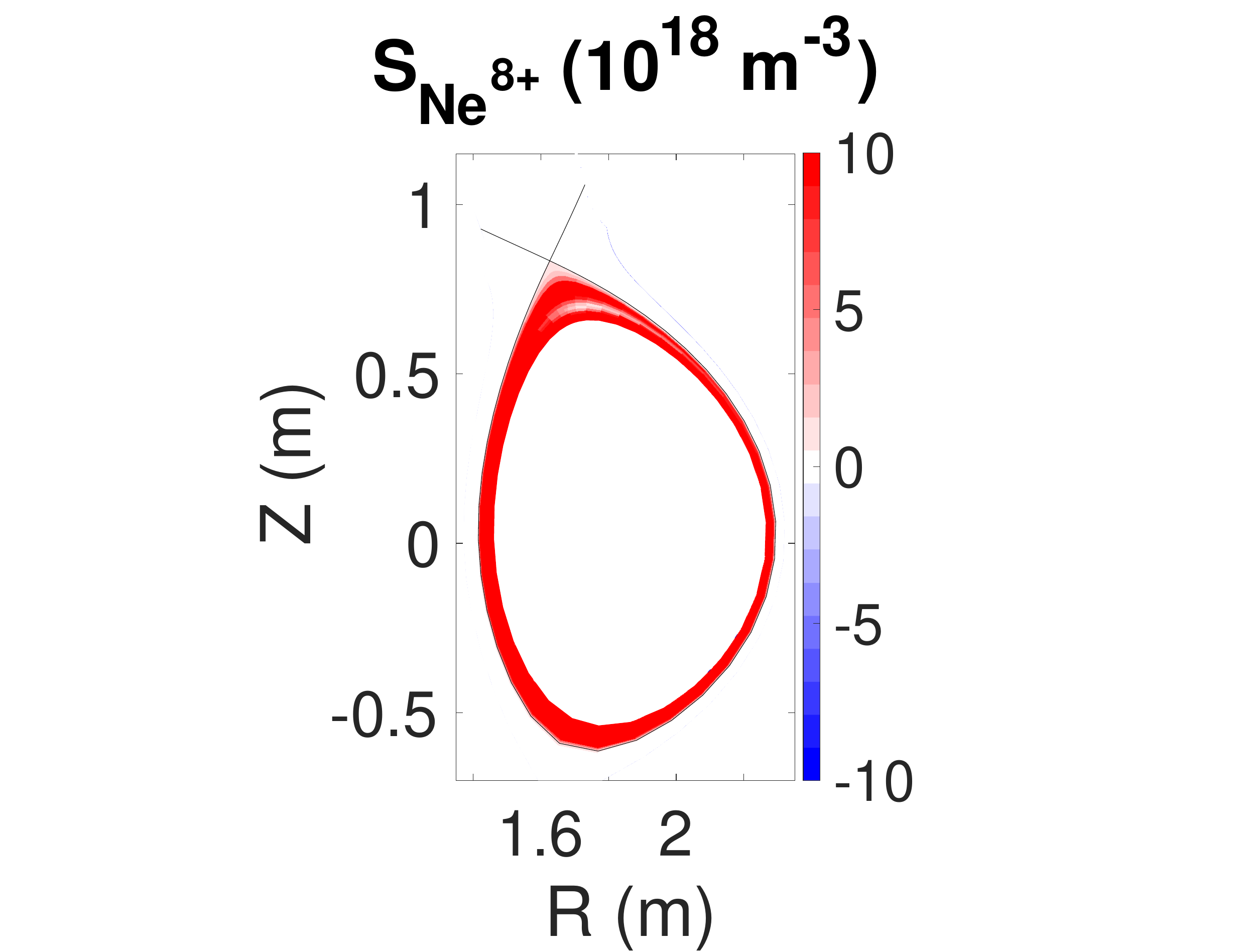}
		\caption{}
		\label{subfig:ionization_Ne8+_Ne1_2D}
	\end{subfigure}
	\begin{subfigure}{6cm}
		\centering
		\medskip
		\includegraphics[clip,height=6cm]{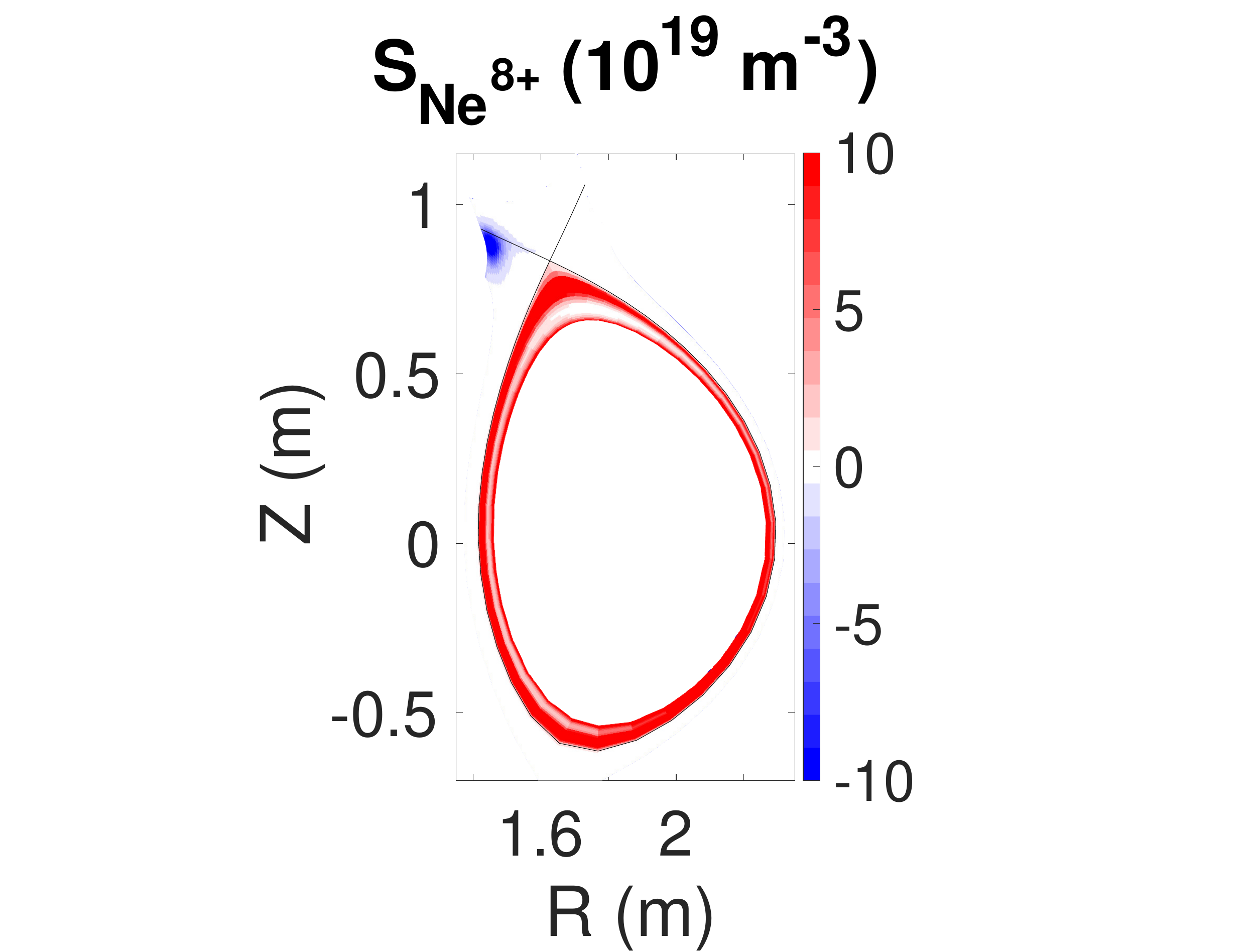}
		\caption{}
		\label{subfig:ionization_Ne8+_Ne2_2D}
	\end{subfigure}
	\caption{The ionization sources from neutral to Ne$^+$ and from Ne$^{7+}$ to Ne$^{8+}$ simulated with SOLPS-ITER for "SOLPS Ne 1" (a,c) and "SOLPS Ne 2" (b,d).}
	\label{fig:ionization_ion_Ne_Temperature_2D}
\end{figure}

The densities for Ne$^+$ and Ne$^{8+}$ are shown in figure \ref{fig:density_ion_Ne_2D}. The density profiles (not exact values) are similar at most locations between the "SOLPS Ne 1" and "SOLPS Ne 2" simulations, apart from the Ne$^{8+}$ density in the vicinity of the target, where there is a peak in density at the UOT for "SOLPS Ne 1", there is rather a peak at the UIT in the "SOLPS Ne 2" simulation. The significant larger parallel velocity of Ne$^{8+}$ shown in figure \ref{fig:u_Ne8} indicates that in the later simulation more of the Ne$^{8+}$ is transported to the inner target. This causes also a different location for the ionization of Ne$^{8+}$. These ionization sources, as well as the ones for  Ne$^+$ are given in figure \ref{fig:ionization_ion_Ne_Temperature_2D}. For the low charge states of neon, the ionization is located in the entire edge, where the higher charge states ionize more towards the core. Figure \ref{fig:u_Ne1} makes clear that there is a large region where the parallel velocity of Ne$^+$ is close to zero. Figure \ref{fig:ionization_ion_Ne_Temperature_2D} shows that ionization of Ne$^+$ is taking place within this region. In that way, the transport mechanism described above will make that Ne$^+$ leaks towards the core. As there the temperature increases, the higher charge states of neon ionizes inside the separatrix as is the case for Ne$^{8+}$.
This leakage in combination with the cooling down of the core due to the ionization of higher charge states of neon explains why it was not possible to perform H-mode experiments with a higher neon injection. 


The small influence of neon on the experimental $T_e$ profiles of figure \ref{subfig:Te_UIT}, on the other hand, indicates that the transport of neon towards the inner target is overestimated in the simulations. The leakage of higher order charge states of Ne is nevertheless confirmed by the EUV spectrometer data from figure \ref{subfig:filterscope}.The chords of the EUV spectrometer are going through the core in order to view the divertor region so they include the core content as well.  

\section{Effect of neon on radiated power fraction}
\label{sec:radiated_power}

The main goal of the injected neon is to radiate a significant power fraction and in that way detach the plasma.

\begin{figure}
	\centering
	\begin{subfigure}{4cm}
		\centering
		\medskip
		\includegraphics[width=4cm]{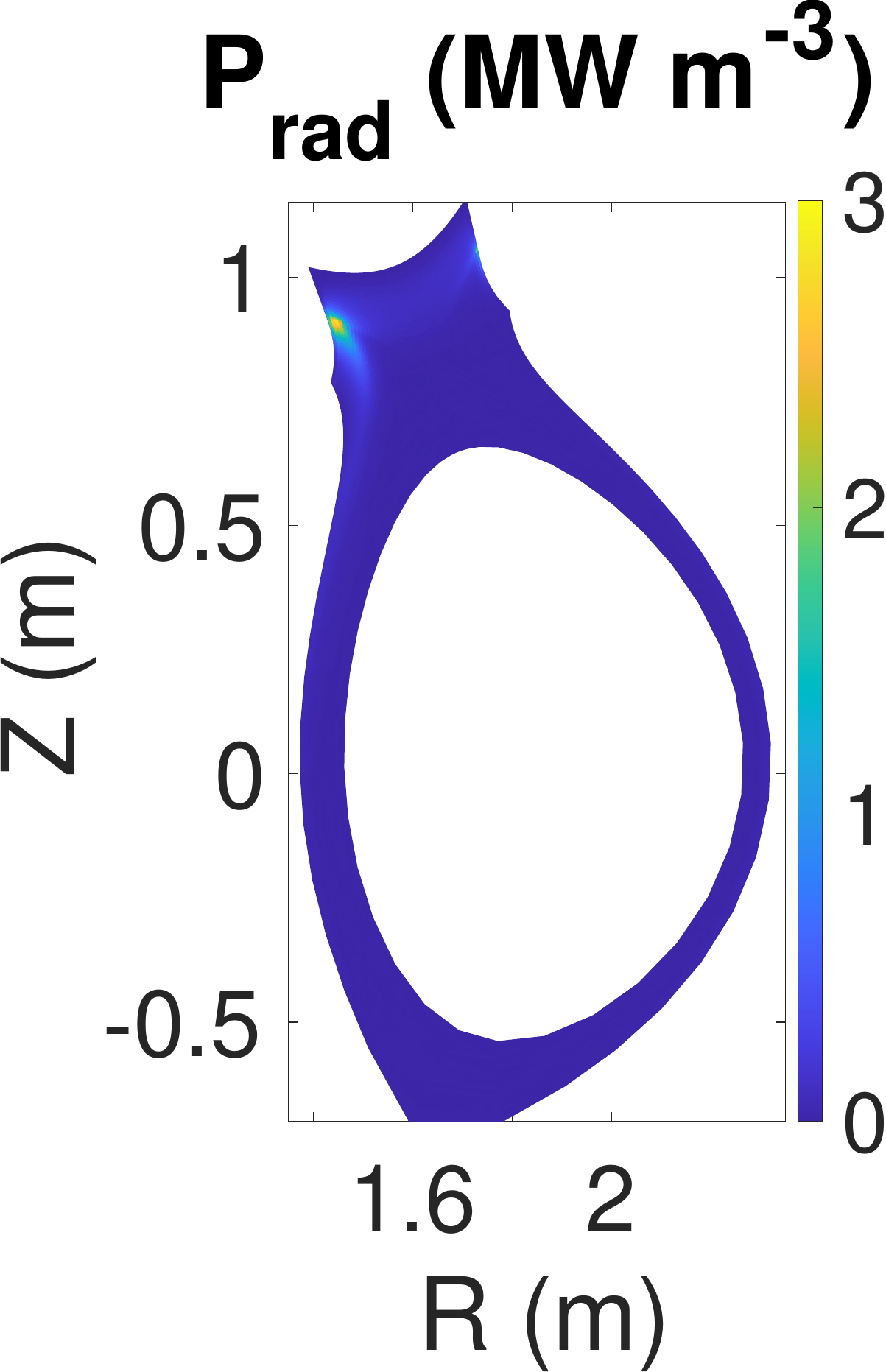}
		\caption{}
		\label{subfig:radiated_power_D2_1}
	\end{subfigure}
	\begin{subfigure}{4cm}
		\centering
		\medskip
		\includegraphics[width=4cm]{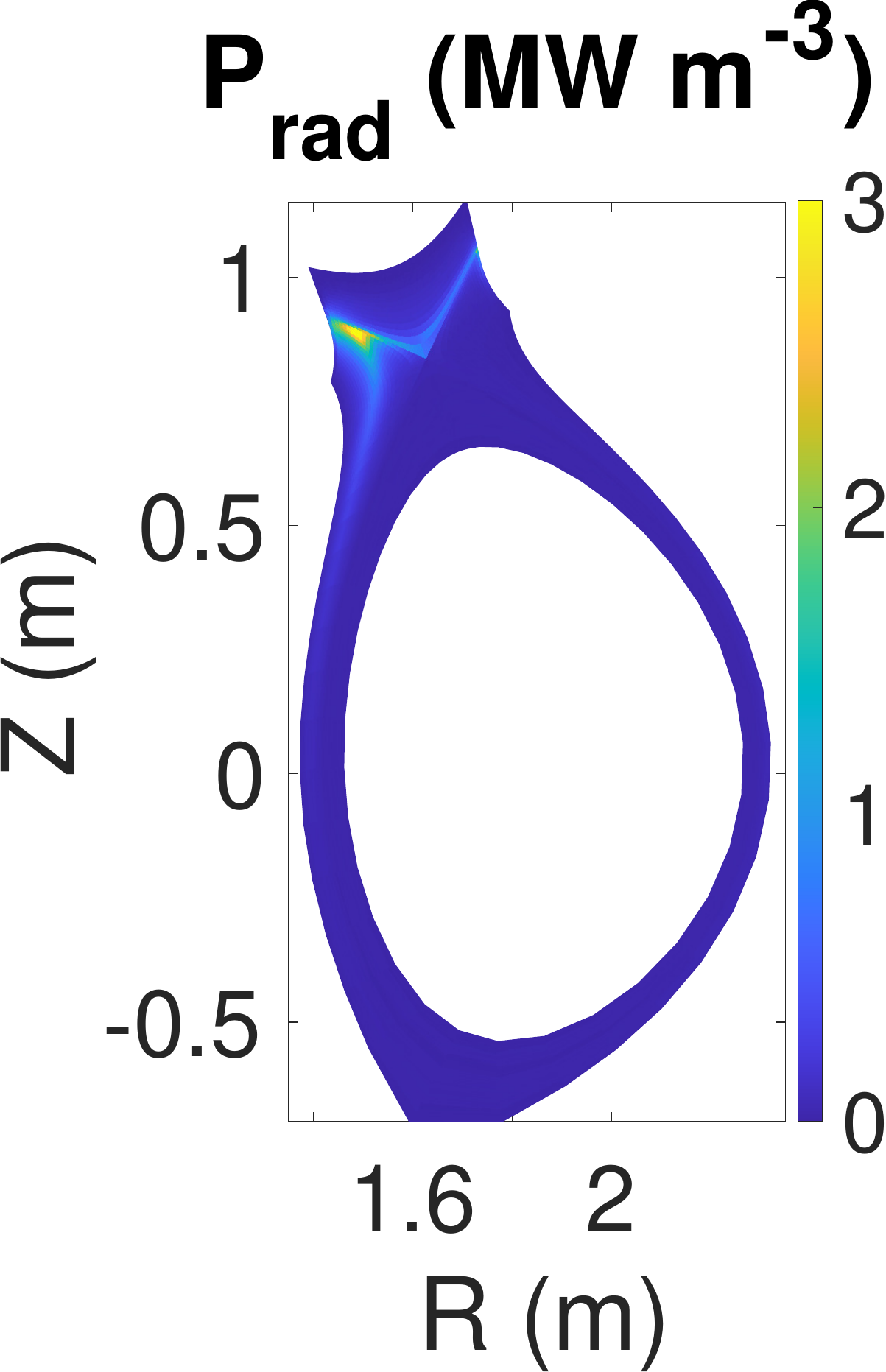}
		\caption{}
		\label{subfig:radiated_power_D2_2}
	\end{subfigure}
	\begin{subfigure}{4cm}
		\centering
		\medskip
		\includegraphics[width=4cm]{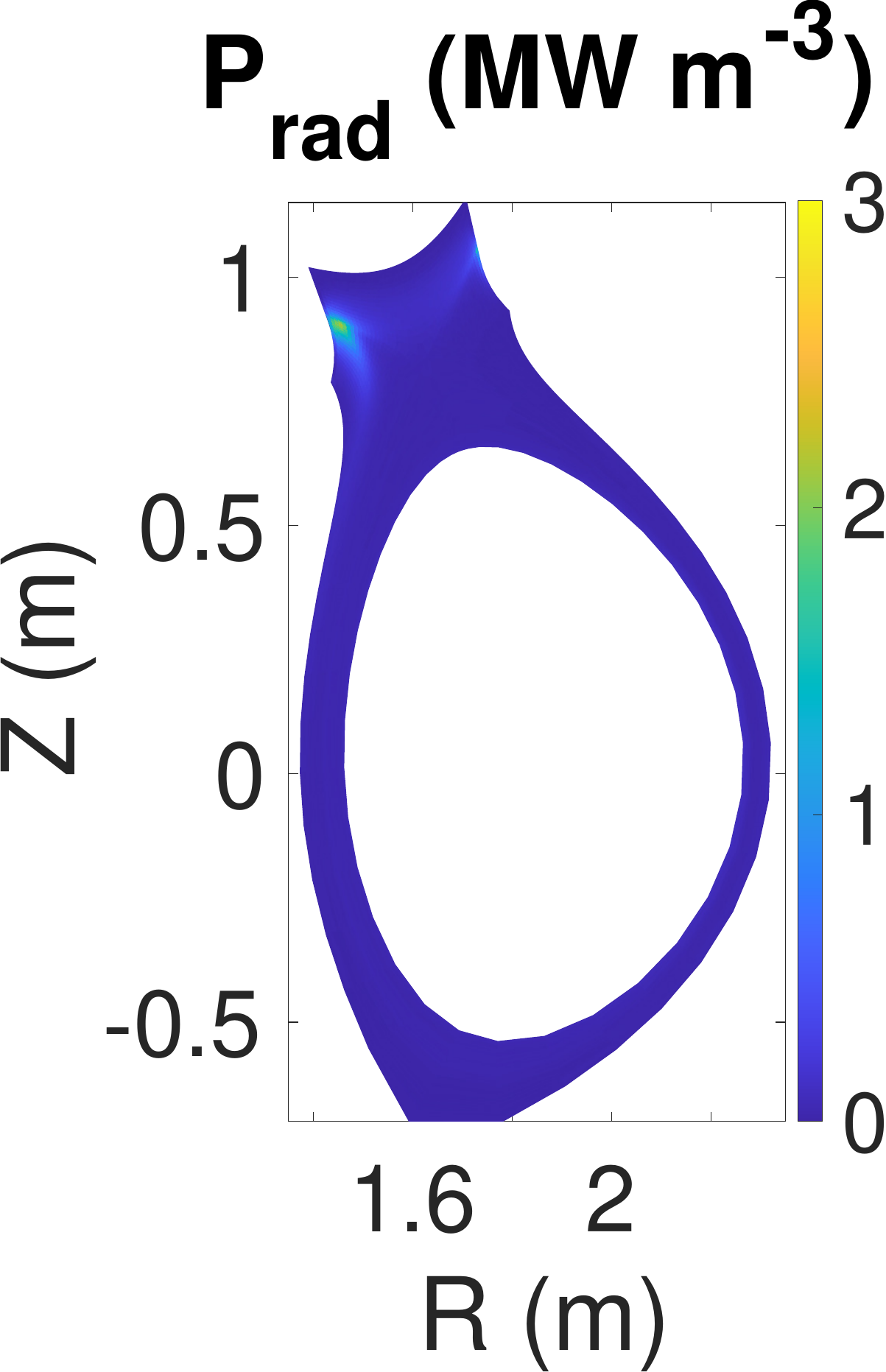}
		\caption{}
		\label{subfig:radiated_power_Ne_1}
	\end{subfigure}
	\begin{subfigure}{4cm}
		\centering
		\medskip
		\includegraphics[width=4cm]{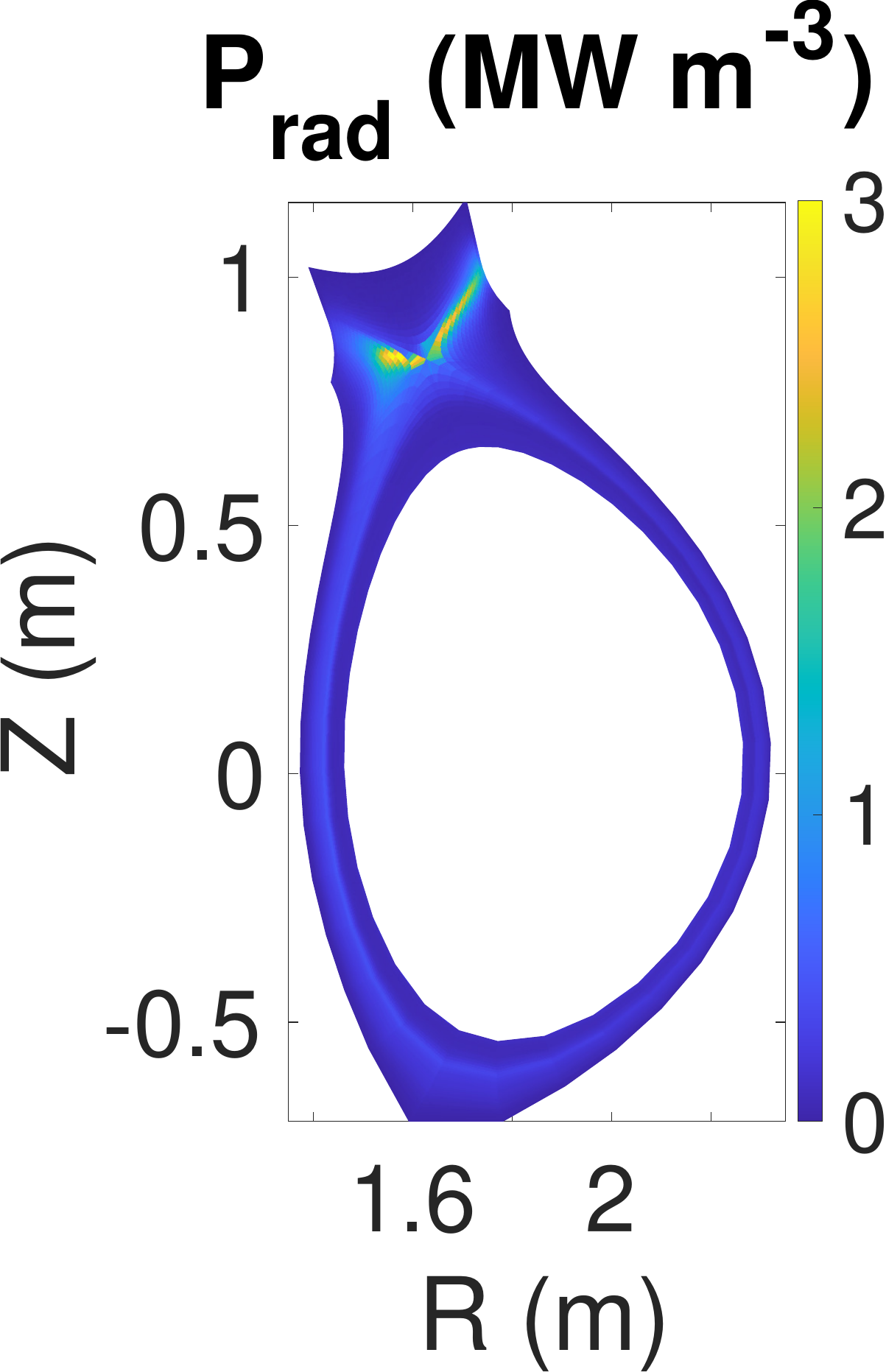}
		\caption{}
		\label{subfig:radiated_power_Ne_2}
	\end{subfigure}
	\caption{The simulated radiated power profiles for the SOLPS D$_2$ 1 (a), SOLPS D$_2$ 2 (b), SOLPS Ne 1 (c) and SOLPS Ne 2 (d) simulations.}
	\label{fig:radiated_power}
\end{figure}

\begin{figure}
	\centering
	\begin{subfigure}{4cm}
		\centering
		\medskip
		\includegraphics[clip,width=4cm]{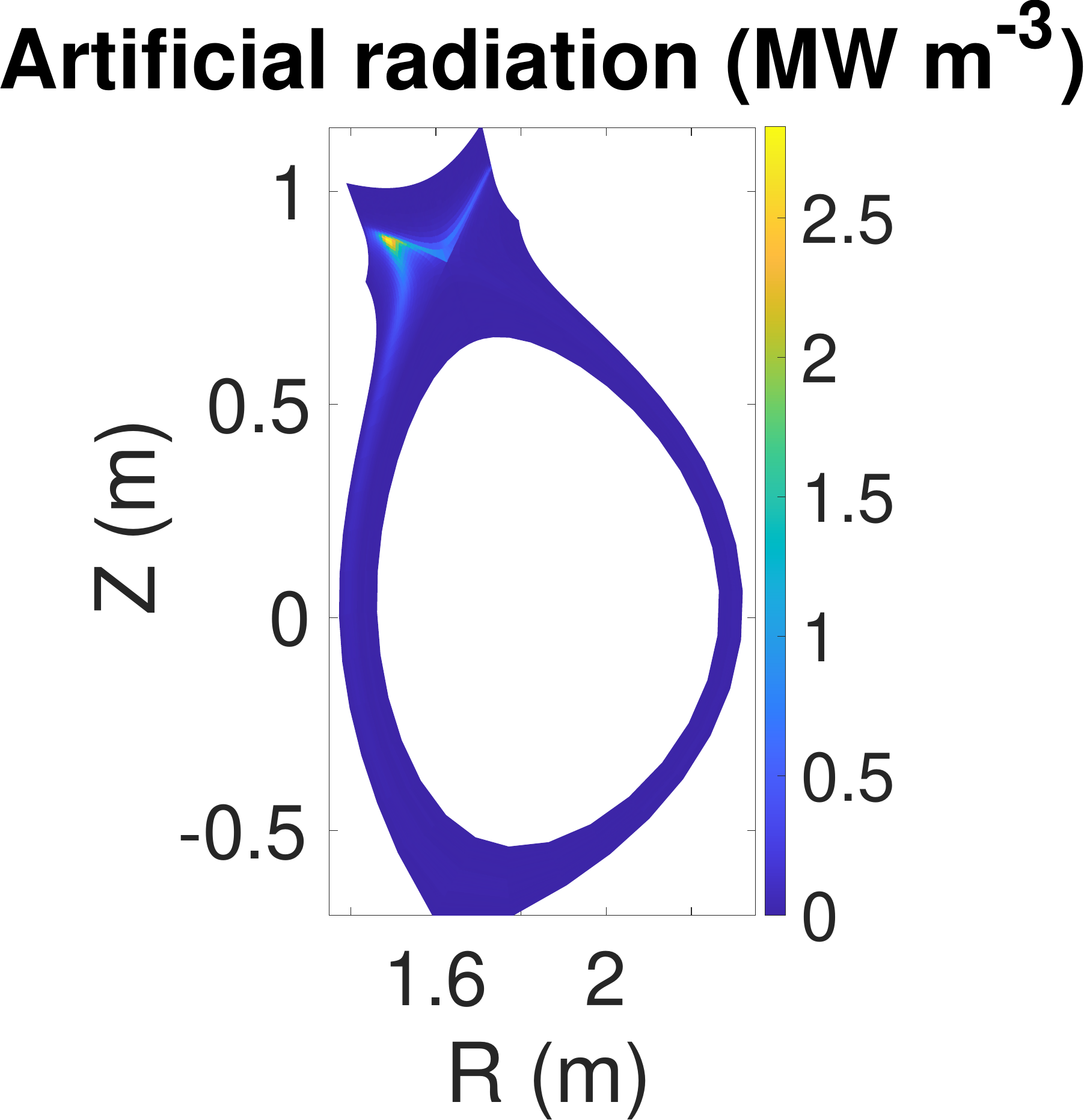}
		\caption{}
		\label{subfig:radiated_power_artificial_radiation_D2_2}
	\end{subfigure}
	\begin{subfigure}{4cm}
		\centering
		\medskip
		\includegraphics[clip,width=4cm]{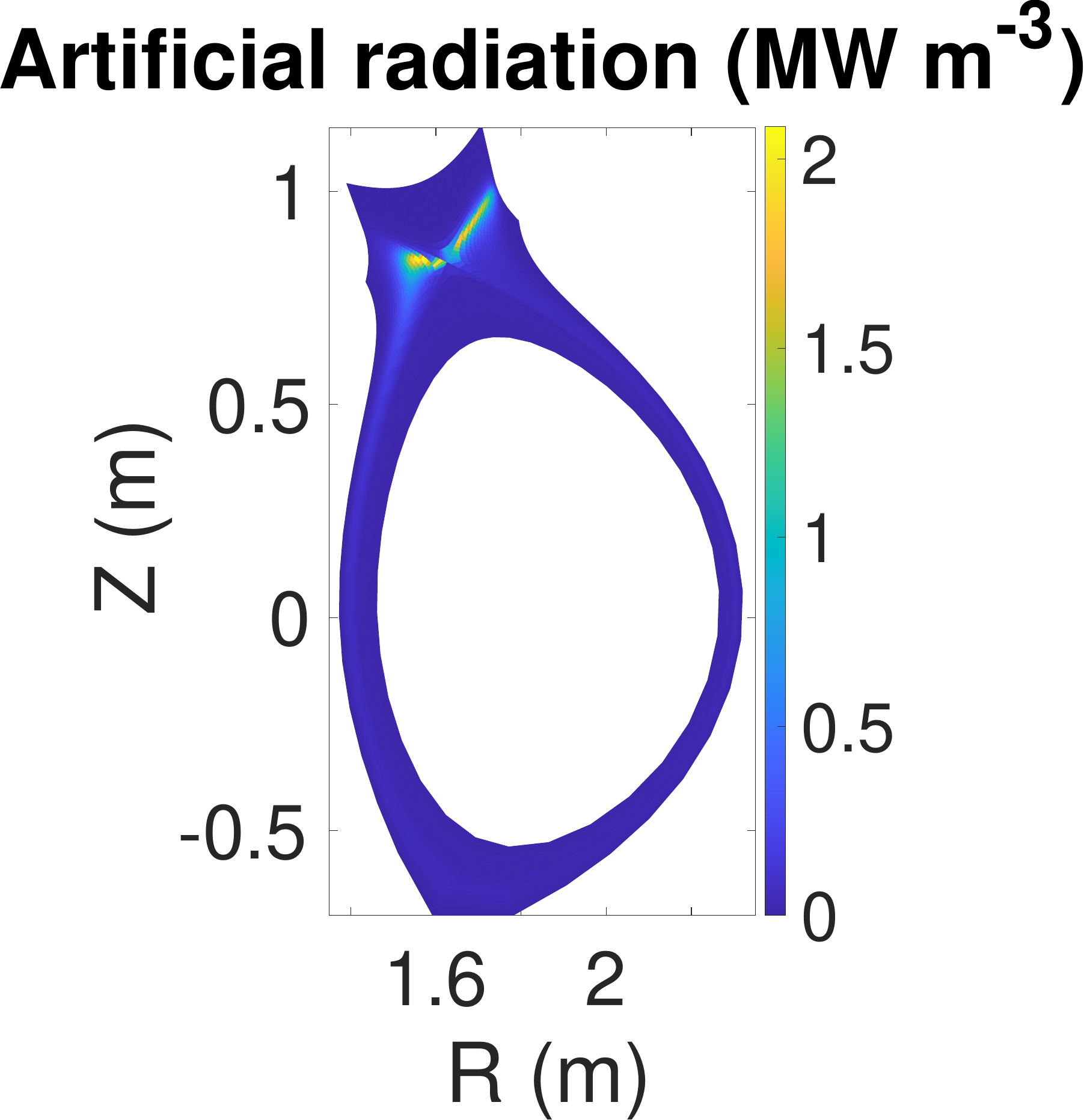}
		\caption{}
		\label{subfig:radiated_power_artificial_radiation_Ne_2}
	\end{subfigure}
	\caption{The simulated artificial radiation profiles for the SOLPS D$_2$ 2 (a) and SOLPS Ne 2 (b) simulations. Notice that the color scales are different.}
	\label{fig:radiated_power_artificial_radiation}
\end{figure}

\begin{figure}
	\centering
	\begin{subfigure}{4cm}
		\centering
		\medskip
		\includegraphics[clip,width=4cm]{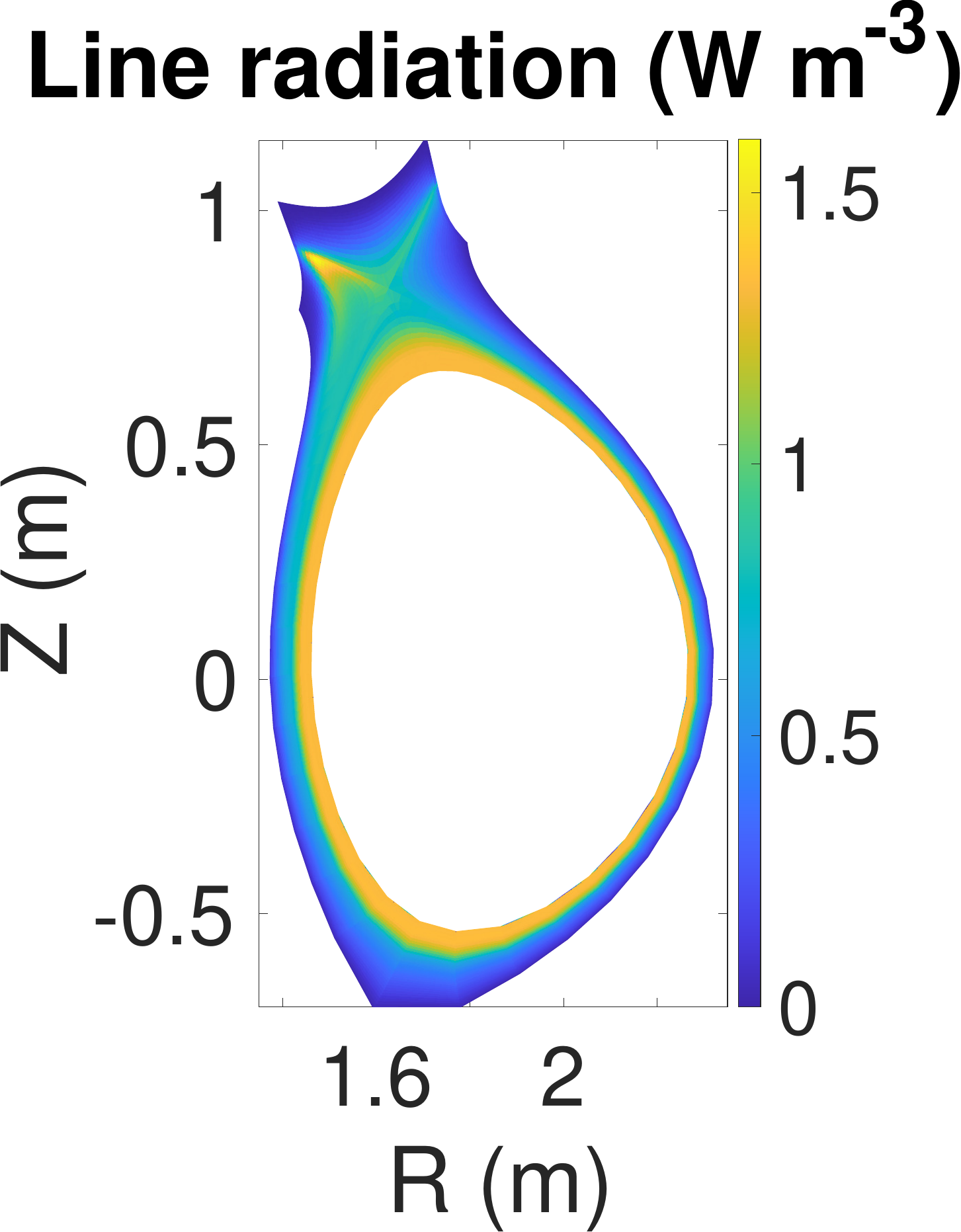}
		\caption{}
		\label{subfig:radiated_power_line_radiation_D2_1}
	\end{subfigure}
	\begin{subfigure}{4cm}
		\centering
		\medskip
		\includegraphics[clip,width=4cm]{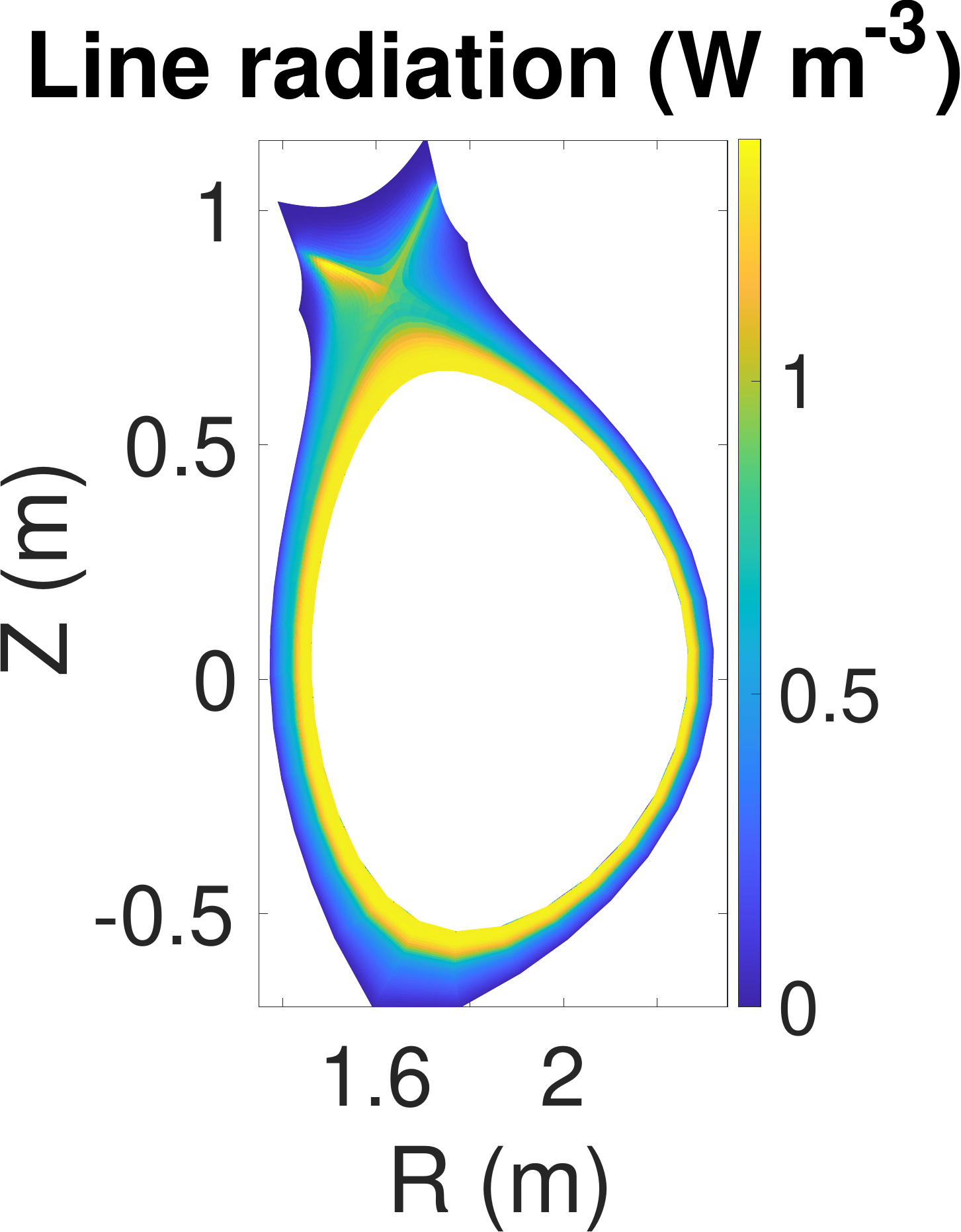}
		\caption{}
		\label{subfig:radiated_power_line_radiation_D2_2}
	\end{subfigure}
	\begin{subfigure}{4cm}
		\centering
		\medskip
		\includegraphics[clip,width=4cm]{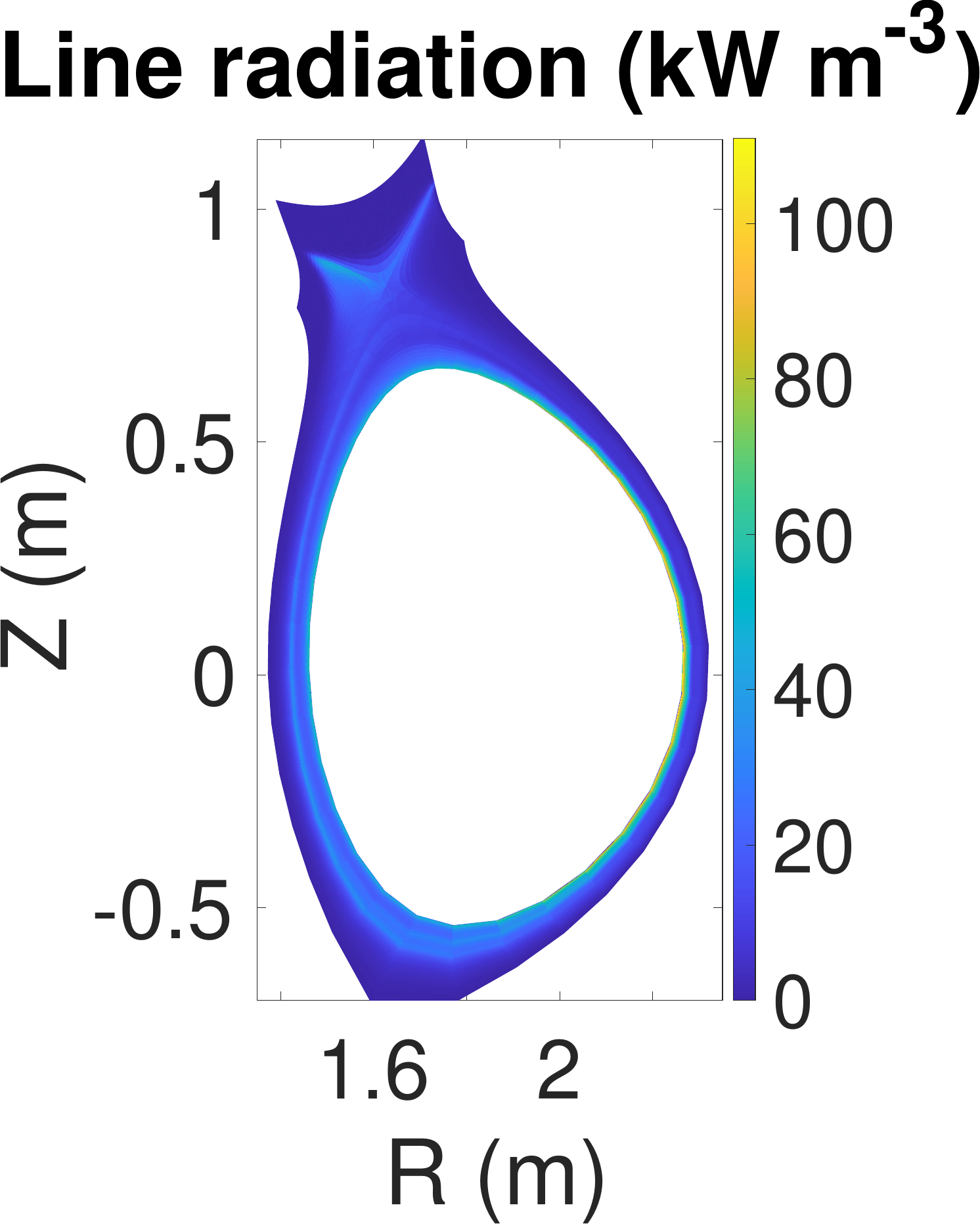}
		\caption{}
		\label{subfig:radiated_power_line_radiation_Ne1}
	\end{subfigure}
	\begin{subfigure}{4.1cm}
		\centering
		\medskip
		\includegraphics[clip,width=4.1cm]{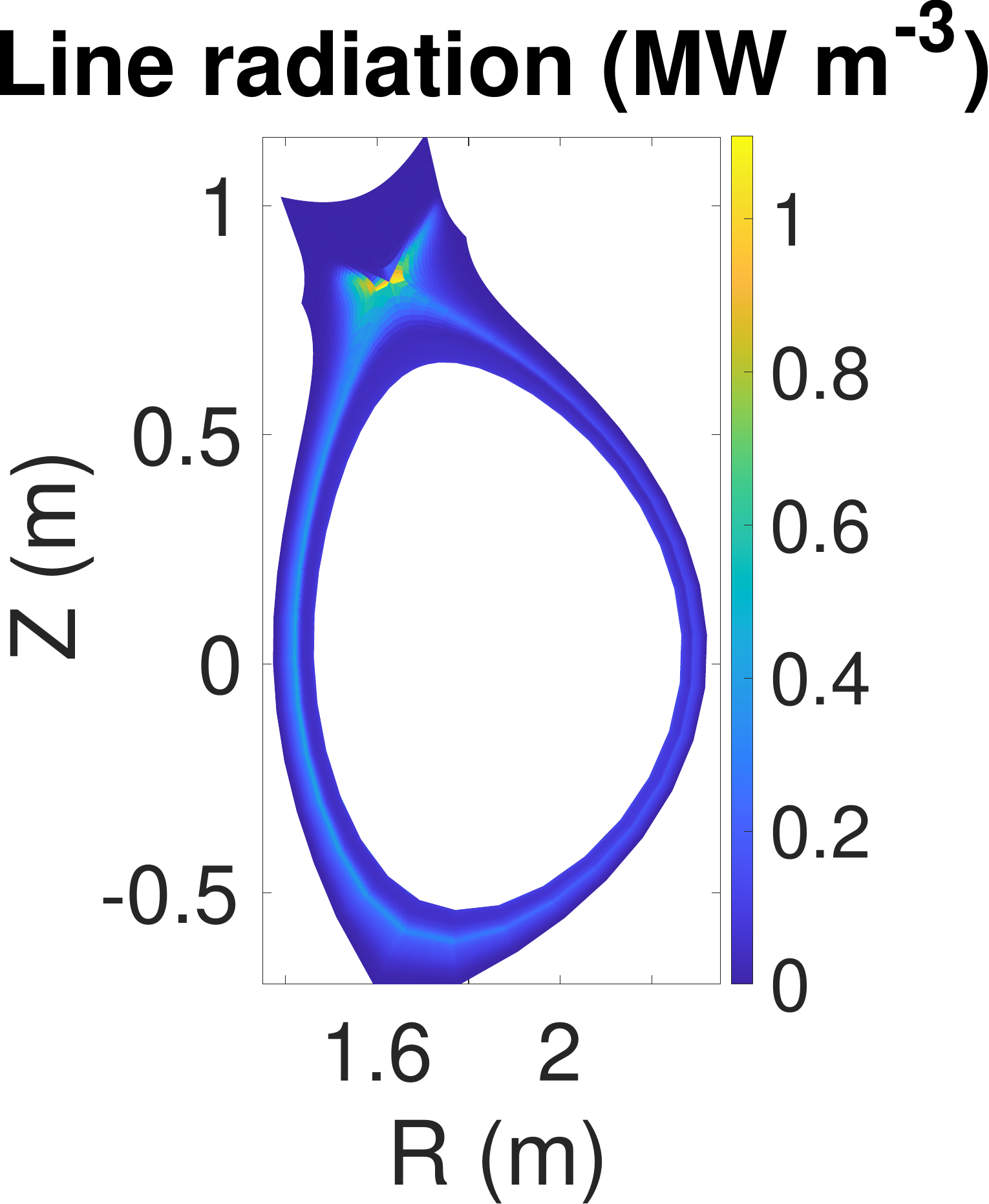}
		\caption{}
		\label{subfig:radiated_power_line_radiation_Ne2}
	\end{subfigure}
	\caption{The simulated line radiation profiles for the SOLPS D$_2$ 1 (a), SOLPS D$_2$ 2 (b)SOLPS Ne 1 (c) and SOLPS Ne 2 (d) simulations. Notice that the color scales are different.}
	\label{fig:radiated_power_line_radiation}
\end{figure}

\begin{figure}
	\centering
	\begin{subfigure}{4cm}
		\centering
		\medskip
		\includegraphics[clip,width=3.5cm]{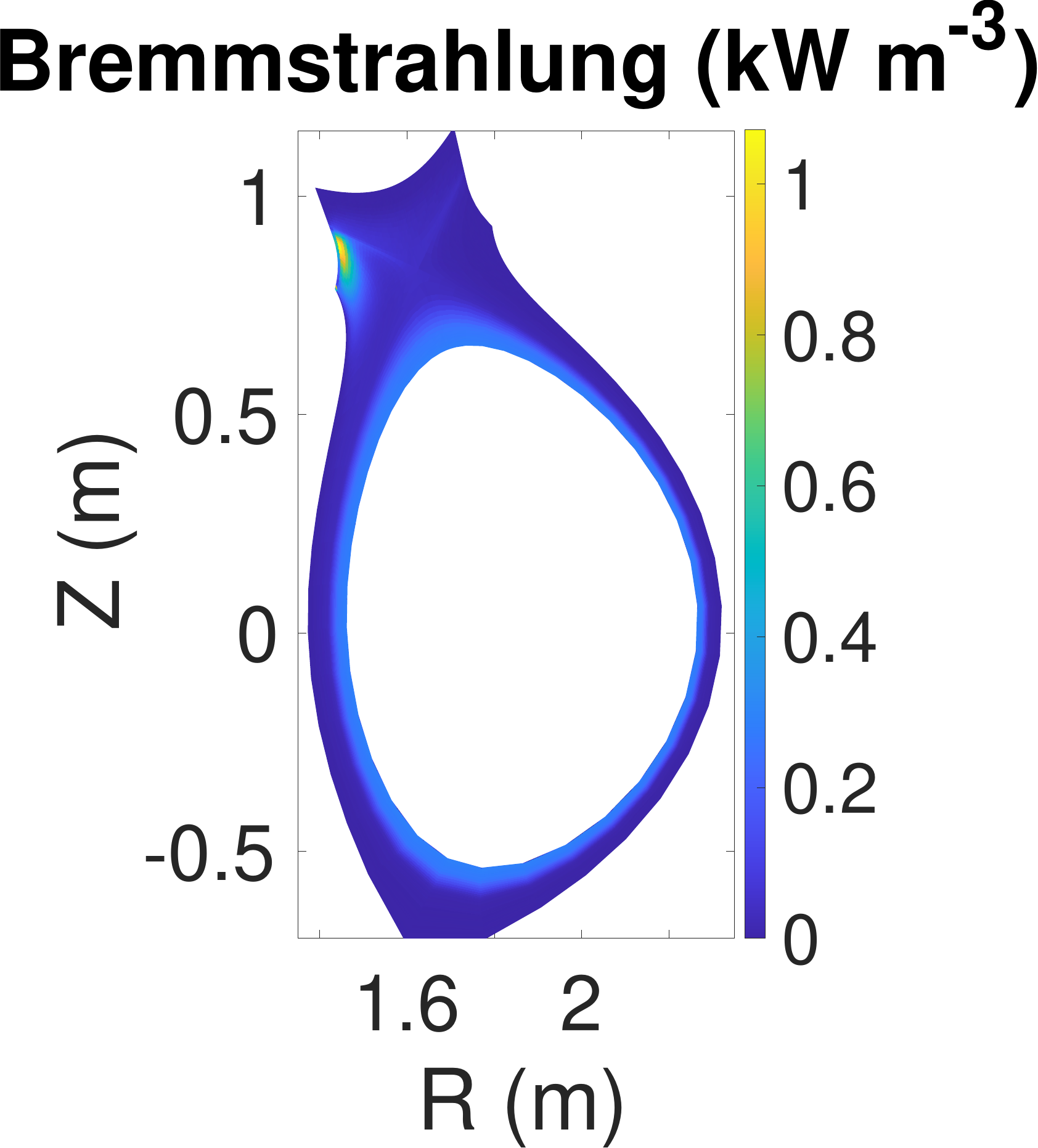}
		\caption{}
		\label{subfig:radiated_power_Bremmstrahlung_D2_1}
	\end{subfigure}
	\begin{subfigure}{4cm}
		\centering
		\medskip
		\includegraphics[clip,width=3.5cm]{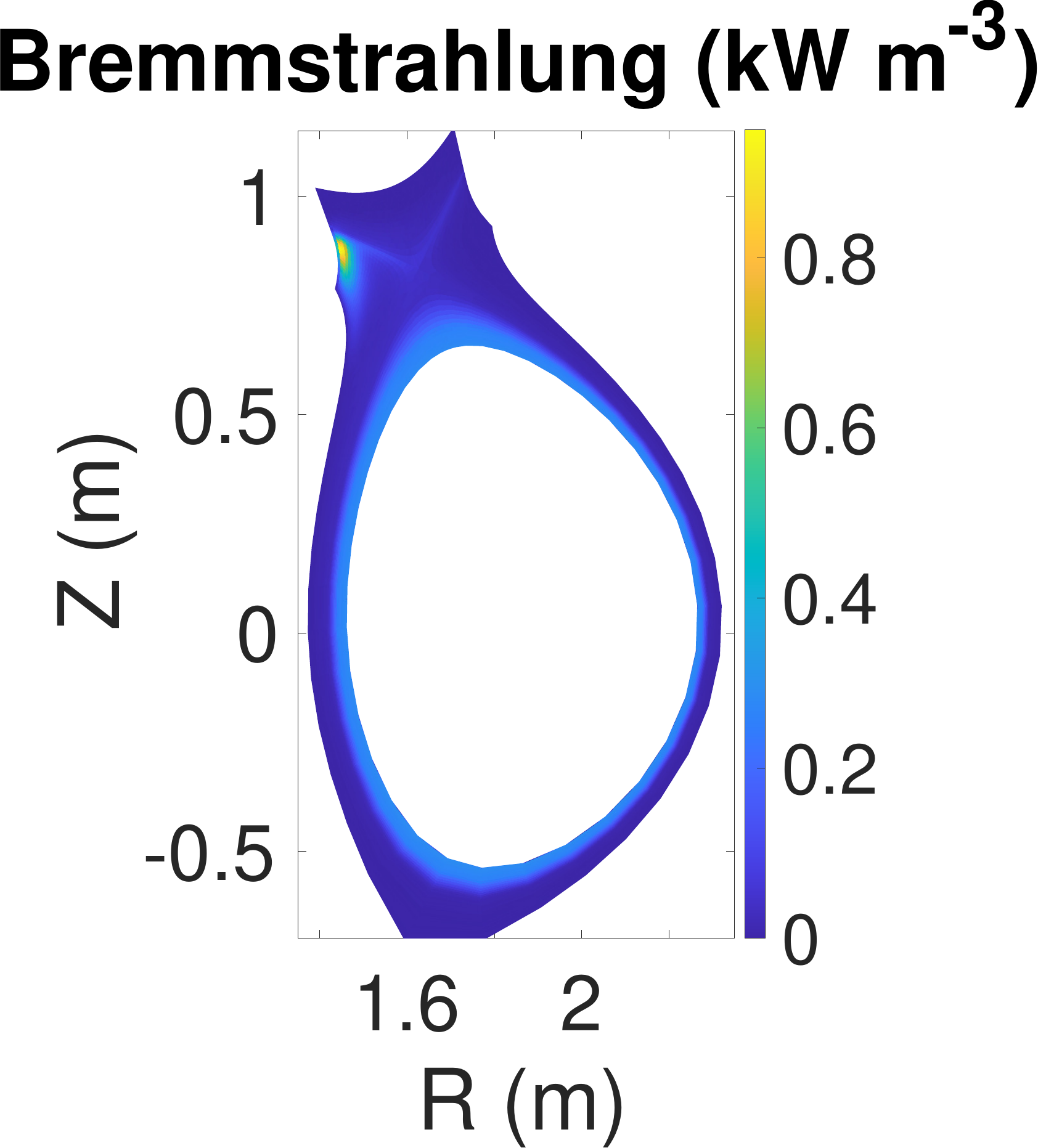}
		\caption{}
		\label{subfig:radiated_power_Bremmstrahlung_D2_2}
	\end{subfigure}
	\begin{subfigure}{4cm}
		\centering
		\medskip
		\includegraphics[clip,width=3.5cm]{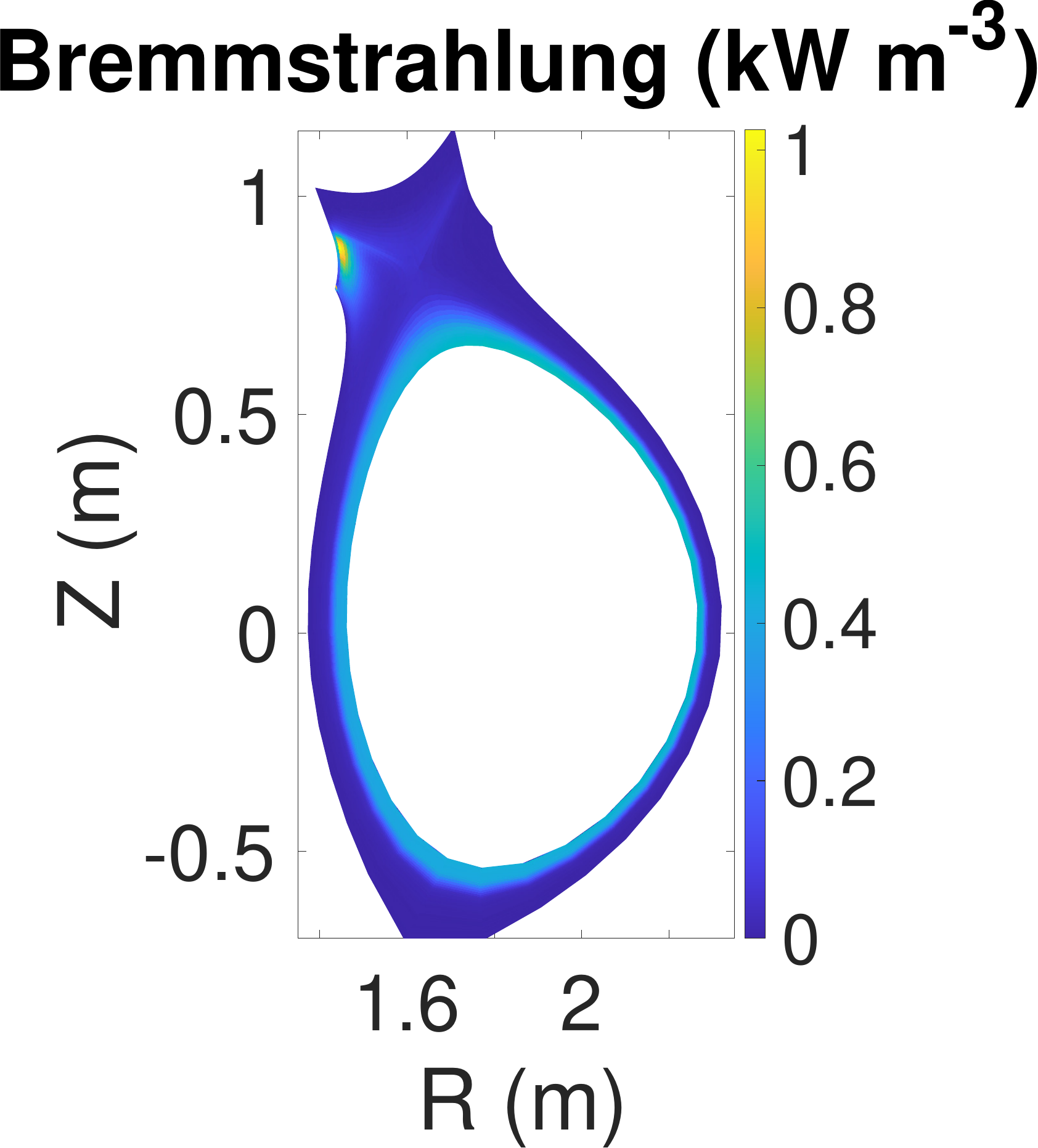}
		\caption{}
		\label{subfig:radiated_power_Bremmstrahlung_Ne_1}
	\end{subfigure}
	\begin{subfigure}{4cm}
		\centering
		\medskip
		\includegraphics[clip,width=3.5cm]{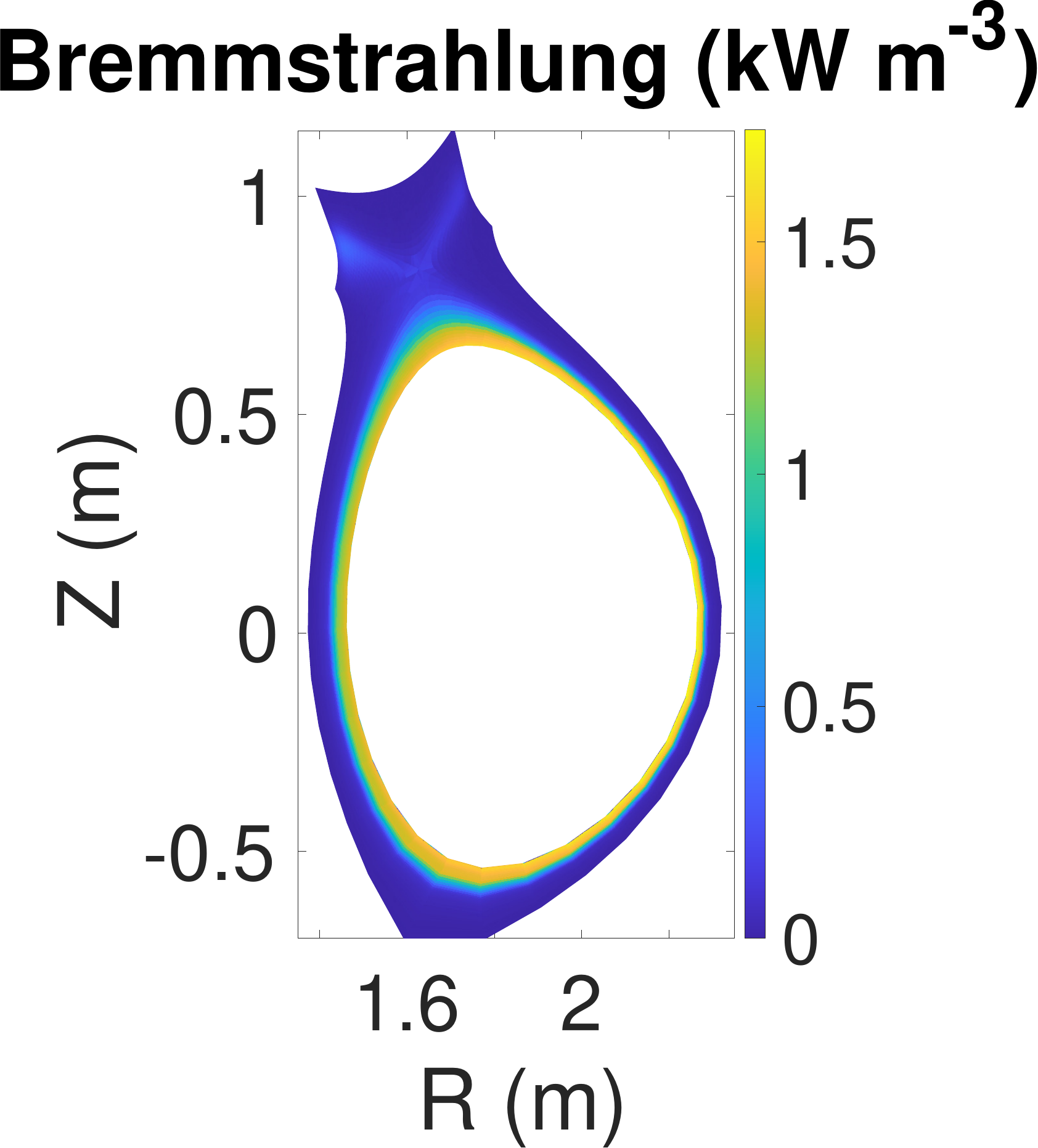}
		\caption{}
		\label{subfig:radiated_power_Bremmstrahlung_Ne2}
	\end{subfigure}
	\caption{The simulated Bremmstrahlung profiles for the SOLPS D$_2$ 1 (a), SOLPS D$_2$ 2 (b), SOLPS Ne 1 (c) and SOLPS Ne 2 (d) simulations. Notice that the color scales are different.}
	\label{fig:radiated_power_Bremmstrahlung}
\end{figure}

\begin{figure}
	\centering
	\begin{subfigure}{4cm}
		\centering
		\medskip
		\includegraphics[clip,width=4cm]{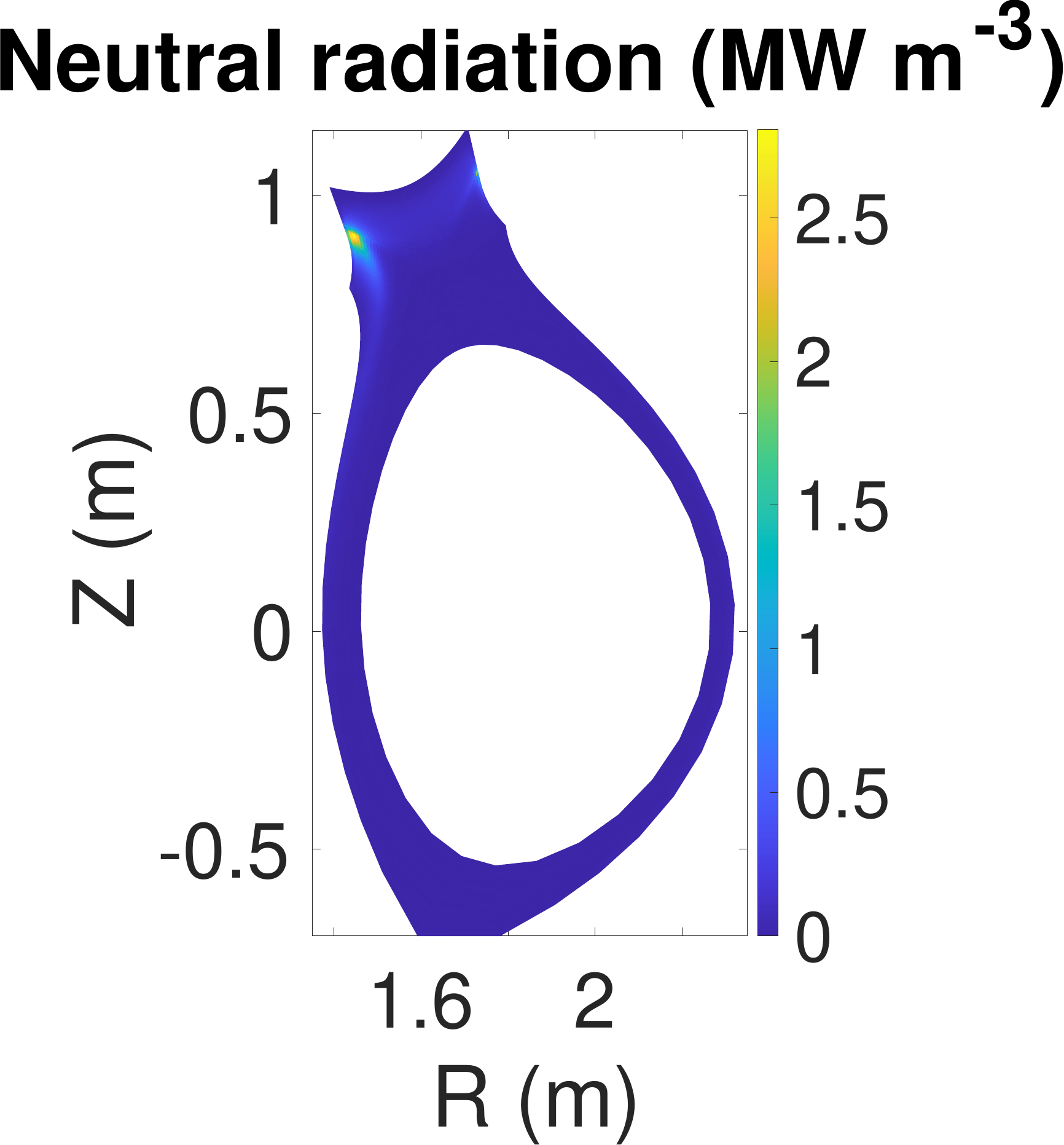}
		\caption{}
		\label{subfig:radiated_power_neutral_radiation_D2_1}
	\end{subfigure}
	\begin{subfigure}{4cm}
		\centering
		\medskip
		\includegraphics[clip,width=4cm]{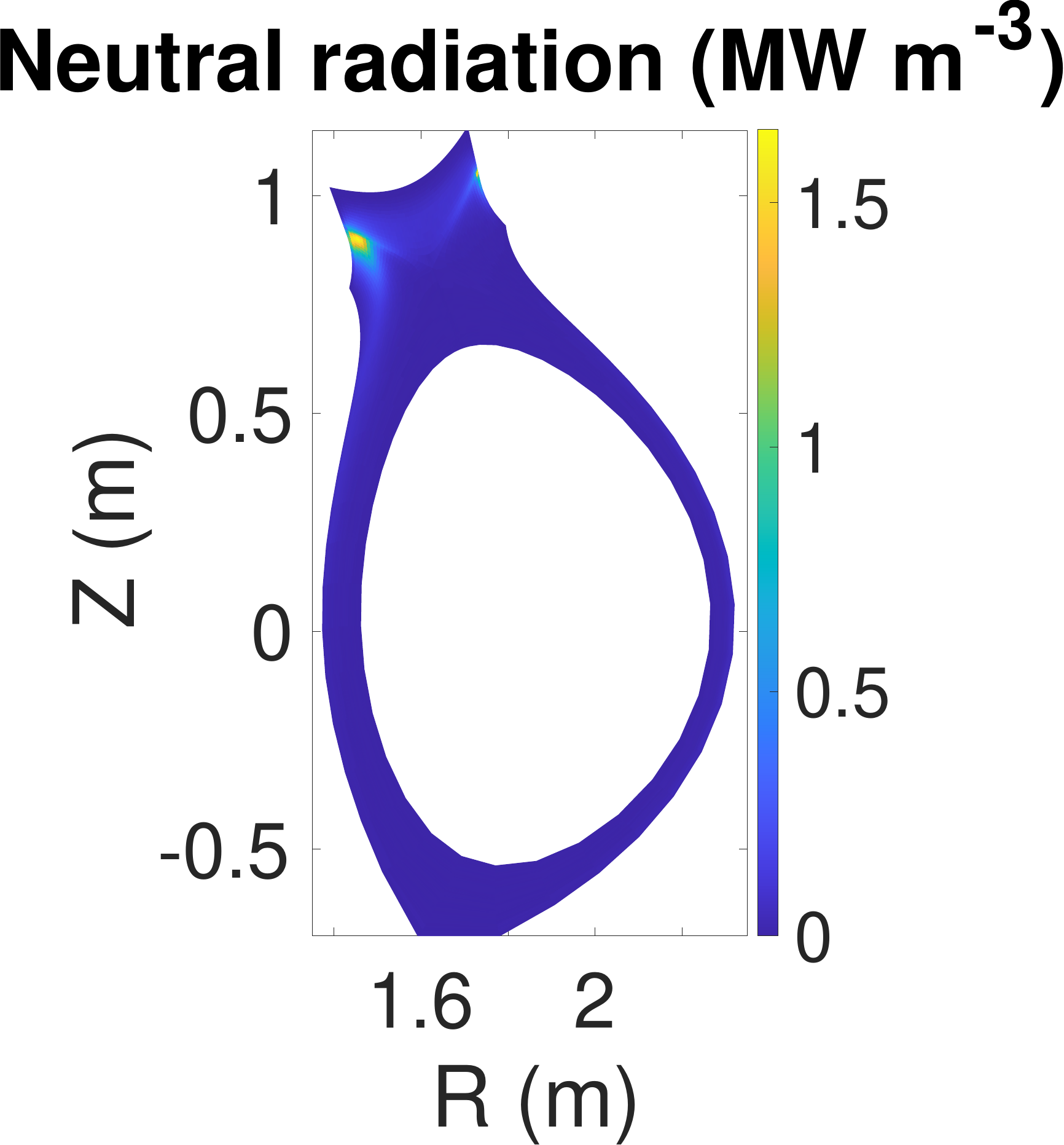}
		\caption{}
		\label{subfig:radiated_power_neutral_radiation_D2_2}
	\end{subfigure}
	\begin{subfigure}{4cm}
		\centering
		\medskip
		\includegraphics[clip,width=4cm]{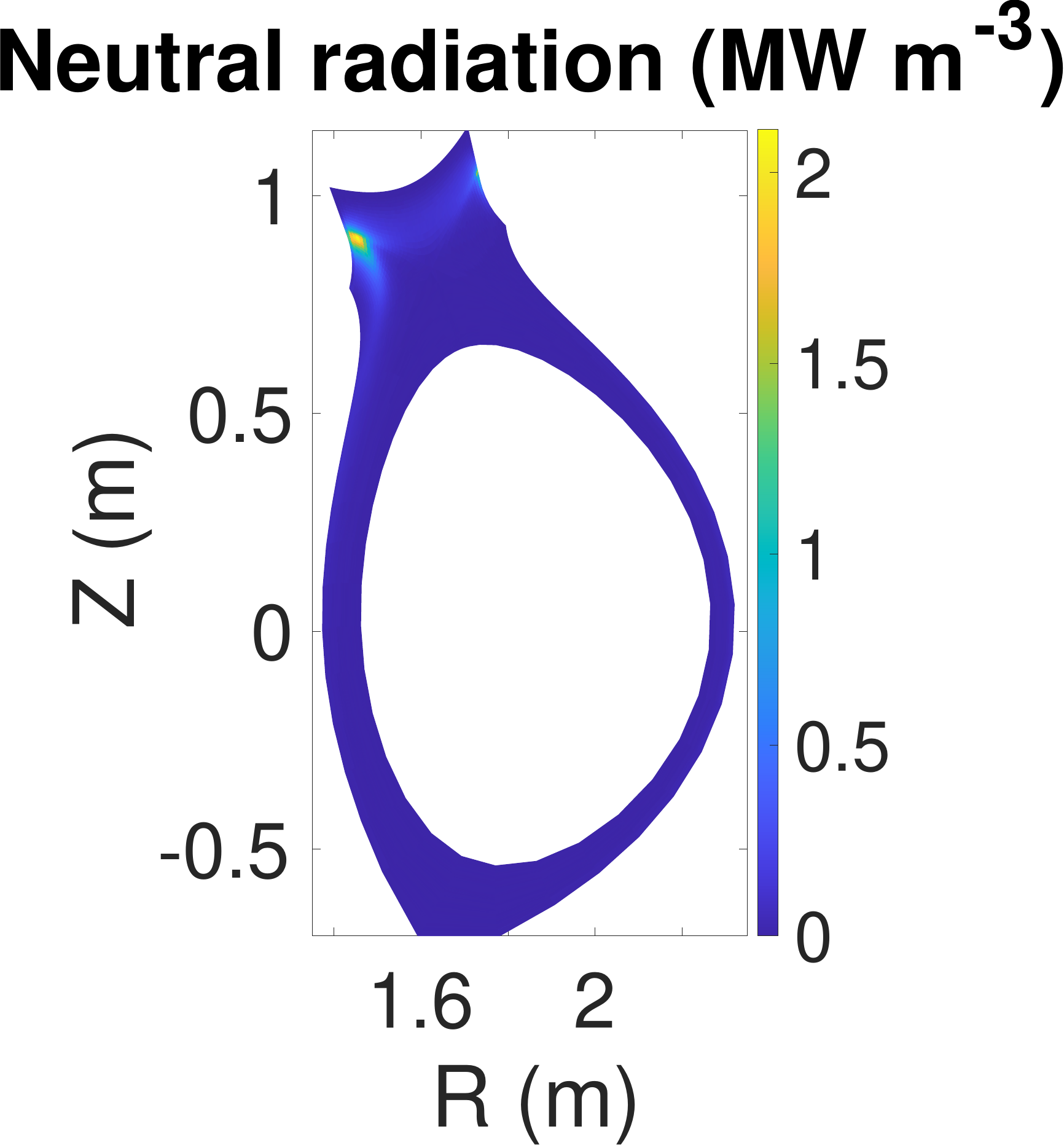}
		\caption{}
		\label{subfig:radiated_power_neutral_radiation_Ne_1}
	\end{subfigure}
	\begin{subfigure}{4cm}
		\centering
		\medskip
		\includegraphics[clip,width=4cm]{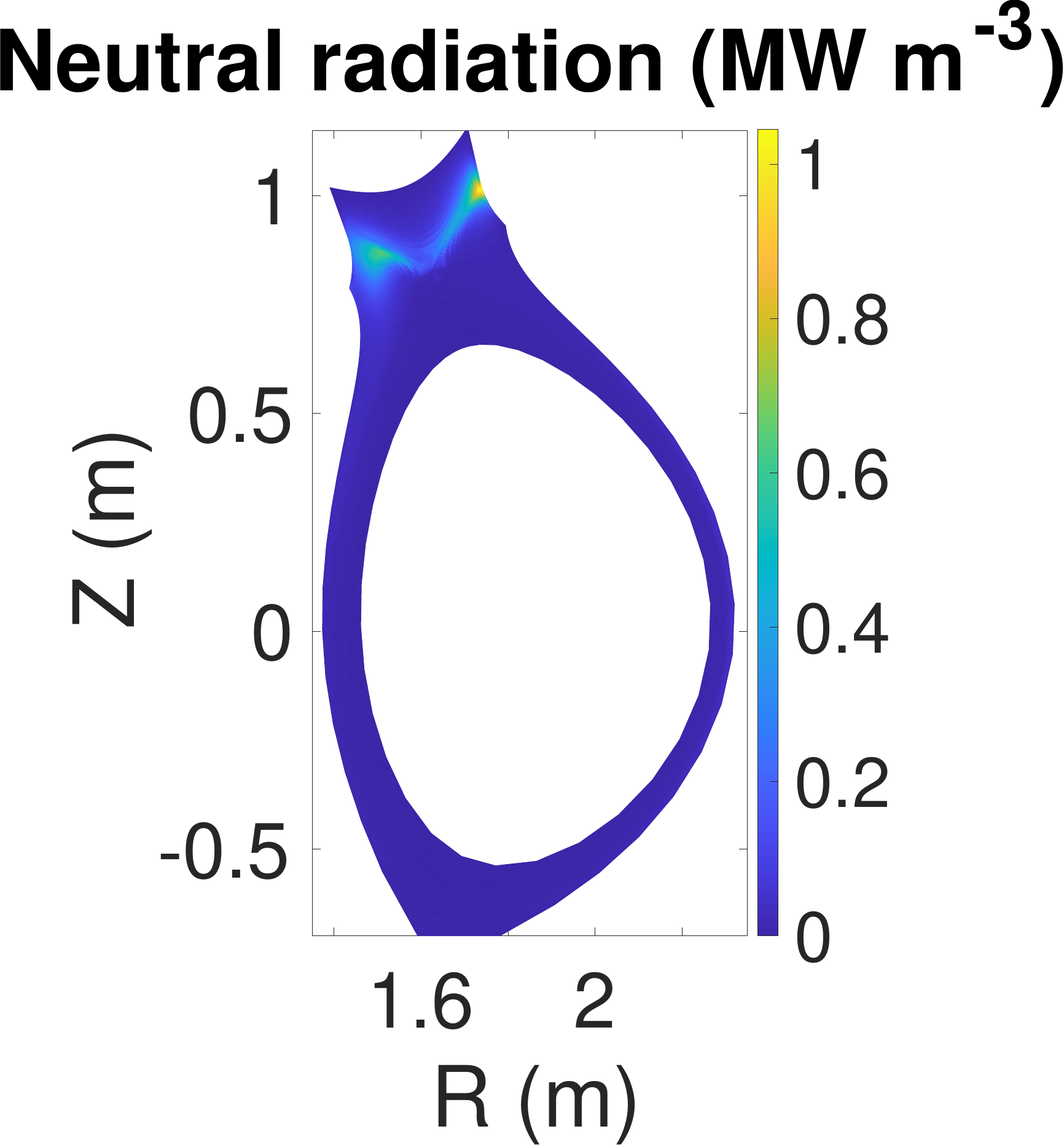}
		\caption{}
		\label{subfig:radiated_power_neutral_radiation_Ne_2}
	\end{subfigure}
	\caption{The simulated neutral radiation profiles for the SOLPS D$_2$ 1 (a), SOLPS D$_2$ 2 (b), SOLPS Ne 1 (c) and SOLPS Ne 2 (d) simulations. Notice that the color scales are different.}
	\label{fig:radiated_power_neutral_radiation}
\end{figure}

The total radiated power for the different simulations is displayed in figure \ref{fig:radiated_power}. This radiated power exists of line radiation (figure \ref{fig:radiated_power_line_radiation}), Bremmstrahlung (figure \ref{fig:radiated_power_Bremmstrahlung}), radiation due to atoms, molecules and molecular ions which are referred to as neutral radiation (figure \ref{fig:radiated_power_neutral_radiation}), and in the case of simulations "SOLPS D$_2$ 2" and "SOLPS Ne 2" also artificial radiation (figure \ref{fig:radiated_power_artificial_radiation}). Where simulation "SOLPS D$_2$ 1" and "SOLPS Ne 1" have only a radiation peak in the vicinity of the inner target, simulations "SOLPS D$_2$ 2" and "SOLPS Ne 2" show a larger radiation region including the region around the X-point. Figure \ref{fig:radiated_power_artificial_radiation} shows that a large contribution is caused by this artificial radiation. As described in section \ref{sec:case_desbription} artificial radiation is a simplified model to describe radiation from impurities which are not included as separate impurity species in the SOLPS-ITER simulation. The experimental divertor profiles at the UOT from figure \ref{fig:DivLP_SOLPS_UOT} could only be obtained by introducing this artificial radiation in the simulation. This suggest that the presence of other radiators besides neon is important to drive the plasma into detachment for an EAST-size device with tungsten wall. This is in agreement with the observation at ASDEX Upgrade that pure neon seeding experiments over there are unstable \cite{bernert2017power} and can only be stabilized when neon injection is combined with another impurity like nitrogen \cite{henderson2023divertor}.

Apart from the other impurities, figure \ref{fig:radiated_power_line_radiation} shows the effect of neon. Especially in the "SOLPS Ne 2" simulation, neon is causing radiation around the X-point and along the outer divertor leg. For the neon itself, line radiation is the main radiation mechanism. This line radiation is the one which is aimed for by impurity cooling. The location of the line radiation for simulation "SOLPS Ne 2" is similar to earlier experimental findings at JET \cite{gloggler2019characterisation,bernert2017power} and SOLPS-ITER modeling for ASDEX Upgrade \cite{sytova2019comparing}, although that there the peak in total radiation is right inside the separatrix where this is just outside the separatrix for the performed EAST simulations. Both at ASDEX Upgrade and at JET, the puff location is located in the PFR where at EAST two puff locations are present: one at the OMP for D$_2$ and one at the upper outer strikeline for the D$_2$/Ne mixture. As shown in ref. \cite{park2024impact} the location of the main ion injection can influence the radiation location. As expected, the effect of line radiation in the "SOLPS D$_2$" simulations is negligible as can be seen in figures \ref{subfig:radiated_power_line_radiation_D2_1} and \ref{subfig:radiated_power_line_radiation_D2_2}.

Bremmstrahlung is in all simulations similar, but stays always small (see figure \ref{fig:radiated_power_Bremmstrahlung}) and is never a major contributor to the total radiated power fraction.

The neutral radiation is the main contributor to the "SOLPS D$_2$ 1" and "SOLPS Ne 1" simulations. This neutral radiation is originating from neutral friction between the deuterium ions and neutral particles (see figure \ref{fig:detachment_physics}) and will happen at temperatures of $T_e \sim 2-5 \mathrm{eV}$ in the divertor region as neutrals are present over there (in contrast to the rest of the SOL). This plasma-neutral friction is important to obtain the necessary conditions for volumetric recombination, but does not contribute to the reduction of the flux to the targets \cite{krasheninnikov2016divertor}. Another way to see where this peak in neutral radiation is to be expected is from the location of deuterium ionization of figure \ref{fig:ionization_detail_drifts} which shows a peak at the same location. For "SOLPS D$_2$ 1", "SOLPS D$_2$ 2" and "SOLPS Ne 1" this is the case in the vicinity of the inner target, where for "SOLPS Ne 2" figure \ref{subfig:Te_UOT} shows that the plasma temperatures are also decently low around the UOT separatrix. This, in combination with the additional gas puff over there, makes that there is not only a peak in neutral radiation around the UIT but also at the UOT.


Figure \ref{fig:radiated_power} has shown the differences in radiation pattern between the different performed simulations. However, in order to avoid overheating of the divertor targets, the remaining question is which percentage of the power entering the SOL is radiated in the edge. This is shown in table \ref{tab:f_rad}. This demonstrates the large influence of the 1 $\%$ "artificial" impurity content on the overall radiation: in the deuterium simulation, this more than doubles the radiated power. As shown in figure \ref{fig:rollover_densityscan}, only the "SOLPS Ne 2" simulation is detached. This also becomes clear from the radiative power fraction. The SOLPS-ITER simulations for ASDEX Upgrade and ITER of ref. \cite{sytova2019comparing} required similar percentages of radiated power fraction to bring the discharge in detachment.

\begin{table}
\begin{center}
	\begin{tabular}{c | c c c c} 
		 & SOLPS D$_2$ 1& SOLPS D$_2$ 2 & SOLPS Ne 1 & SOLPS Ne 2 \\
		\hline
		P$_{rad}$ & 0.174 MW & 0.375 MW & 0.232 MW & 0.986 MW\\ 
		P$_{in}$ & 2.05 MW & 2.05 MW & 1.8 MW & 1.77 MW\\
		\hline
		f$_{rad}$ & $8.5 \, \%$ & $18 \, \%$ & $13 \, \%$ & $55.7 \, \%$\\
	\end{tabular}
	\caption{The radiated power fractions for the performed SOLPS-ITER simulations.}
	\label{tab:f_rad}
\end{center}
\end{table}


As the AXUV measurement of figure \ref{subfig:AXUV} is mainly looking to core radiation and as all filterscope chords of figure \ref{subfig:filterscope} pass through the core, it is difficult to make a dedicated radiation analysis based on the available experimental data. Nevertheless, the DivLP profiles at the UOT (figure \ref{fig:DivLP_SOLPS_UOT}) and the $D_\alpha$ signal over there from the filterscope show that there was an influence of neon during the experiment.


\section{Influence of neutral transport}
\label{sec:neutral_transport}

Neutral processes play an important role in the divertor region where the plasma temperatures are low, plasma densities are high, and where there is an influx of neutrals from recycling and gas injection \cite{krasheninnikov2017physics, janev1984survey}. Table \ref{tab:EIRENE_neutral_reactions} shows the neutral reactions which are included in the performed SOLPS-ITER modeling. The molecular hydrogen originates from the gas puff at the OMP and at the UOT strikepoint for the neon-seeded simulations, and from recycling at the walls. Apart from such surface recombination, the atomic hydrogen results from elastic collisions, ionizing dissociation or charge exchange. In EIRENE, the recycling of deuterium  (and neon) at the tungsten divertor target, or the carbon first wall, is calculated using the TRIM database \cite{eckstein1987data}. Apart from this PWI, table \ref{tab:EIRENE_neutral_reactions} lists the resulting products of a particular reaction. In the neon seeded simulations, an additional neon ionization reaction from the AMJUEL database \cite{reiter2000data} and neon recombination from the ADAS database \cite{summers2011atomic} are included. The neutral neon originates from the neon injection at the UOT strikeline, and due to recombination of Ne ions.

\begin{figure}
	\centering
	\begin{subfigure}{4cm}
		\centering
		\medskip
		\includegraphics[trim={15cm 0cm 15cm 0cm},clip,width=4cm]{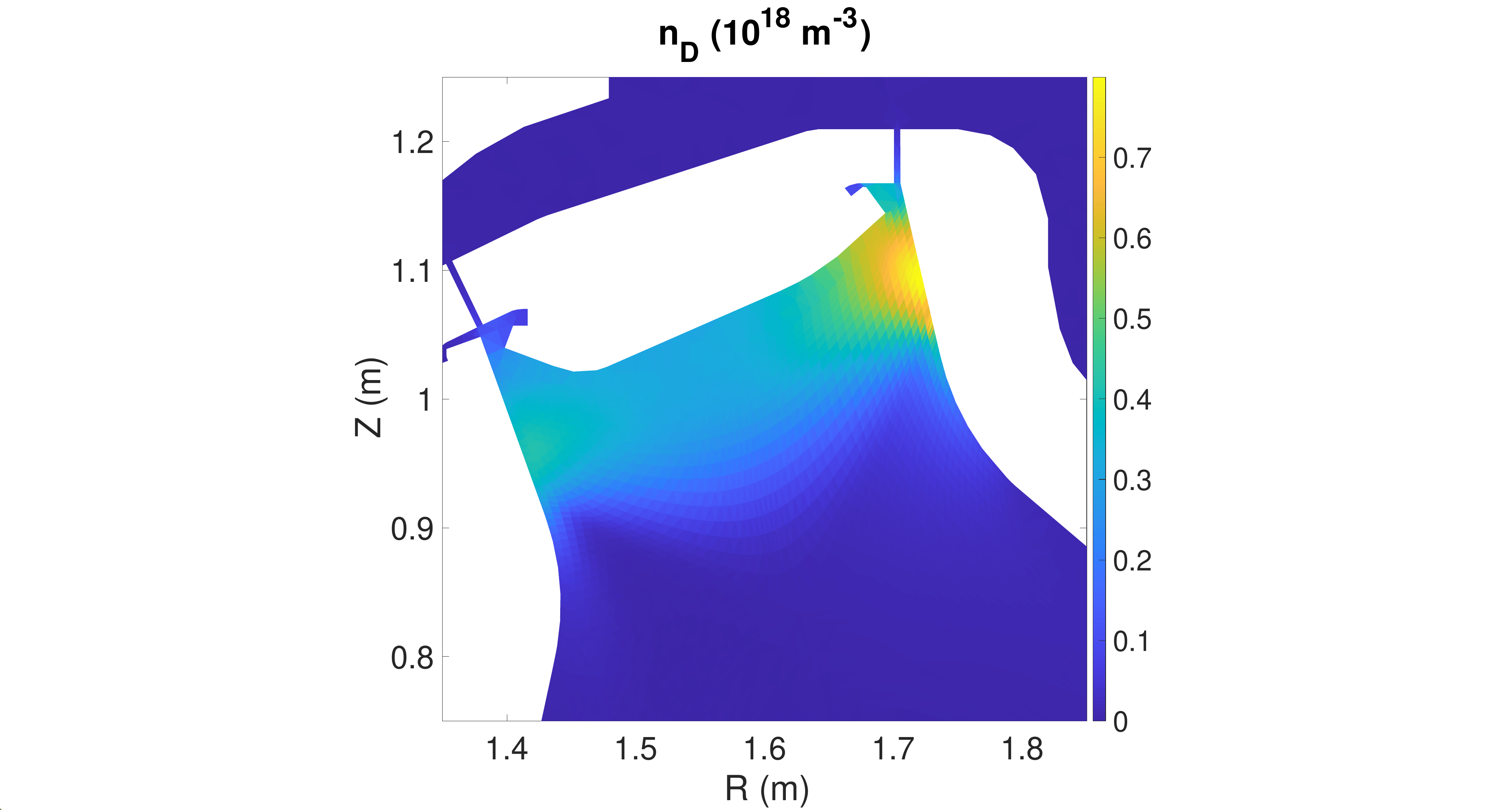}
		\caption{}
		\label{subfig:n_D_D2_div}
	\end{subfigure}
	\begin{subfigure}{4cm}
		\centering
		\medskip
		\includegraphics[trim={15cm 0cm 15cm 0cm},clip,width=4cm]{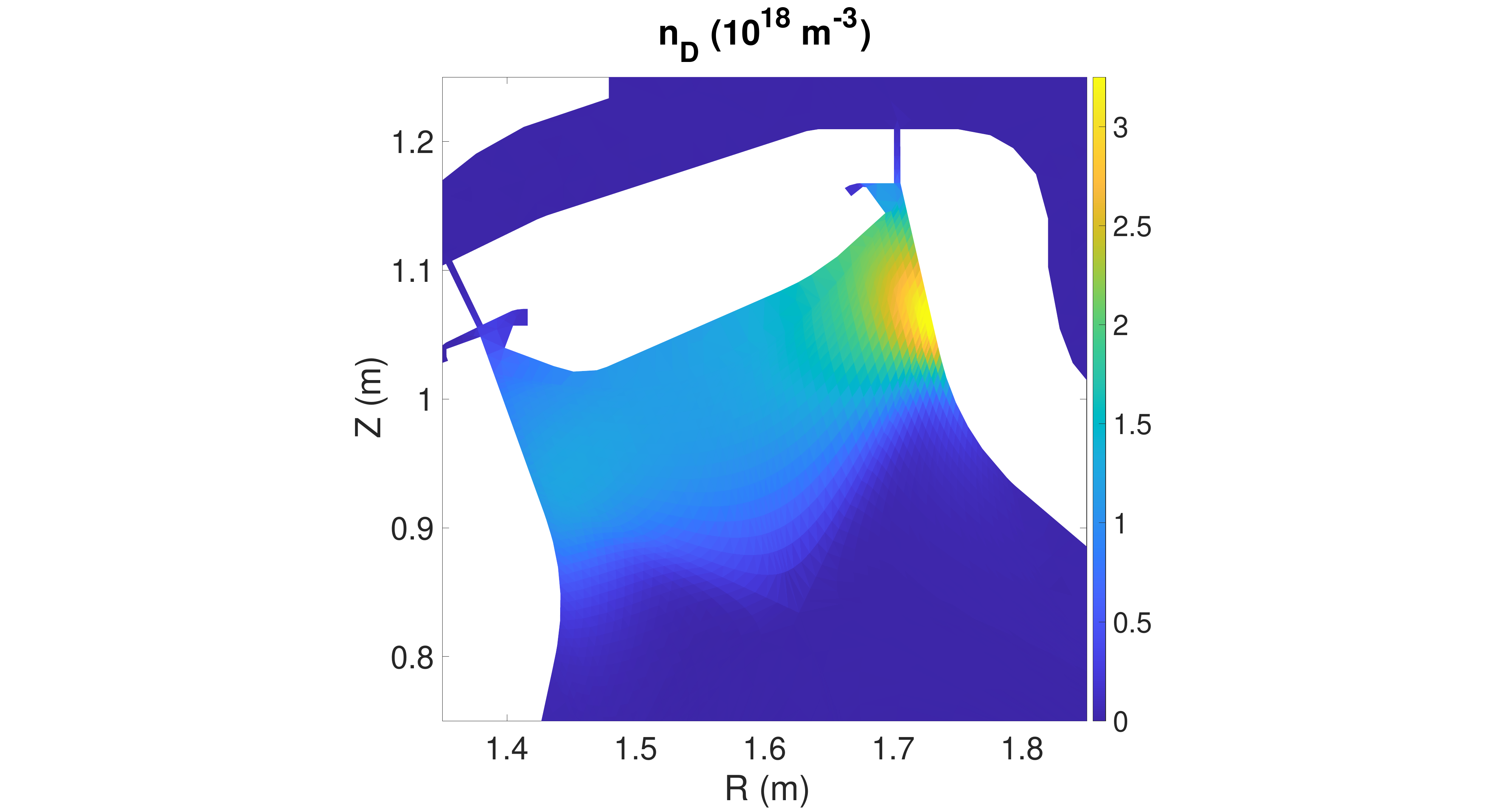}
		\caption{}
		\label{subfig:n_D_Ne_div}
	\end{subfigure}
	\caption{The neutral deuterium atom density in the "SOLPS D$_2$ 1" (a)  and "SOLPS Ne 2" (b) simulations in the divertor region. Remark the difference in the colorbar between the two density profiles.}
	\label{fig:n_D}
\end{figure}

Some of these reactions have a larger influence on the plasma behavior than others. In section \ref{sec:drifts} it was already shown that ionization plays an important role in explaining the plasma behavior. The ionization location of deuterium is mainly driven by a combination of PWI and the plasma quantities in the divertor region. In figure \ref{fig:n_D} the neutral deuterium atom density is shown for the "SOLPS D$_2$ 1" and "SOLPS Ne 2" simulations. The atom densities from the other simulations are similar to the ones of the "SOLPS D$_2$ 1" simulation. Where the atom density is largely increasing due to the added injection of a deuterium/neon mixture in the vicinity of the UOT, the main change in ionization takes place at the UIT (see figure \ref{fig:ionization_detail_drifts}). In the absence of the D$_2$-Ne injection, the main driver of deuterium atom generation is PWI. 

\begin{figure}
	\centering
	\begin{subfigure}{4cm}
		\centering
		\medskip
		\includegraphics[trim={15cm 0cm 15cm 0cm},clip,width=4cm]{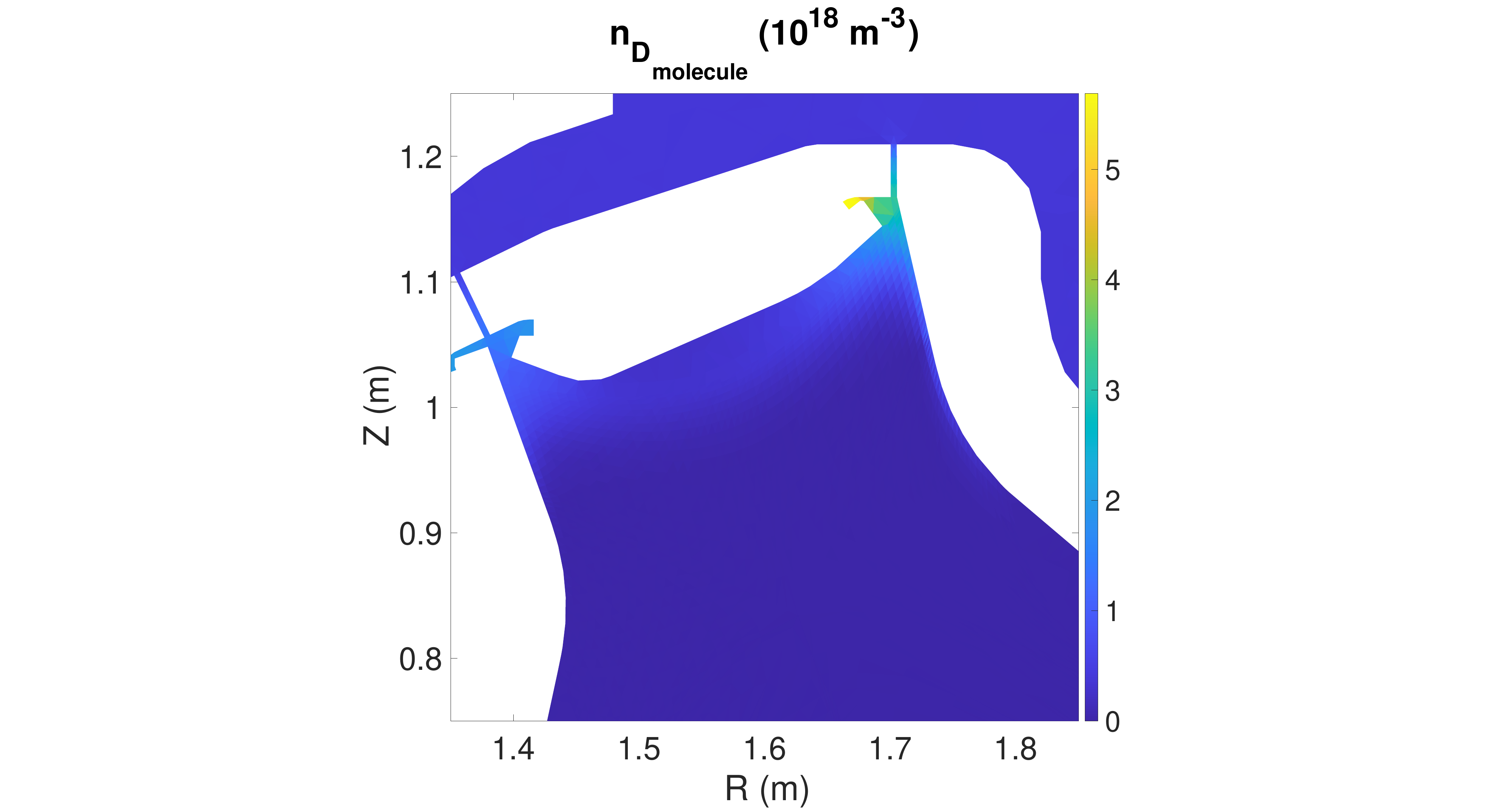}
		\caption{}
		\label{subfig:n_Dmol_D2_1_div}
	\end{subfigure}
	\begin{subfigure}{4cm}
		\centering
		\medskip
		\includegraphics[trim={15cm 0cm 15cm 0cm},clip,width=4cm]{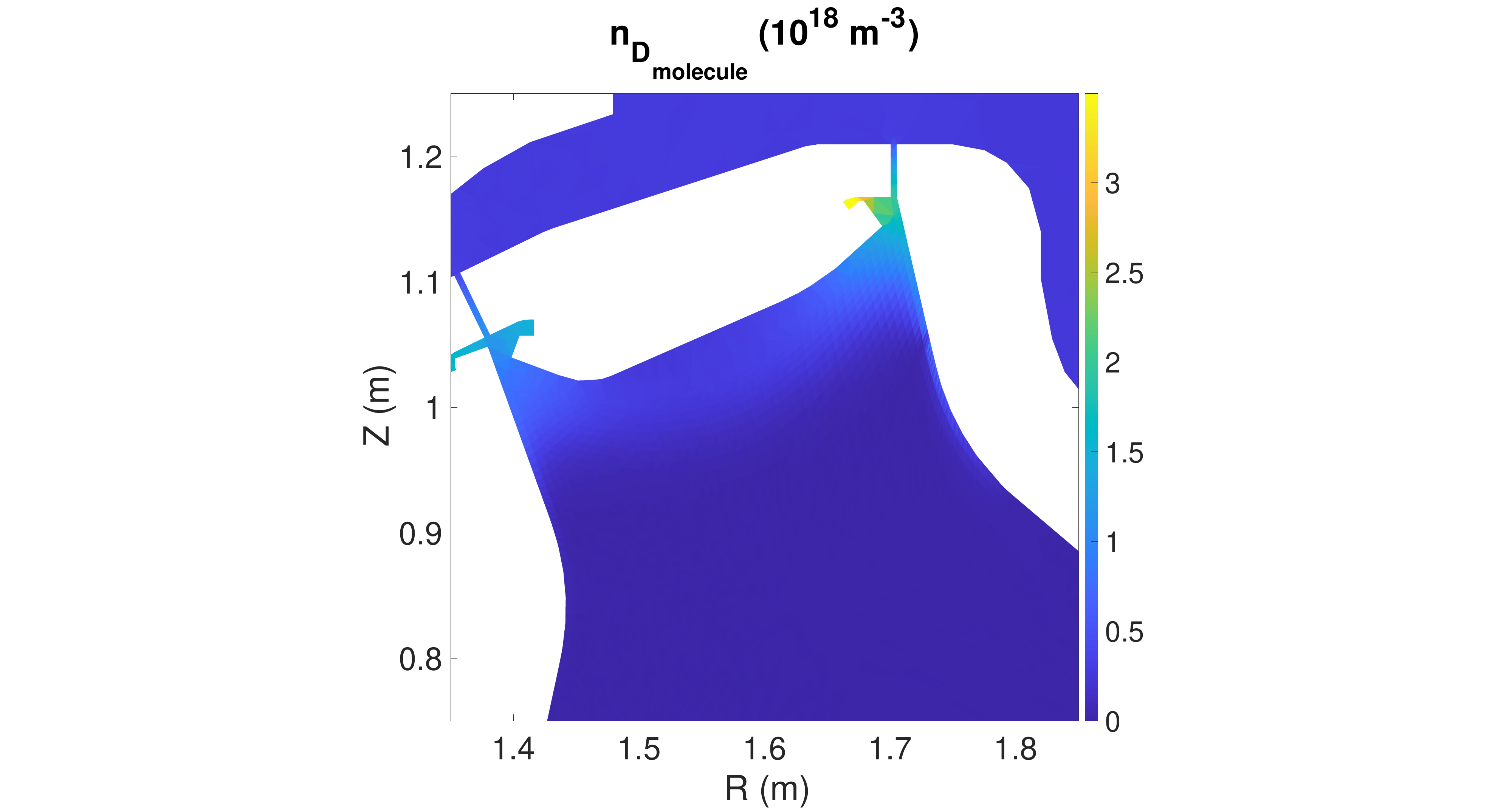}
		\caption{}
		\label{subfig:n_Dmol_D2_2_div}
	\end{subfigure}
	\begin{subfigure}{4cm}
		\centering
		\medskip
		\includegraphics[trim={15cm 0cm 15cm 0cm},clip,width=4cm]{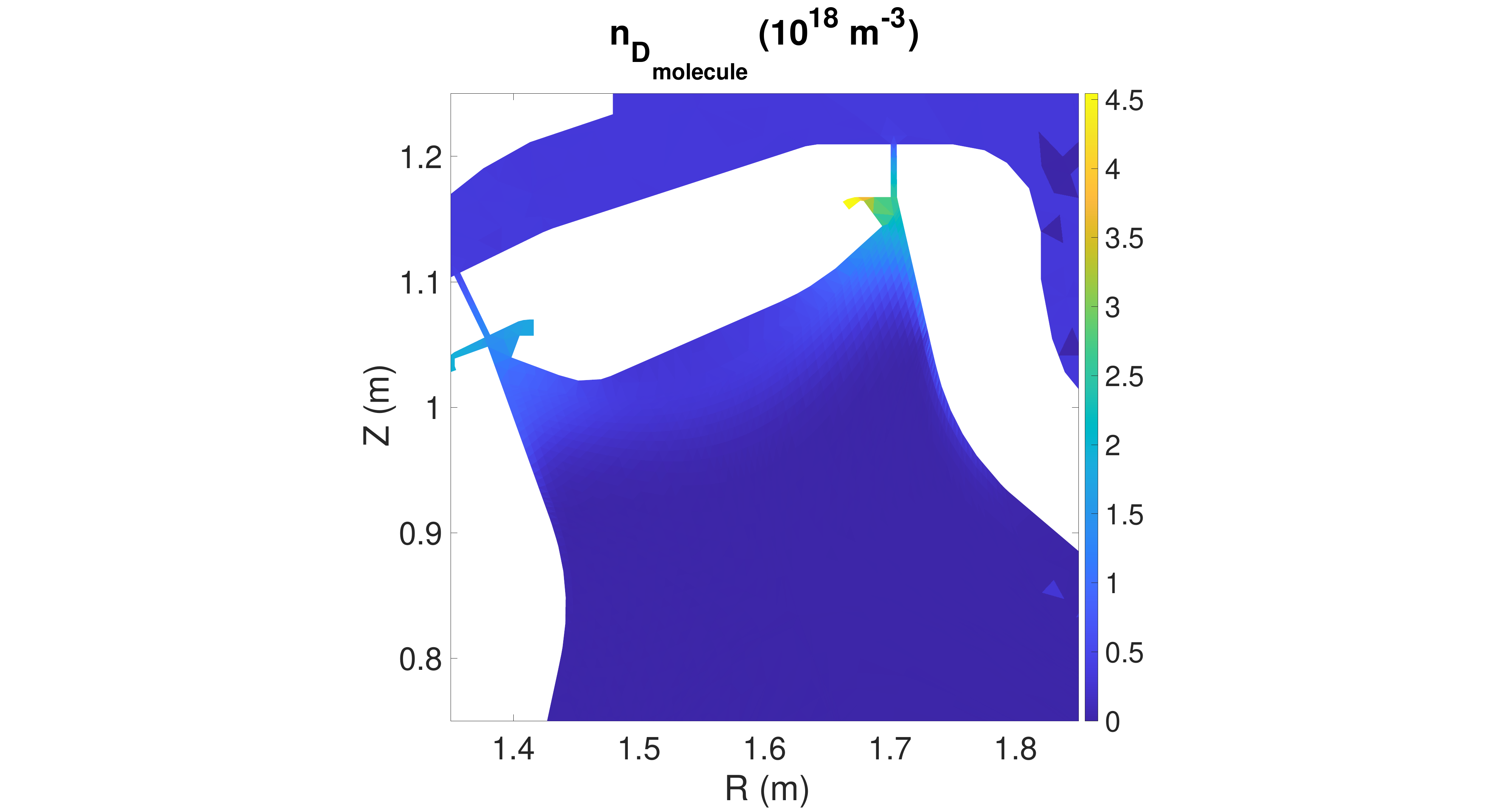}
		\caption{}
		\label{subfig:n_Dmol_Ne_1_div}
	\end{subfigure}
	\begin{subfigure}{4cm}
		\centering
		\medskip
		\includegraphics[trim={15cm 0cm 15cm 0cm},clip,width=4cm]{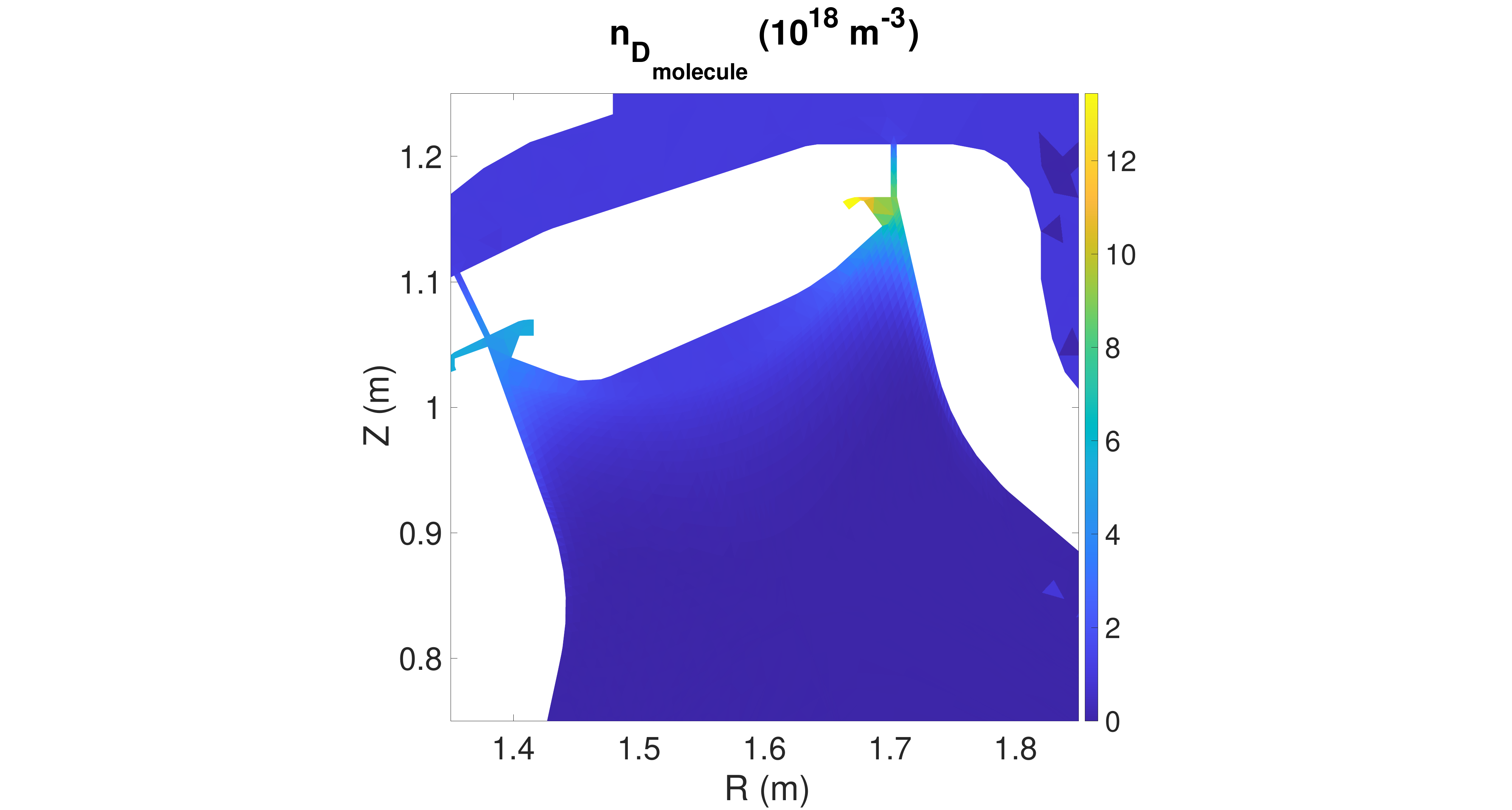}
		\caption{}
		\label{subfig:n_Dmol_Ne_2_div}
	\end{subfigure}
	\caption{The neutral deuterium molecule density in the "SOLPS D$_2$ 1" (a), "SOLPS D$_2$ 2" (b), "SOLPS Ne 1" (c)  and "SOLPS Ne 2" (d) simulations in the divertor region. Remark the difference in colorbar between the two density profiles.}
	\label{fig:n_Dmol}
\end{figure}

Where the atom density is similar for all simulations apart from "SOLPS Ne 2", the differences in the hydrogen molecular density are larger as can be seen in figure \ref{fig:n_Dmol}: the location of largest density stays similar, but the maximum one increases with a factor $\sim$4 between the "SOLPS D$_2$ 1" and "SOLPS Ne 2" simulations. Also remark that the maximum molecular density is always higher than the maximum atom one but is located further away from the strike line towards the subdivertor region where it cannot react with plasma anymore. Looking to the table of included neutral reactions (table \ref{tab:EIRENE_neutral_reactions}) it is concluded that the PWI is the driver for the deuterium molecule density.

The elastic collision reactions have a small impact due to the size of EAST. As the mean free path of neutral-neutral reactions is long in comparison to the tokamak size, they do not contribute much to the plasma physics \cite{boeyaert2022numerical}. Also the atomic and plasma particle and energy source terms originating from charge exchange reactions appeared to be very small in the studied simulations.

\begin{figure}
	\centering
	\medskip
	\includegraphics[height=5cm]{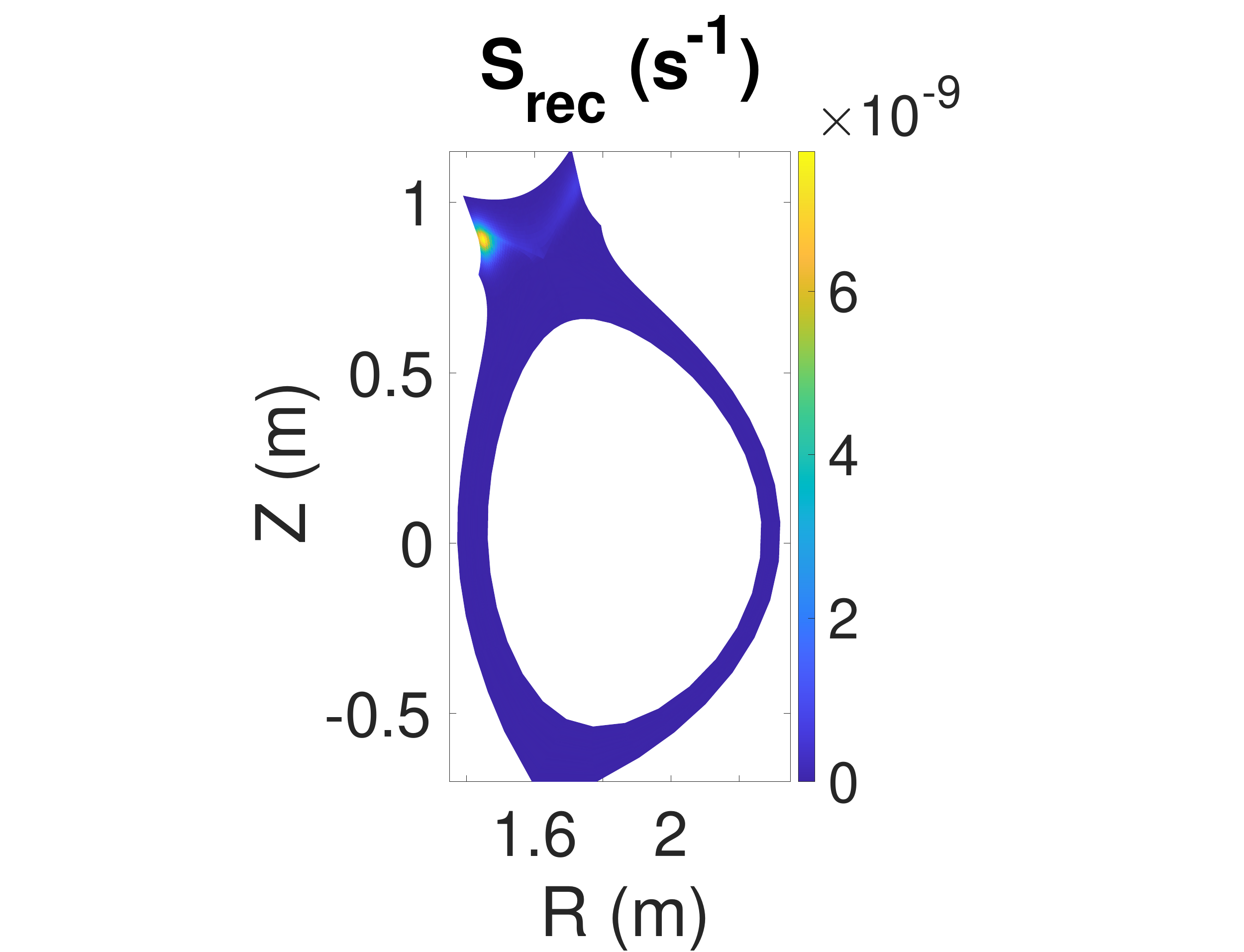}
	\caption{The neutral source rate of deuterium atoms in the "SOLPS Ne 2" simulation due to recombination.}
	\label{fig:recombination_neon}
\end{figure}

\begin{figure}
	\centering
	\begin{subfigure}{8cm}
		\centering
		\medskip
		\includegraphics[width=8cm]{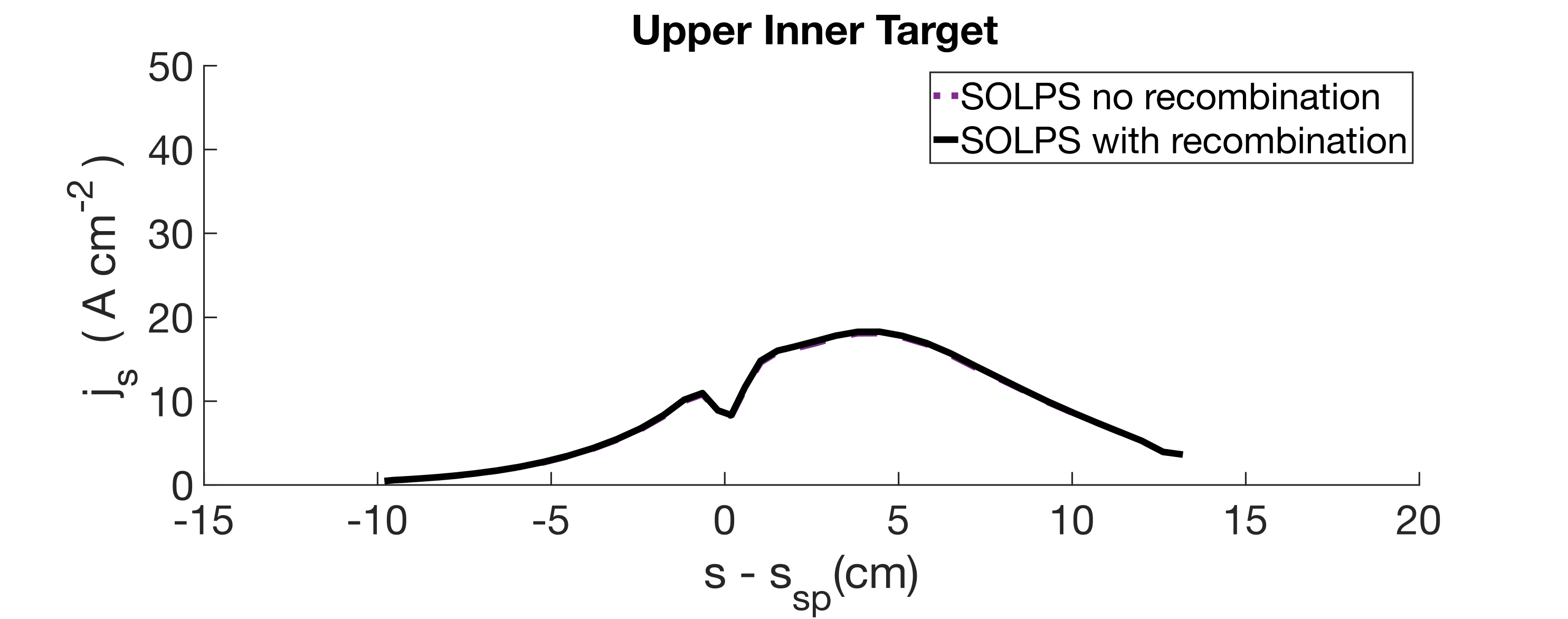}
		\caption{}
		\label{subfig:js_UIT_noRC}
	\end{subfigure}
	\begin{subfigure}{8cm}
		\centering
		\medskip
		\includegraphics[width=8cm]{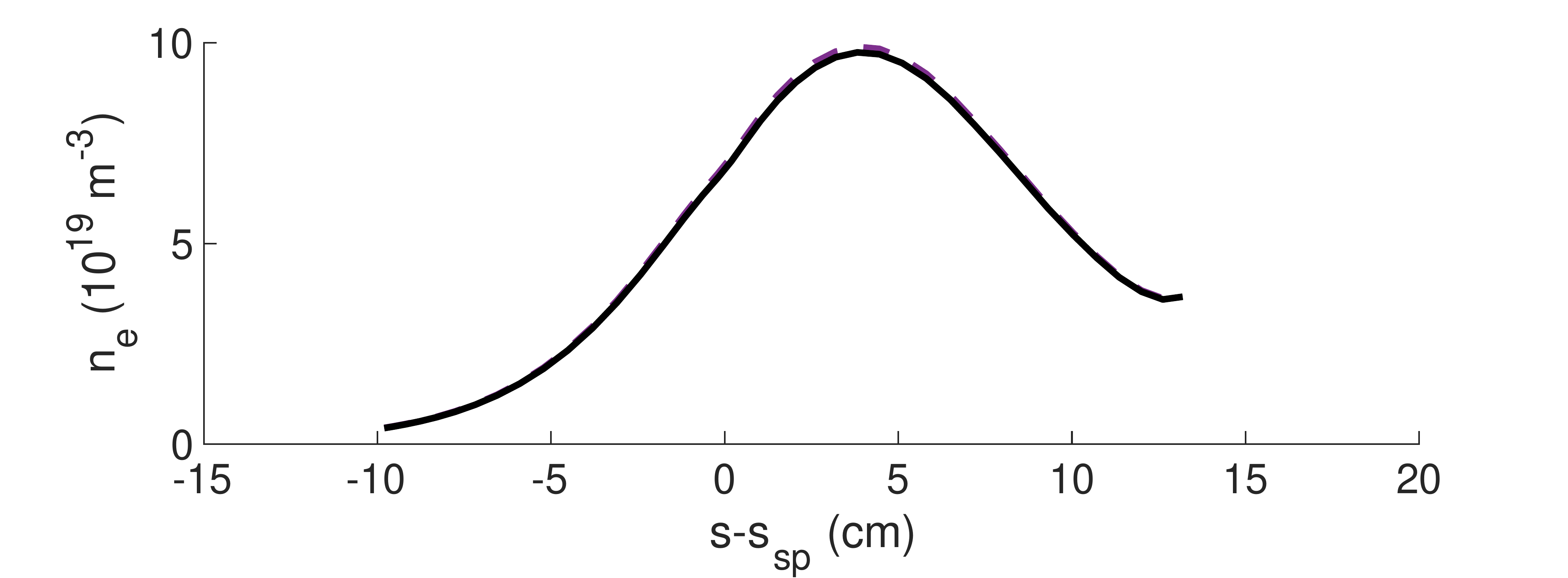}
		\caption{}
		\label{subfig:ne_UIT_noRC}
	\end{subfigure}
	\caption{The limited impact of the inclusion of recombination reactions on the $j_s$ (a) and $n_e$ profiles at the UIT.}
	\label{fig:UIT_noRC}
\end{figure}

The recombination of deuterium has a negligible impact, both in the purely deuterium and in the neon seeded simulation. This is shown for the particle sources for atomic deuterium due to recombination in the "SOLPS Ne 2" simulation in figure \ref{fig:recombination_neon}. Here it can be seen that recombination only plays a small role at the UIT. 
Comparison of the target profiles of a "SOLPS Ne 2" simulation with and without recombination switched on confirms the small effect of recombination as indicated in figure \ref{fig:UIT_noRC} for the $j_s$ and $n_e$ profiles at the UIT. This is in agreement with what can be expected from the target profiles of figures \ref{fig:DivLP_SOLPS_UOT} and \ref{fig:DivLP_SOLPS_UIT}: the lowest electron temperature is obtained at the UIT, and even this one is $\sim$ 1.8 eV. In figure \ref{fig:detachment_physics} it was indicated that temperatures below $\sim$ 1 eV are required to allow recombination to play an effective role. The atomic data of refs. \cite{reiter2000data,sawada1995effective} show that the cross-section for the density levels obtained in the "SOLPS Ne 2" simulation result in $\sim$ $\mathrm{5 \, 10^{-13} cm^{3}s^{-1}}$ where for an electron temperature of 0.1 eV, this would result in $\sim$ $\mathrm{10^{-9} cm^3s^{-1}}$.

The small effect of deuterium recombination is also confirmed by the low line radiation of deuterium. The recombination of neon in the "SOLPS Ne 1" and especially in the "SOLPS Ne 2" simulations, on the other hand, has a larger effect resulting in larger levels of line radiation in the vicinity of the targets where the temperature is low as shown in figures \ref{subfig:radiated_power_line_radiation_Ne1} and \ref{subfig:radiated_power_line_radiation_Ne2}. Due to the low density of neon in comparison with deuterium, their impact on the plasma profiles is, however, limited.

From the above analysis, it is clear that ionization is the driving reaction for the plasma-neutral interactions in the analyzed simulations. This also explains the significant impact of the neutral radiation on the total radiated power as shown in figure \ref{fig:radiated_power_neutral_radiation}.

\section{Discussion}
\label{sec:degree_of_detachment}

The main question to be answered in this work is the influence of neon on the plasma edge transport. Edge transport in SOLPS-ITER is dominated by four mechanisms: anomalous transport which is originating from turbulence, drifts, radiation, and neutral interactions \cite{schneider2006plasma}. These mechanisms are the drivers for the physics that particles can undergo and which was shown in figure \ref{fig:detachment_physics}. We now want to know the influence of neon on these mechanisms and on the physical processes.

\begin{figure}
	\centering
	\medskip
	\includegraphics[width=8cm]{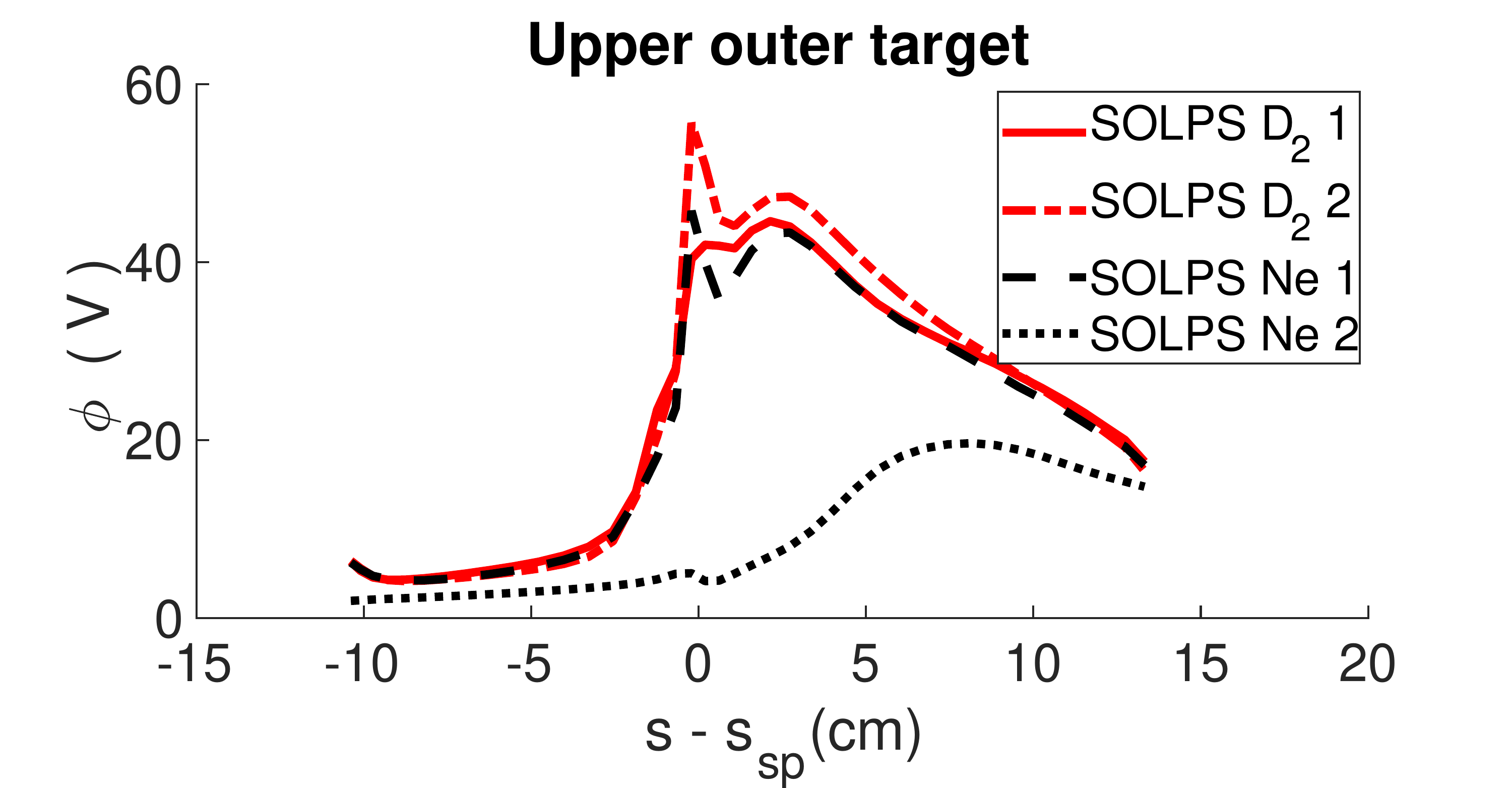}
	\caption{The potential at the UOT for all the performed SOLPS-ITER simulations.}
	\label{fig:po_UOT}
\end{figure}

Figure \ref{fig:DivLP_SOLPS_UOT} shows that only with modified anomalous transport in comparison with purely deuterium simulations it is possible to match the UOT profiles of the neon seeded experiment. For the performed neon simulation which matched best the experimental conditions at the UOT ("SOLPS Ne 2"), the transport was reduced in the SOL, but increased in the divertor and PFR regions. This increased transport in the divertor and PFR region causes a decrease in potential in the vicinity of the targets and especially around the separatrix as shown for the UOT in figure \ref{fig:po_UOT}. This makes that in the detached neon seeded simulation ("SOLPS Ne 2") the double peaking for the target profiles vanished.

\begin{figure}{6cm}
	\centering
	\medskip
	\includegraphics[clip,width=6cm]{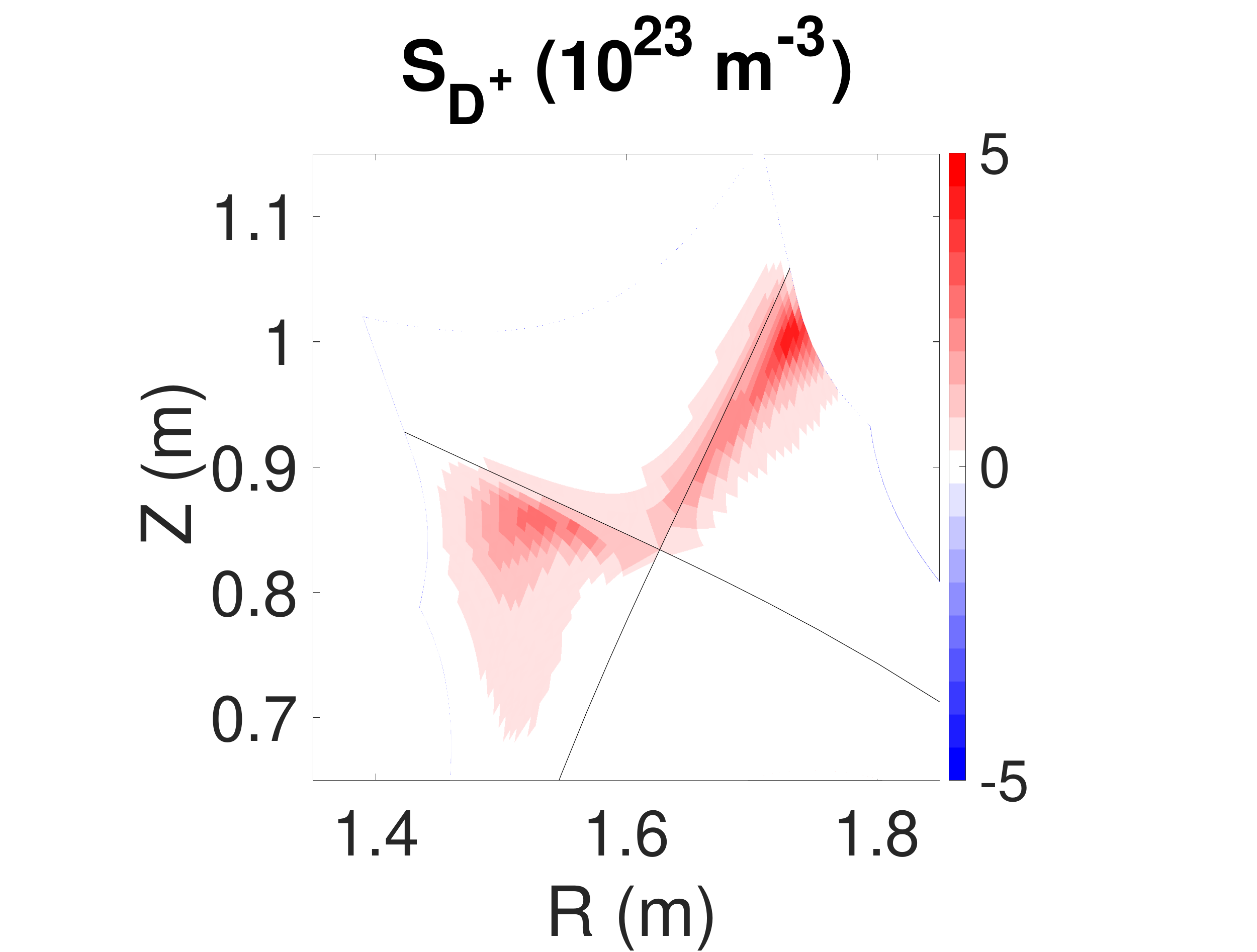}
	\caption{}
	\caption{The ionization source for D$^+$ in the vicinity of the divertor region in the "SOLPS Ne 2" simulation.}
	\label{fig:ionization_D_Ne2_detail}
\end{figure}

Apart from the divertor target profiles, the clearest influence of neon is observed in the radiation data. Due to a large increase in line radiation, the total radiation and in that way the energy losses the ions experience on their way towards the divertor target increases in comparison with the purely deuterium case. Comparing simulations "SOLPS D$_2$ 2" and "SOLPS Ne 2" on the other hand shows that the neutral radiation around the UIT is decreasing. This is caused by the reduced deuterium ionization which is clear from comparing figure \ref{fig:ionization_D_Ne2_detail} with figure \ref{subfig:ionization_D_D2_detail}.

To come back to figure \ref{fig:detachment_physics}, the neon ions radiate around the X-point and cause in that way additional loss terms to the energy equation. When looking to the outer target, for which the simulations are matched better with experimental data in comparison with the inner target, there will be a larger influence of the neutral friction as the deuterium ionization sources increase. Right in front of the target, the recombination of neon causes further radiation losses which are visible in the line radiation in figure \ref{subfig:radiated_power_line_radiation_Ne2}. The influence of deuterium recombination, on the other hand, is very small because of the larger temperatures.

To summarize, analysis of neon in the vicinity of the UOT shows that neon influences the radiated power fraction, decreases the deuterium ionization, increases the neutral friction, but does not manage to cause deuterium recombination.

\section{Summary and conclusion}

In the presented paper neon seeding experiments, performed at EAST in the campaign of 2019 in upper single null configuration, are taken as a starting point for SOLPS-ITER modeling to evaluate the effect of neon seeding on plasma edge transport in EAST. 

The performed SOLPS-ITER simulations have similar profiles as the experiments at the upper outer target and at the outer midplane. At the upper inner target, on the other hand, the simulated temperatures are an order of magnitude smaller than in the experiments as no varying anomalous transport between inner and outer divertor is used. Therefore, the focus of the performed analysis is on the neon effects at the midplane and outer divertor region. Different anomalous transport had to be used between the deuterium and the neon-seeded simulations to match the experimental data. This indicate a different influence of turbulence. The increased anomalous transport in the PFR and divertor region in combination with the large amounts of neon in the neon-seeded simulation, decrease the effect of drifts.

The ExB drifts explain the observed profiles at the targets, especially  the double peaking which is also observed in other devices, and the shift in peak value. Furthermore, drifts are also essential to determine the location where the ionization will take place - which is the most important neutral reaction in the performed analysis - and in that way determine the particle sources. Drifts increase the ionization sources on the inboard side due to an increased particle flow towards the inner target. 

By analyzing the behavior of the Ne ions, it is shown that Ne$^+$ leaks towards the core. This makes Ne-seeding experiments with sufficient Ne to bring the discharge into full detachment challenging and causes that higher order states of neon ionize and in that way radiate in the core.

Due to neon, a shift in radiated power from neutral radiation to line radiation is observed. On top, the radiated power fraction more than triples for the simulation which fits best to the experimental data. The analysis suggests that the presence of other radiative species besides Ne is needed to explain the detachment behavior of EAST-size devices.

At the upper outer target, there is a decrease in temperature and an increase in density, but the cross-section of the deuterium recombination reaction stays very low. At the inner target, the simulations experience a very small influence of recombination in the target profiles. 

This analysis shows that neon has a strong impact on the plasma edge transport, but that it remains difficult to confirm the simulated advantages with experiments.

\section*{Acknowledgements}
The authors would like to thank Niels Horsten and Petra Börner for their help with all EIRENE-related questions.

The authors would like to thank Xavier Bonnin for his help with SOLPS-ITER-related questions.

This work has partially been carried out within the framework of the EUROfusion Consortium, funded by the European Union via the Euratom Research and Training Programme (Grant Agreement No 101052200 — EUROfusion). Views and opinions expressed are however those of the author(s) only and do not necessarily reflect those of the European Union or the European Commission. Neither the European Union nor the European Commission can be held responsible for them.

This work was funded in parts by the 
College of Engineering at UW Madison, WI, USA

This work is supported by the National Key R\&D Program of China under Contract No.  2024YFE03270700,  and program of National Natural Science Foundation of China under Contract No. 12305250.

The SOLPS-ITER simulations were performed at the Marconi supercomputer from the National Supercomputing Consortium CINECA.

\bibliography{papers_assessmentNe}

\end{document}